\documentclass[fleqn,usenatbib]{mnras}

\usepackage{newtxtext,newtxmath}

\usepackage[T1]{fontenc}

\DeclareRobustCommand{\VAN}[3]{#2}
\let\VANthebibliography\thebibliography
\def\thebibliography{\DeclareRobustCommand{\VAN}[3]{##3}\VANthebibliography}

\usepackage{subcaption}
\usepackage{graphicx}	
\usepackage{amsmath}	
\usepackage{tikz}
\usepackage{blindtext}
\usepackage{placeins}
\usepackage{xcolor}

\usepackage{hyperref}

\usepackage{booktabs,array}
\usepackage{supertabular}
\usepackage{threeparttable}
\newcount\rowc

\makeatletter
\def\ttabular{%
\hbox\bgroup
\let\\\cr
\def\rulea{\ifnum\rowc=\@ne \hrule height 0pt \fi}
\def\ruleb{
\ifnum\rowc=1\hrule height 1pt \else
\ifnum\rowc=3\hrule height 0.5pt \else
\ifnum\rowc=4\hrule height 0.5pt \else
\ifnum\rowc=5\hrule height 0.5pt \else
\hrule height 0pt \fi\fi\fi\fi}
\valign\bgroup
\global\rowc\@ne
\rulea
\hbox to 10em{\strut \hfill##\hfill}%
\ruleb
&&%
\global\advance\rowc\@ne
\hbox to 10em{\strut\hfill##\hfill}%
\ruleb
\cr}
\def\endttabular{%
\crcr\egroup\egroup}

\title[Measuring Glitch Recoveries]{Measuring glitch recoveries and braking indices with Bayesian model selection}

\author[Y. Liu et al.]{Y. Liu,$^{1}$\thanks{E-mail: yang.liu-50@postgrad.manchester.ac.uk}
M. J. Keith,$^{1}$\thanks{E-mail: michael.keith@manchester.ac.uk}
D. Antonopoulou,$^{1}$
P. Weltevrede,$^{1}$
B. Shaw,$^{1}$
B. W. Stappers,$^{1}$
\newauthor A. G. Lyne,$^{1}$
M. B. Mickaliger$^{1}$ and A. Basu$^{1}$
\\
$^{1}$Jodrell Bank Centre for Astrophysics, Department of Physics and Astronomy, The University of Manchester, Manchester M13 9PL, UK\\
}

\date{Accepted XXX. Received YYY; in original form ZZZ}

\pubyear{2024}

\begin{document}
\label{firstpage}
\pagerange{\pageref{firstpage}--\pageref{lastpage}}
\maketitle

\begin{abstract}
For a selection of 35 pulsars with large spin-up glitches ($\Delta{\nu}/\nu\geq10^{-6}$), which are monitored by the Jodrell Bank Observatory, we analyse 157 glitches and their recoveries. All parameters are measured consistently and we choose the best model to describe the post-glitch recovery based on Bayesian evidence. We present updated glitch epochs, sizes, changes of spin down rate, exponentially recovering components (amplitude and corresponding timescale) when present, as well as pulsars' second frequency derivatives and their glitch associated changes if detected. We discuss the different observed styles of post-glitch recovery as well as some particularly interesting sources. Several correlations are revealed between glitch parameters and pulsar spin parameters, including a very strong correlation between a pulsar's interglitch $|\ddot{\nu}|$ and $\dot{\nu}$, as well as between the glitch-induced spin-down rate change $\Delta\dot{\nu}_{\rm p}$ that does not relax exponentially and $\dot{\nu}$. We find that the ratio $\left|\Delta \dot{\nu}_{\mathrm{p}}/\ddot{\nu}\right|$ can be used as an estimate of glitch recurrence times, especially for those pulsars for which there are indications of a characteristic glitch size and interglitch waiting time. We calculate the interglitch braking index $n$ and find that pulsars with large glitches typically have $n$ greater than $3$, suggesting that internal torques dominate the rotational evolution between glitches. The external torque, e.g. from electromagnetic dipole radiation, could dominate the observed $\ddot{\nu}$ for the youngest pulsars ($\lesssim10^{4}\;\mathrm{yr}$), which  may be expected to display $n\sim3$. 

\end{abstract}

\begin{keywords}
stars: neutron -- pulsars: general -- methods: statistical -- methods: data analysis
\end{keywords}



\section{Introduction}\label{sec: intro}

Pulsars, rapidly rotating neutron stars, are detected in various wavelengths through their pulsed emission, which is used to measure their relatively stable rotational periods \citep{gold_rotating_1968}. However, the periods of pulsars are still changing to some extent. The angular momentum of a pulsar is decreasing through the process of magnetic dipole radiation and particle outflow \citep{davies_changing_1969, goldreich_pulsar_1969}. As a consequence, the rotation of the pulsar gradually slows down. Apart from this slowdown due to magnetic dipole braking, other processes, such as glitches, may also affect the rotational status of the pulsar \citep{radhakrishnan_detection_1969, reichley_observed_1969}. In addition, stochastic variations -- another type of timing irregularity known as timing noise, contribute noise in the pulse arrival times \citep{hobbs_analysis_2010}.

Glitches are positive changes in spin frequency $\nu$, generally small ($10^{-12}\lesssim\Delta\nu/\nu\lesssim10^{-5}$) and happen abruptly. They are commonly attributed to the transfer of angular momentum between two distinct components of the pulsar, i.e., the star's crust and the interior neutron superfluid \citep{haskell_models_2015, antonopoulou_pulsar_2022}. The rotation of this superfluid is supported by vortices of quantised circulation whose density defines the superfluid rotation rate. Whereas the neutron star's solid crust and charged components of the core are coupled and spin down together under the external electromagnetic torque, the superfluid can follow only if vortices migrate outwards and their density reduces. As first proposed by \citet{anderson_pulsar_1975}, vortices can become pinned in some stellar regions, i.e. immobilised by their interactions with other components such as the inner crust nuclei. Where this happens, the superfluid cannot lose angular momentum at the same rate as the rest of the star and hence acts as a hidden angular momentum reservoir. Glitches are the result of a rapid exchange of angular momentum between this reservoir and the observed crust, for example due to a catastrophic unpinning event which enables a fast decrease of vortex density. The glitch spin-up sizes and time intervals between consecutive glitches can be used to infer the fractional moment of inertia of the superfluid component that causes a glitch \citep{link_pulsar_1999, andersson_pulsar_2012}.

Following a glitch, the spin-down rate increases, driving the spin frequency towards the pre-glitch values. Many glitches show some transient recovery, often described by one or more exponential terms \citep{lyne_massive_1987, basu_observed_2020}. The post-glitch relaxation to the pre-glitch rotational state is often not complete; most commonly, persisting changes, $\Delta{\dot{\nu}}_{\mathrm{p}}$, are observed \citep{antonopoulou_pulsar_2018}. Additionally, interglitch values of the frequency second order derivative $\ddot{\nu}$ often lead to much higher braking indices ($n=\nu\ddot{\nu}/\dot{\nu}^2$) than those expected from known spin-down mechanisms. Nonetheless, the cumulative effect of the permanent glitch changes, $\Delta{\dot{\nu}}_{\mathrm{p}}$, counteracts the high interglitch $\ddot{\nu}$ so that the long term evolution of glitching pulsars is typically closer to a canonical $n_{\mathrm{b}}=3$ braking \citep{shapiro_black_1983}, often with $n\lesssim2$ \citep{lyne_very_1996, espinoza_new_2017}.
 
In order to undertake statistical studies of glitches, it is vital to have  consistent glitch measurements. Large catalogues have published the parameters of hundreds of glitches, but often lack detailed measurements of the transient recoveries \citep{espinoza_study_2011, basu_jodrell_2022}. \citet{yu_detection_2013} presented measurements for the exponential transient recoveries, but their work does not account for the effect of timing noise. Recently \citet{lower_impact_2021} applied Bayesian techniques to measure the properties of glitches, including exponential recoveries, whilst also modelling timing noise. The glitch parameters were determined in a systematic fashion, and Bayesian inference was used to determine if exponential terms should be included in the model of the post-glitch recovery.

The Jodrell Bank observatory (JBO) routinely monitors the rotation of more than 800 pulsars and records all detected glitches \citep{basu_jodrell_2022}. In this paper we adopt a similar technique to \citet{lower_impact_2021} and apply it to JBO-monitored pulsars. Since it is difficult to distinguish transient recoveries in small glitches, we implement a cut off on the glitch size when selecting our sample. We focus on 35 pulsars which displayed large glitches, $\Delta{\nu}/{\nu}\geq10^{-6}$, during our observation span. These sources have a total of 157 detected glitches, which we analyse together with their respective recoveries. The paper is structured as follows: in section \ref{sec: observ} we briefly introduce the observational program at the JBO. In section \ref{sec: method}, we explain the method we adopted for measuring pulsar glitch parameters, whilst the fitting results are discussed in section \ref{sec: result}. We finally present a further analysis for the statistics of glitch parameters and their correlations in section \ref{sec: analysis}, followed by a summary of our findings in section \ref{sec: conclu}.

\section{Observations}\label{sec: observ}
The 76-m Lovell telescope at the Jodrell Bank Observatory (JBO) regularly observes around 800 pulsars. Typically, observations have a duration of $\sim6 \,\mathrm{min}$ depending on the pulsar’s flux density, and are divided into a series of subintegrations lasting from 1 to 3 minutes \citep{hobbs_long-term_2004}. Data before year 2009 are recorded with an analogue filterbank (AFB), but in 2009 this was updated to a digital filterbank (DFB). The bandwidth for the AFB observations is 32~MHz, centered on 1400 MHz. A small number of observations are carried out at 500, 600 or 925 MHz. A bandwidth of 400 MHz, centred on 1520 MHz was used for DFB observations \citep{perera_international_2019}.

In each observation, \textsc{psrchive} \citep{van_straten_pulsar_2012} is used to average the remaining sub-integrations, frequency channels, and polarisations to form a single integrated profile. The pulse time-of-arrival (ToA) can be determined by cross-correlating the integrated profile with a noise-free template profile representing an idealized realization of the expected pulse profile in the time domain. The obtained ToAs are subsequently adjusted to the approximately inertial Solar system barycenter, employing the DE436 planetary ephemeris\footnote{https://naif.jpl.nasa.gov/pub/naif/JUNO/kernels/spk/de436s.bsp.lbl}. This correction results in Barycentric arrival times, utilized in the subsequent timing analysis discussed below. Details on the observing setup and the JBO glitch database can be found in \citet{basu_jodrell_2022}. 

The collective dataset of glitches detected by JBO was presented in \citet{basu_jodrell_2022}. It was shown therein that the distribution of all glitch sizes $\Delta\nu$ can be modelled as two Gaussian components: a relatively narrow one centered at large glitch sizes, and a wide component spanning from small to intermediate/large glitches. For this work we focus on pulsars that have at least one large glitch, as these tend to have clearly detectable transient components in their recoveries. We therefore set a threshold based on the Gaussian mixture model of \citet{basu_jodrell_2022} and select a subset of 35 pulsars that have at least one glitch with $\Delta{\nu}/{\nu}\geq10^{-6}$. Details for the observational period and the total number of glitches for these pulsars are given in Table~\ref{tab: toa}. Of these glitches, a subset -- consisting primarily of intermediate to large size glitches -- was used to explore post-glitch recovery signatures. 

\begin{table}
\begin{center}
\centering
\caption{The pulsars used in this work, the MJD range over which they were observed at JBO, number of ToAs in this range, and the total number of glitches that occurred in this period. In the last column we show the number of glitches for which we analysed the post-glitch recovery.}
\label{tab: toa}
\begin{tabular}{lrrrrr}
\hline
\hline
Pulsar name & Start &  Finish &  No. of &  No. of &  Glitches \\
PSR & (MJD) & (MJD) & ToAs & glitches & modelled \\
\hline
J0205$+$6449  &  53725 &   59241 &    778 &               7 &               5 \\
B0355$+$54    &  41807 &   48717 &    271 &               2 &               2 \\
J0611$+$1436  &  54516 &   58500 &    298 &               1 &               1 \\
J0631$+$1036  &  49994 &   59331 &   1641 &              17 &               5 \\
J0729$-$1448  &  50758 &   58526 &    635 &               6 &               1 \\
B0919$+$06    &  43586 &   58523 &   1335 &               1 &               1 \\
B1727$-$33    &  48416 &   59153 &    913 &               3 &               2 \\
B1737$-$30    &  46300 &   58533 &   1619 &              36 &              11 \\
J1737$-$3137  &  50758 &   59335 &    243 &               6 &               3 \\
J1740$+$1000  &  54224 &   59565 &    883 &               2 &               1 \\
B1754$-$24    &  47156 &   59221 &    352 &               1 &               1 \\
B1757$-$24    &  48329 &   58528 &   1103 &               6 &               5 \\
B1800$-$21    &  47364 &   59240 &   1776 &               6 &               5 \\
J1806$-$2125  &  50820 &   58500 &    329 &               1 &               1 \\
J1809$-$1917  &  50820 &   59236 &    311 &               1 &               1 \\
B1821$-$11    &  46612 &   58435 &    358 &               1 &               1 \\
B1823$-$13    &  46612 &   59518 &   1796 &               6 &               4 \\
B1830$-$08    &  46449 &   59224 &    978 &               3 &               1 \\
J1837$-$0604  &  51153 &   58494 &    264 &               3 &               1 \\
J1841$-$0345  &  51507 &   59241 &    431 &               1 &               1 \\
J1841$-$0524  &  52607 &   59248 &    175 &               8 &               5 \\
J1842$+$0257  &  52608 &   58528 &    188 &               1 &               1 \\
J1850$-$0026  &  54028 &   58462 &    286 &               4 &               1 \\
B1853$+$01    &  47576 &   59236 &    825 &               2 &               2 \\
J1856$+$0245  &  53877 &   59170 &    443 &               1 &               1 \\
B1859$+$01    &  46724 &   59141 &    252 &               3 &               1 \\
J1907$+$0631  &  55729 &   58485 &    165 &               2 &               2 \\
J1909$+$0749  &  53692 &   58923 &    409 &               1 &               1 \\
J1909$+$0912  &  51252 &   59227 &    224 &               2 &               1 \\
J1921$+$0812  &  53083 &   58495 &    279 &               1 &               1 \\
B1930$+$22    &  46860 &   58503 &   1150 &               1 &               1 \\
B1951$+$32    &  47028 &   59191 &   1535 &               6 &               1 \\
J2021$+$3651  &  52589 &   59243 &    370 &               5 &               5 \\
J2229$+$6114  &  51976 &   59720 &   1120 &               9 &               8 \\
B2334$+$61    &  46717 &   58502 &    989 &               1 &               1 \\
\hline
\end{tabular}
\end{center}
\end{table}

\section{Methodology}\label{sec: method}
In pulsar timing, we use the truncated Taylor series expansion to model the rotational phase as:
\begin{equation}\label{eq: phase_model}
\phi(t)=\phi_{0}+\nu_{0}\left(t-t_{0}\right)+\frac{1}{2} \dot{\nu}_{0}\left(t-t_{0}\right)^2+\frac{1}{6} \ddot{\nu}_0\left(t-t_{0}\right)^3
\end{equation}
where $\phi_{0}$, $\nu_{0}$, $\dot{\nu}_{0}$, and $\ddot{\nu}_{0}$ are the phase, rotation frequency, and its time derivatives at some reference epoch $t_{0}$. This timing model approximately describes the rotational phase of a pulsar over the full data span. The presence of glitches and timing noise introduces deviations from Equation~\ref{eq: phase_model}. For each pulsar in our sample, an initial timing solution and preliminary estimates of glitch parameters are taken from \citet{basu_jodrell_2022}. Phase residuals are calculated as the differences between measured ToAs and those predicted by that model. Residuals are minimised by fitting for the rotational parameters of pulsars as well as any astrometric, interstellar medium (ISM), and (where applicable) binary parameters. 

In this section, we introduce the glitch model and noise models we used to describe pulsar rotation. To investigate the post-glitch recoveries, we used Bayesian model selection to compare evidence of various models, such as models with a varying number of exponential relaxation terms.

\subsection{Bayesian model selection}\label{sub: selection}
Bayesian inference offers a consistent approach to the estimation of a set of parameters $\Theta$ describing a model or hypothesis $H$, given the data $D$. The core of Bayesian analysis is Bayes’ theorem, which states that
\begin{equation}\label{eq: bayes_theorem}
P(\Theta \mid D, H)=\frac{P(D \mid \Theta, H) P(\Theta \mid H)}{P(D \mid H)}
\end{equation}
where $P(\Theta \mid D, H)$ is the posterior probability distribution of the parameters, $P(D \mid \Theta, H)$ is the likelihood, $P(\Theta \mid H)$ is the prior probability distribution, and $P(D \mid H) \equiv Z$ is the Bayesian evidence. (See \citet{antoniadis_second_2023} for a detailed explanation on Bayesian inference.)

Our aim is to work out the most plausible glitch and post-glitch recovery model (see equation \ref{eq: glitch_phase}). We tested for post-glitch exponential relaxation and glitch-induced permanent step changes in frequency second order derivative, and measure the posteriors of parameters. The 7 different post-glitch recovery models explored are summarised in Table \ref{tab: models}.
To decide on the appropriate model we adopt a slightly modified version of Jeffreys' scale \citep{jeffreys_theory_1939}, and use the logarithm of Bayes factor (ratio of the Bayesian evidence between model 2 and model 1)
\begin{equation}\label{eq: log_factor}
\ln{B_{21}} = \ln{Z_{2}} - \ln{Z_{1}}
\end{equation}
to access the strength of evidence. A value of $\ln{B_{21}} < 2.5$ is interpreted as model 2 having insufficient evidence over model 1, $2.5 \leq \ln{B_{21}} < 5.0$ is regarded as moderate evidence favouring model 2 over model 1, $5.0 \leq \ln{B_{21}} < 10.0$ as strong evidence preferring model 2, and $\ln{B_{21}} \geq 10.0$ as decisive evidence for model 2. Since the prior odds ratio is unity among models of a specific glitch, the posterior odds ratio is equivalent to the Bayes factor. For example, a logarithmic Bayes factor $\ln{B_{21}}\geq2.5$ means the odds ratio of model 2 over model 1 is larger than $e^{2.5}\simeq12$. Note that only the ratios of the evidence between models matter, so one model might have much larger evidence than the other (for example a double recovery model compared to a model without any recovery terms), but both models may be incorrect.

In order to generate a single result for each glitch/pulsar, and to make plots, we select a single `best' model for each glitch. If there is a single model that has at least moderate evidence over all other models then we select this as the best model. However, in the case where several models have similar evidence, we inspect all models with logarithmic Bayes factor $\ln{B_{21}}>-2.5$ and choose from amongst these. In principle the statistical modelling suggests all these models are plausible, and we usually select the simplest model (the model labelled with smallest number), but in two cases (PSR J0631$+$1036 and the 1st glitch of PSR J1907$+$0631) we selected a different model based on visual inspection of the residuals. These special cases will be discussed in Section \ref{sub: view}.

\subsection{Glitch Model}\label{sub: glitch}
We can describe the additional pulse phase change induced by a glitch (only applied to post-glitch data) by:
\begin{equation}\label{eq: glitch_phase}
\phi_{\mathrm{g}}=\Delta \nu_{\mathrm{p}} \Delta t+\frac{1}{2} \Delta \dot{\nu}_{\mathrm{p}}\Delta t^{2}+\frac{1}{6} \Delta \ddot{\nu}_{\mathrm{p}}\Delta t^{3}+\sum_{i=1}^{N_{\mathrm{e}}\leq3} \left(1-e^{-\Delta t / \tau_{i}}\right) \Delta \nu_{\mathrm{d}_{i}}\tau_{i}
\end{equation}
where $\Delta t=t-t_{\mathrm{g}} \geq 0$, and $t_{\mathrm{g}}$ is the glitch epoch. Besides permanent changes in rotational parameters (denoted with index $\mathrm{p}$), we assume that any transient post-glitch components can be described as a sum of  $N_{\mathrm{e}}$ exponentials, with amplitudes $\Delta \nu_{\mathrm{d}_{i}}$ and timescales $\tau_{i}$. In principle, one can include the quantity $\Delta \phi \equiv \phi_{0} - \phi'_{0}$ to account for any discontinuity in phase due to uncertainties in the glitch epoch. However, in this work, we stipulate $\Delta \phi = 0$ to avoid the strong covariance between glitch epoch $t_{g}$ and $\Delta \phi$. From the glitch model in Equation~\ref{eq: glitch_phase}, the change in rotation frequency at the time of a glitch is $\Delta \nu = \Delta \nu_{\mathrm{p}} + \sum_i \Delta \nu_{\mathrm{d}_{i}}$ and the change in its first derivative is $\Delta \dot{\nu} = \Delta \dot{\nu}_{\mathrm{p}} - \sum_i \Delta \nu_{\mathrm{d}_{i}}/\tau_i$.

\begin{table*}
\begin{center}
\centering
\caption{The details of post-glitch models we fitted for in model selection.}
\label{tab: models}
\renewcommand{\arraystretch}{1.5}
\begin{tabular}{llll}
\hline
\hline
Model & No. of exponentials & Fit for $\Delta \ddot{\nu}_{\mathrm{p}}$ & Expression of $\phi_{\mathrm{g}}$ \\
\hline
1 & 0 & No & $\Delta \nu_{\mathrm{p}} \Delta t+\frac{1}{2} \Delta \dot{\nu}_{\mathrm{p}}\Delta t^{2}$ \\
2 & 0 & Yes & $\Delta \nu_{\mathrm{p}} \Delta t+\frac{1}{2} \Delta \dot{\nu}_{\mathrm{p}}\Delta t^{2}+\frac{1}{6} \Delta \ddot{\nu}_{\mathrm{p}}\Delta t^{3}$ \\
3 & 1 & No & $\Delta \nu_{\mathrm{p}} \Delta t+\frac{1}{2} \Delta \dot{\nu}_{\mathrm{p}}\Delta t^{2}+\left(1-e^{-\Delta t / \tau_{1}}\right) \Delta \nu_{\mathrm{d}_{1}}\tau_{1}$ \\
4 & 1 & Yes & $\Delta \nu_{\mathrm{p}} \Delta t+\frac{1}{2} \Delta \dot{\nu}_{\mathrm{p}}\Delta t^{2}+\frac{1}{6} \Delta \ddot{\nu}_{\mathrm{p}}\Delta t^{3}+\left(1-e^{-\Delta t / \tau_{1}}\right) \Delta \nu_{\mathrm{d}_{1}}\tau_{1}$ \\
5 & 2 & No & $\Delta \nu_{\mathrm{p}} \Delta t+\frac{1}{2} \Delta \dot{\nu}_{\mathrm{p}}\Delta t^{2}+\left(1-e^{-\Delta t / \tau_{1}}\right) \Delta \nu_{\mathrm{d}_{1}}\tau_{1}+\left(1-e^{-\Delta t / \tau_{2}}\right) \Delta \nu_{\mathrm{d}_{2}}\tau_{2}$\\
6 & 2 & Yes & $\Delta \nu_{\mathrm{p}} \Delta t+\frac{1}{2} \Delta \dot{\nu}_{\mathrm{p}}\Delta t^{2}+\frac{1}{6} \Delta \ddot{\nu}_{\mathrm{p}}\Delta t^{3}+\left(1-e^{-\Delta t / \tau_{1}}\right) \Delta \nu_{\mathrm{d}_{1}}\tau_{1}+\left(1-e^{-\Delta t / \tau_{2}}\right) \Delta \nu_{\mathrm{d}_{2}}\tau_{2}$ \\
7 & 3 & No & $\Delta \nu_{\mathrm{p}} \Delta t+\frac{1}{2} \Delta \dot{\nu}_{\mathrm{p}}\Delta t^{2}+\left(1-e^{-\Delta t / \tau_{1}}\right) \Delta \nu_{\mathrm{d}_{1}}\tau_{1}+\left(1-e^{-\Delta t / \tau_{2}}\right) \Delta \nu_{\mathrm{d}_{2}}\tau_{2}+\left(1-e^{-\Delta t / \tau_{3}}\right) \Delta \nu_{\mathrm{d}_{3}}\tau_{3}$ \\
\hline
\end{tabular}
\end{center}
\end{table*}

Permanent step changes $\Delta \nu_{\mathrm{p}}$ and $\Delta \dot{\nu}_{\mathrm{p}}$ are fitted for in all our glitch models of Table~\ref{tab: models}, whilst the inclusion of other terms varies. No exponential relaxation term is considered in Model 1 and 2. We fit for one exponential relaxation term in Model 3 and 4, two exponential relaxation terms in Model 5 and 6, and three exponential relaxation terms in Model 7. A permanent change in the second order frequency derivative $\Delta \ddot{\nu}_{\mathrm{p}}$ is fitted for in Model 2, 4, and 6. We always consider models 1-4, for every glitch. When we find evidence for exponential relaxation, then we also test for models 5 and 6. In rare cases when there is evidence in favour of multiple exponential recoveries, Model 7 is tested for as well.

\subsection{Noise Model}\label{sub: noise}
In addition to the deterministic timing model (equations \ref{eq: phase_model} and \ref{eq: glitch_phase}) and the time-uncorrelated instrumental noise, stochastic processes also contribute additional time-correlated noise in the observed ToAs \citep{cordes_measurement_2010}. In this work, we adopt the noise model described in \citet{lentati_hyper-efficient_2013} and solve for both the pulsar timing model, white noise parameters, and a stochastic Gaussian-process noise model simultaneously. 

For the time-correlated timing noise, we use a Fourier basis Gaussian process, where the power spectral density (PSD) at a frequency $f$ is given by
\begin{equation}\label{eq: red_enterprise}
P(f)=\frac{A^{2}}{12 \pi^{2}}\left(\frac{f}{1 \mathrm{yr}^{-1}}\right)^{-\gamma} \mathrm{yr}^{3},
\end{equation}
where A is a dimensionless amplitude \citep{lentati_hyper-efficient_2013}. White noise is modelled such that a ToA with error $\sigma_{i}$ is weighted as if it had an error, 
\begin{equation}\label{eq: white_parameters}
\hat{\sigma}_{i}= \sqrt{\left(\mathrm{EFAC} \,\sigma_{i}\right)^{2}+\mathrm{EQUAD}^{2}},
\end{equation}
where EFAC and EQUAD represent a scaling and quadrature addition term \citep{lentati_temponest_2014}. EFAC and EQUAD values are estimated for the AFB and DFB data separately as the systematic noise processes may be different for the higher resolution DFB data.

\subsection{Implementation}\label{sub: implementation}
In order to fit for the glitch parameters we have added them to \textsc{run\_enterprise} \citep{run_enterprise}, a single pulsar Bayesian toolkit based on the \textsc{enterprise} framework \citep{ellis_enterprise_2019}, which can also simultaneously fit the noise model parameters. We adopt \textsc{multinest} \citep{feroz_multinest_2009} as the primary solver for sampling and calculating Bayesian evidence. 

To do the fitting, prior ranges for the fitting parameters need to be defined. We find that a uniform prior in $\Delta\nu_{\mathrm{p}}$ and $\Delta\dot{\nu}_{\mathrm{p}}$ results in high covariance among $\Delta\nu_{\mathrm{p}}$,  $\Delta\dot{\nu}_{\mathrm{p}}$, and $\Delta\nu_{\mathrm{d}}$. Further, the majority of the parameter space yields phase predictions that are extremely divergent from the observed residuals, leading to numerical instabilities in the fitting, particularly as the time-correlated noise component tries to model the resulting residuals. Therefore, in order that the priors include only phase models sufficiently close to the observed residuals, we adopt an alternative parameterization for the glitches, replacing $\Delta\nu_{\mathrm{p}}$ and $\Delta \dot{\nu}_{\mathrm{p}}$ with $\Delta\nu$, the instantaneous change in the spin frequency, and $\Delta\nu_{\scriptscriptstyle\Delta T}$, the additional glitch-induced change in spin frequency between the moment after the glitch and a time $\Delta T$ later. The choice of $\Delta T$ is arbitrary, but should be chosen such that the spin frequency at $\Delta T$ can be well measured from the residuals. The relationship between the new and old parameterisation is
\begin{equation}\label{eq: reparameterization}
\begin{aligned}
\Delta\nu & = \Delta\nu_{\mathrm{p}} + \sum_{i=1}^{N_{\mathrm{e}}\leq3} \Delta\nu_{\mathrm{d}_{i}} \\
\Delta\nu_{\scriptscriptstyle\Delta T} & = \Delta\dot{\nu}_{\mathrm{p}} \Delta T + \frac{1}{2} \Delta\ddot{\nu}_{\mathrm{p}} \Delta T^{2} + \sum_{i=1}^{N_{\mathrm{e}}\leq3} \Delta\nu_{\mathrm{d}_{i}} \left( e^{-\frac{\Delta T}{\tau_{i}}} -1 \right) \text{.}
\end{aligned}
\end{equation}
The posterior distribution for the traditional glitch parameterisation, $\Delta\nu_{\mathrm{p}}$ and $\Delta \dot{\nu}_{\mathrm{p}}$, can be reconstructed after the sampling is complete.

We choose a $\Delta T$ in the range $50\,\textrm{days}\lesssim \Delta T \lesssim 700\,\textrm{days}$, where we find that the density of ToAs will allow for a good estimate of $\Delta\nu_{\scriptscriptstyle\Delta T}$. Hence, we can estimate $\Delta\nu$ and $\Delta\nu_{\scriptscriptstyle\Delta T}$ directly from the gradient of the residuals before the glitch, immediately after the glitch, and at $\Delta T$, in order to define a reasonable prior range on $\Delta\nu_{\scriptscriptstyle\Delta T}$ and $\Delta\nu$. Based on the estimate for the mean value of $\Delta\nu$ and $\Delta\nu_{\scriptscriptstyle\Delta T}$, and the standard deviation $\sigma$ for the gradient of the residuals, we set a uniform prior range around their mean value with a size of $\pm100\sigma$. This range is usually narrow enough to exclude any implausible models that cause numerical issues but also wide enough as to not overly constrain the fitting.

We perform a preliminary parameter fit using \textsc{tempo2} \citep{hobbs_tempo2_2006}, and set uniform prior ranges for other parameters (glitch epoch, $\Delta\nu_{\mathrm{d}_{i}}$, and $\Delta\ddot{\nu}_{\mathrm{p}}$ if applicable) around their estimated values. We pick a default prior range of $\pm2$ days for the glitch epoch, $\pm0.8\Delta\nu_{\mathrm{p}}$ for the amplitude of the exponential transients $\Delta\nu_{\mathrm{d}_{i}}$, and $\pm10\ddot{\nu}$ for $\Delta\ddot{\nu}_{\mathrm{p}}$.

For $\tau_{i}$, we take a uniform prior in log space between 10 and 1000 days. A lower limit of 10 days is chosen because we do not have enough ToAs to constrain an exponential relaxation shorter than 10 days. The upper limit of 1000 days is set as it is hard to constrain an extremely long exponential relaxation. We stipulate $\tau_{1}<\tau_{2}<\tau_{3}$, and split the prior ranges into two or three parts for different relaxation components tested in models 5-7, so that $10<\tau_{1}<c_{1}<\tau_{2}<c_{2}<\tau_{3}<1000$, where $c_{1}$ and $c_{2}$ are constants used to split the priors, in order to avoid covariance among exponential relaxations.

In some cases we find that the default prior bounds can still be too large and again lead to numerical issues, especially when fitting for a large number of parameters. If this occurs we selectively reduce the prior bounds to ensure that the fit converges cleanly. It is important to ensure that the priors are the same for all models of that glitch (for evidence computation) and for all cases we verify that the posterior distribution is small ($\lesssim10\%$) compared to the prior. Furthermore, the glitch parameters are not expected to be multi-modal, so further widening of the priors should not influence the posteriors.

In the model selection procedure, we normally include only ToAs around the target glitch. However, when the previous or next glitch is too close to the target glitch, or when nearby glitches have considerable effects on the timing evolution around the target glitch, we extend the ToA span to include these glitches as well. Small glitches without modelled recoveries are treated as step changes in $\nu$ and $\dot{\nu}$ only, and all the remaining timing model parameters are extracted from the maximum likelihood solution using \textsc{tempo2}.

After the model selection procedure is completed for all target glitches in a pulsar, we produce a single timing solution that includes the maximum likelihood solution for the best model for each glitch. We then re-fit the red and white noise model parameters with \textsc{multinest}, and the basic timing model parameters, $\nu_{0}$, $\dot{\nu}_{0}$, $\ddot{\nu}_{0}$, astrometric parameters RAJ, DECJ, parallax, proper motion, dispersion measure with \textsc{tempo2}. We also allow the parameters of the smaller glitches (for which no model selection was performed and are simply modelled as step changes in $\nu$ and $\dot{\nu}$) to vary during this process. The final results consist of the time series of $\nu$ and $\dot{\nu}$ and the best solution for all fitted glitch parameters (glitch epoch, $\Delta\nu$, $\Delta\nu_{\scriptscriptstyle\Delta T}$, $\Delta\ddot{\nu}_{\mathrm{p}}$, $\Delta\nu_{\mathrm{d}_{i}}$, $\tau_{i}$), the Gaussian process noise model parameters ($A$, $\gamma$), and an EFAC and EQUAD for each observing system in the dataset. The permanent step changes $\Delta\nu_{\mathrm{p}}$, $\Delta\dot{\nu}_{\mathrm{p}}$, and the recovery ratio $Q=\frac{\Delta\nu_{\mathrm{d}}}{\Delta\nu}$ are derived afterwards.

\section{Results}\label{sec: result}

\begin{table*}
	\centering
	\caption{The spin frequency, its derivatives, and fit epoch, obtained from fitting the polynomial timing model (Equation~\ref{eq: phase_model}), and the derived physical properties for the pulsars in our sample derived from the data introduced in Table~\ref{tab: toa}. The braking indices of pulsars with unconstrained $\ddot{\nu}$ are omitted.}
	\label{tab: pulsars_parameters}
    \begin{tabular}{lrrrrrrrrr}
\hline
\hline
Pulsar name & $t_{0}$ & $\nu_{0}$ & $\dot{\nu}_{0}$ & $\ddot{\nu}_{0}$ & $\mathrm{log}_{10}\dot{E}$ & $\mathrm{log}_{10}\tau_{\mathrm{c}}$ & $\mathrm{log}_{10}B_{\mathrm{S}}$ & $n$ & $A_{g}$ \\
PSR & (MJD) & ($\mathrm{s}^{-1}$) & ($10^{-12}\;\mathrm{s}^{-2}$) & ($10^{-23}\;\mathrm{s}^{-3}$) & ($\mathrm{erg/s}$) & ($\mathrm{yr}$) & ($\mathrm{G}$) &  & ($10^{-12}\;\mathrm{s}^{-2}$) \\
\hline
 J0205$+$6449 & 55935.3 &      15.209975(1) &      $-$44.70(1) &   80(10) &   37.4 &    3.7 &   12.6 &          5.70 &   0.2749 \\
   B0355$+$54 & 50514.0 &    6.394538994(2) &    $-$0.17927(1) &   0.028(2) &   34.7 &    5.8 &   11.9 &         55.19 &  0.04737 \\
 J0611$+$1436 & 56319.6 &     3.69918760(4) &     $-$0.0518(1) &    $-$1.0(6) &   33.9 &    6.1 &   12.0 &     -- &  0.05993 \\
 J0631$+$1036 & 53854.0 &     3.47472633(7) &     $-$1.2635(3) &     0.5(1) &   35.2 &    4.6 &   12.7 &          9.85 &  0.02247 \\
 J0729$-$1448 & 54229.7 &     3.97319312(7) &     $-$1.7880(4) &     0.3(1) &   35.4 &    4.5 &   12.7 &          4.34 &  0.03981 \\
   B0919$+$06 & 50907.3 &    2.322218844(1) &   $-$0.073931(2) &   0.020(2) &   33.8 &    5.7 &   12.4 &         82.91 & 0.002246 \\
   B1727$-$33 & 53784.0 &     7.16915570(6) &     $-$4.3404(1) &     6.9(2) &   36.1 &    4.4 &   12.5 &         26.01 &  0.04216 \\
   B1737$-$30 & 52263.0 &     1.64802125(8) &     $-$1.2589(2) &    1.57(9) &   34.9 &    4.3 &   13.2 &         16.16 &   0.0152 \\
 J1737$-$3137 & 55046.0 &     2.21986396(3) &     $-$0.6830(1) &    0.31(3) &   34.8 &    4.7 &   12.9 &         14.70 &  0.01829 \\
 J1740$+$1000 & 56177.0 &     6.48948025(1) &    $-$0.90308(8) &    0.18(2) &   35.4 &    5.1 &   12.3 &         14.07 &  0.04103 \\
   B1754$-$24 & 53188.0 &    4.271601499(2) &   $-$0.235691(3) &   0.015(4) &   34.6 &    5.5 &   12.2 &         11.34 &    0.032 \\
   B1757$-$24 & 53085.7 &      8.0044111(4) &      $-$8.062(1) &      31(1) &   36.4 &    4.2 &   12.6 &         36.63 &   0.1137 \\
   B1800$-$21 & 52243.0 &      7.4822949(2) &     $-$7.4083(4) &    22.6(3) &   36.3 &    4.2 &   12.6 &         30.14 &   0.1512 \\
 J1806$-$2125 & 54266.5 &     2.07545998(6) &     $-$0.5042(1) &     0.3(3) &   34.6 &    4.8 &   12.9 &         -- &  0.04853 \\
 J1809$-$1917 & 55028.0 &   12.083913554(7) &   $-$3.719966(4) &    2.81(3) &   36.2 &    4.7 &   12.2 &         24.47 &  0.02699 \\
   B1821$-$11 & 52404.6 &    2.294841524(3) &   $-$0.018713(5) &   0.004(6) &   33.2 &    6.3 &   12.1 &        -- &  0.00646 \\
   B1823$-$13 & 52243.0 &      9.8546837(2) &     $-$7.2431(4) &    16.3(4) &   36.5 &    4.3 &   12.4 &         30.29 &   0.1044 \\
   B1830$-$08 & 52836.0 &    11.72521619(2) &    $-$1.26044(4) &  $-$0.168(5) &   35.8 &    5.2 &   12.0 &        $-$12.43 &  0.01983 \\
 J1837$-$0604 & 54637.0 &     10.3836264(1) &     $-$4.8607(5) &     4.3(7) &   36.3 &    4.5 &   12.3 &         19.07 &   0.0231 \\
 J1841$-$0345 & 55374.0 &    4.899877232(2) &   $-$1.387537(1) &   0.650(8) &   35.4 &    4.7 &   12.5 &         16.57 &  0.02872 \\
 J1841$-$0524 & 55372.0 &     2.24310956(2) &    $-$1.17350(8) &    1.07(3) &   35.0 &    4.5 &   13.0 &         17.44 &  0.01246 \\
 J1842$+$0257 & 55476.0 & 0.323806716085(4) & $-$0.00310395(1) & 0.00025(3) &   31.6 &    6.2 &   13.0 &         85.51 & 0.007637 \\
 J1850$-$0026 & 56127.0 &     6.00095231(1) &     $-$1.4067(1) &     1.2(1) &   35.5 &    4.8 &   12.4 &         36.72 &  0.02053 \\
   B1853$+$01 & 53405.0 &      3.7382506(2) &     $-$2.9024(2) &     3.5(3) &   35.6 &    4.3 &   12.9 &         15.64 &  0.05611 \\
 J1856$+$0245 & 56052.0 &    12.35899860(7) &     $-$9.4629(3) &    24.6(7) &   36.7 &    4.3 &   12.4 &         34.05 &   0.0724 \\
   B1859$+$01 & 52415.0 &    3.469574279(2) &   $-$0.028377(7) &  $-$0.002(2) &   33.6 &    6.3 &   11.9 &       -- & 0.003543 \\
 J1907$+$0631 & 57107.0 &    3.089504537(7) &    $-$4.31229(5) &     6.3(2) &   35.7 &    4.1 &   13.1 &         10.54 &  0.09207 \\
 J1909$+$0749 & 58161.4 &    4.215772386(6) &     $-$2.6945(2) &     2.3(1) &   35.7 &    4.4 &   12.8 &         13.32 &   0.0271 \\
 J1909$+$0912 & 55239.0 &    4.485092254(3) &    $-$0.72000(2) &    0.11(1) &   35.1 &    5.0 &   12.5 &          9.87 &  0.04867 \\
 J1921$+$0812 & 55614.4 &    4.747222300(3) &   $-$0.120865(6) &    0.05(2) &   34.4 &    5.8 &   12.0 &        -- &  0.03689 \\
   B1930$+$22 & 52681.0 &     6.92153331(4) &    $-$2.74958(7) &    2.00(8) &   35.9 &    4.6 &   12.5 &         18.29 &  0.03081 \\
   B1951$+$32 & 52622.1 &     25.2955817(1) &     $-$3.7384(7) &     0.9(2) &   36.6 &    5.0 &   11.7 &         15.63 &   0.0364 \\
 J2021$+$3651 & 55916.0 &      9.6384518(3) &     $-$8.7926(3) &      42(2) &   36.5 &    4.2 &   12.5 &         51.91 &   0.1675 \\
 J2229$+$6114 & 56422.5 &      19.359969(4) &      $-$28.05(1) & 340(20) &   37.3 &    4.0 &   12.3 &         76.56 &   0.1497 \\
   B2334$+$61 & 52471.9 &    2.018791215(3) &   $-$0.780826(2) &   0.335(5) &   34.8 &    4.6 &   13.0 &         11.10 &  0.04061 \\
\hline
    \end{tabular}
\end{table*}

In this section we present the parameters of 157 glitches in the 35 pulsars of our sample. The basic pulsar parameters are given in Table~\ref{tab: pulsars_parameters}: the second to fifth column are the $t_{0}$, $\nu_{0}$, $\dot{\nu}_{0}$, and $\ddot{\nu}_{0}$ in the phase model Equation~\ref{eq: phase_model} for each pulsar. Table~\ref{tab: pulsars_parameters} also shows the spin-down luminosity calculated by 
\begin{equation}\label{eq: E_dot}
\dot{E} = 3.95\times10^{31}\;\mathrm{erg/s} \left(\frac{\nu}{\mathrm{s}^{-1}}\right)\left(\frac{-\dot{\nu}}{10^{-15}\;\mathrm{s}^{-2}}\right)
\end{equation}
the pulsar's characteristic age \begin{equation}\label{eq: Tau_c}
\tau_{\mathrm{c}} = 1.58\times10^{7}\;\mathrm{yr} \left(\frac{\nu}{\mathrm{s}^{-1}}\right)\left(\frac{-\dot{\nu}}{10^{-15}\;\mathrm{s}^{-2}}\right)^{-1},
\end{equation}
the inferred surface magnetic field strength 
\begin{equation}\label{eq: B_sur}
B_{\mathrm{S}} = 1.0\times10^{12}\;\mathrm{G} \left(\frac{\nu}{\mathrm{s}^{-1}}\right)^{-\frac{3}{2}}\left(\frac{-\dot{\nu}}{10^{-15}\;\mathrm{s}^{-2}}\right)^{\frac{1}{2}},
\end{equation}
and the braking index $n=\nu\ddot{\nu}/\dot{\nu}^{2}$.
We also include the glitch activity parameter, $A_{\mathrm{g}}=\frac{1}{T}\sum\Delta \nu$, which is the cumulative spin-up due to multiple glitches over a total observation span $T$ (see \citet{basu_jodrell_2022} and \citet{antonopoulou_pulsar_2022} for discussions of this parameter).

\subsection{Overview}\label{sub: view}

\begin{table*}
\begin{center}
\centering
\caption{The glitch epochs, sizes, recovery amplitudes, and corresponding time scales of 85 glitches in 35 pulsars from our
sample for which model selection was performed. The data span used in the model selection for each glitch is given in the ToA span column, with the corresponding number of ToAs in the next column. Errors are given in parentheses in units of the last quoted digit. For glitches with exponential recoveries, we list the amplitude, timescale, and the recovery ratio $Q$ of the recovery components in the last three columns. The values quoted are the mean of posterior with standard deviation. The five small glitches with asterisks are also included in the model selection process.}
\label{tab: glitch_parameters}
\begin{threeparttable}
\begin{tabular}{lrrrrrrrrrrrr}
\hline
\hline
Pulsar name & Gl & Ref. & Epoch & ToA span & ToAs & $\Delta\nu/\nu$ & $\Delta\dot{\nu}/\dot{\nu}$ & $\Delta\ddot{\nu}$ & $\Delta\nu_{\mathrm{d}}$ & $\tau_{\mathrm{d}}$ & Q \\
PSR & No. &    No. & (MJD) & (MJD) & No. & ($10^{-9}$) & ($10^{-3}$) & ($10^{-23}\;\mathrm{s}^{-3}$) & ($10^{-8}\;\mathrm{s}^{-1}$) & (days) &  \\       
\hline
J0205$+$6449 & 1  & [1] & 54906.97(3) & 53725-55831 & 462  & 1764(6) & 11.2(7) & +500(100) &                 -- &                 -- &           -- \\
         & 2  & [1] & 55722.1(2) & 54924-56826 & 228  & 123(4) & 3.5(7) &            &                 -- &                 -- &           -- \\
         & 3  & [1] & 55841.092(7) & 54924-56826 & 228  & 2927(3) & 13.6(7) &            &                 -- &                 -- &           -- \\
         & 5  & [1] & 57358.15(2) & 55841-58224 & 248  & 542(2) & 5.7(7) &            &                 -- &                 -- &           -- \\
         & 7  & [1] & 58320.8(4) & 57358-59241 & 165  & 3170(30) & 22.1(8) & +500(100) &                 -- &                 -- &           -- \\
B0355$+$54 & 1*  & [2,3] & 46082.9(7) & 41807--47066 & 247  & 6.4(2) & 9(2) &   &           2.5(2) &            130(20) &     0.60(6) \\
         & 2  & [2,3] & 46469.92(2) & 45500--48717 & 271  & 4414(9) & 1900(400) &   &          30(6) &          $\lesssim11.8$ &    0.011(2) \\
         &    &    &              &              &      &          &           &   &           3.3(1) &          75(5) &  0.00117(5) \\
         &    &    &              &              &      &          &           &  &           2.8(5) &       600(100) &   0.0010(2) \\        
J0611$+$1436 & 1  & [1] & 55833.515(6) & 54516-58500 & 298  & 5575.3(6) & $-$10(20) &  &                 -- &                 -- &         -- \\
J0631$+$1036 & 4  & [2] & 50730.384(3) & 49994-51909 & 336  & 1662.6(5) & 3.1(9) &  &             30(40) &       300(300) &    0.05(6) \\
         & 6*  & [2,4] & 52852.7(1) & 51857-53848 & 287  & 17.6(2) & 2.6(4) & +1.5(9) &           1.8(6) &            120(80) &    0.30(9) \\
         & 12* & [2] & 54632.46(6) & 53633-55622 & 237  & 43.3(4) & 1.5(8) &            &                 -- &                 -- &           -- \\
         & 15 & [1] & 55701.9231(6) & 54703-58342 & 769  & 3278.1(2) & 5.6(8) & $-$2(1) &           7(5) &            150(90) &   0.006(5) \\
         & 17* & [1,5] & 58352.12(2) & 57358-59331 & 327  & 96.8(2) & 2.7(6) &            &           5(4) &       200(100) &     0.1(1) \\
J0729$-$1448 & 5  & [2,6,7] & 54694.230(6) & 53697--58227 & 387  & 6658(3) & 60(10) &  &          13(1) &          18(5) &   0.0048(5) \\
B0919$+$06 & 1  & [1] & 55151.655(9) & 43586-58523 & 1335 & 1249(1) & -330(60) &  &            $-$2.2(4) &            $\lesssim12.4$ &  $-$0.008(1) \\
B1727$-$33 & 1  & [2,7] & 52069.2(1) & 48416-55878 & 769  & 3197(4) & 5.2(4) &  &                 -- &                 -- &         -- \\
         & 2  & [1,5] & 55923.49(1) & 52274-58217 & 654  & 2252(1) & 9(1) &         &           40(50) &           200(200) &    0.02(3) \\
B1737$-$30 & 1  & [2,8,9] & 46985.3(2) & 46300-47646 & 73   & 423(4) & 0.9(10) &  &                 -- &                 -- &           -- \\
         & 5  & [2,8,9] & 47669.9(6) & 47028-48180 & 131  & 597(7) & 0.5(20) &            &                 -- &                 -- &           -- \\
         & 7  & [2,9,10] & 48192.21(4) & 47028-49225 & 260  & 667(2) & 0.9(6) &            &                 -- &                 -- &           -- \\
         & 11 & [2,9,10] & 49239.03(7) & 48268-50236 & 203  & 169(1) & 0.8(2) &            &                 -- &                 -- &           -- \\
         & 14 & [2,9,11] & 50574.811(9) & 49567-51986 & 319  & 441.5(5) & 1.6(2) &            &                 -- &                 -- &           -- \\
         & 15 & [2,7,9,11,12] & 50936.796(3) & 49567-51986 & 319  & 1444.0(5) & 1.3(2) &            &                 -- &                 -- &           -- \\
         & 21 & [2,7,9,13] & 52346.7(1) & 50948-53023 & 319  & 158(1) & 0.7(4) &            &                 -- &                 -- &           -- \\
         & 26 & [2,7,9] & 53023.509(4) & 52349-54449 & 403  & 1849.3(6) & 1.2(3) &            &                 -- &                 -- &           -- \\
         & 32 & [1,7] & 55211.65(3) & 54225-56191 & 327  & 2663(2) & 1.1(2) &            &                 -- &                 -- &           -- \\
         & 35 & [1,14,15] & 57468.59(1) & 55953-58533 & 306  & 229.3(4) & 1.6(3) &            &                 -- &                 -- &           -- \\
         & 36 & [1,15] & 58240.781(3) & 55953-58533 & 306  & 837.6(5) & 1.8(3) &            &                 -- &                 -- &           -- \\
J1737$-$3137 & 2  & [2] & 53030.41(7) & 51564-54340 & 82   & 236.5(3) & 2.8(1) &  &           5(2) &            300(60) &   0.09(4) \\
         & 3  & [2,6,7] & 54348.42(2) & 53063-57063 & 143  & 1341.7(5) & 1.6(2) &            &                 -- &                 -- &       -- \\
         & 5  & [1,5] & 58146.77(1) & 54355-59335 & 127  & 4498.2(9) & 1.4(2) &            &                 -- &                 -- &       -- \\
J1740$+$1000 & 2  & [1] & 56159.544(6) & 54769--59565 & 841  & 2916(2) & 40(20) &   &           7(1) &          29(8) &   0.0036(5) \\
         &    &    &              &              &      &          &           &   &          10(3) &       400(200) &    0.005(2) \\
B1754$-$24 & 1  & [1] & 55697.928(4) & 47156-59221 & 352  & 7807.9(7) & 64(6) &  &             6.7(4) &              65(7) &  0.0020(1) \\
         &         &    &              &             &      &          &       &  &             7.9(7) &           500(100) &  0.0024(2) \\
B1757$-$24 & 1  & [2,7,16] & 49475.89(2) & 48329-50618 & 175  & 1987(3) & 4.6(7) &  &                 -- &                 -- &         -- \\
         & 2  & [2,7,16] & 50651.06(3) & 49514-51986 & 241  & 1245(2) & 4.5(5) &         &                 -- &                 -- &         -- \\
         & 3  & [2,7] & 52055.2(4) & 50673-53027 & 225  & 3730(10) & 7(1) &         &                 -- &                 -- &         -- \\
         & 5  & [2,6,7] & 54661.1(1) & 53060-56943 & 584  & 3114(5) & 19(2) &         &              15(1) &              16(3) &  0.0060(5) \\
         & 6  & [1,5] & 56942.98(4) & 54663-58528 & 410  & 2413(2) & 5.3(4) &         &                 -- &                 -- &         -- \\
B1800$-$21 & 1  & [2,10] & 48244.929(3) & 47364--50262 & 305  & 4079.6(9) & 11.1(4) &   &          31(3) &             100(10) &    0.010(1) \\
         & 3  & [2,7,16] & 50776.743(3) & 49800--53428 & 527  & 3190(1) & 13(1) &   &           9(2) &          25(6) &   0.0037(6) \\
         &    &    &              &              &      &          &           &   &             40(40) &       300(200) &     0.02(2) \\
         & 4  & [2,4,7] & 53428.6875(6) & 50785--55772 & 952  & 3948.0(4) & 33(1) &   &          14.8(5) &          $\lesssim11.3$ &   0.0050(2) \\
         &    &    &              &              &      &          &           &   &          23(4) &             90(10) &    0.008(1) \\
         & 5  & [5] & 55774.749(4) & 53430--58919 & 930  & 4808.9(8) & 48(2) &   &          23.4(8) &          $\lesssim11.5$ &   0.0065(2) \\
         &    &    &              &              &      &          &           &   &          32(3) &          77(8) &   0.0088(9) \\
         & 6  & [JBO] & 58920.04(2) & 55778--59240 & 456  & 4696(4) & 11(1) &   &             50(50) &       100(100) &     0.01(2) \\
J1806$-$2125 & 1  & [2,17] & 52200(1) & 50820-58500 & 329  & 15510(60) & 6(5) &  &                 -- &                 -- &         -- \\
J1809$-$1917 & 1  & [2] & 53255.60(1) & 50820-59236 & 311  & 1624(1) & 9(2) &  &               9(1) &           60(10) &  0.0045(5) \\
         &         &    &              &             &      &          &       &  &              16(6) &           400(200) &   0.008(3) \\
B1821$-$11 & 1  & [2] & 54305.648(6) & 46612-58435 & 358  & 2875.6(4) & $-$1(4) &  &                 -- &                 -- &         -- \\
\hline
\end{tabular}
\end{threeparttable}
\end{center}
\end{table*}
\begin{table*}
\begin{center}
\centering
\contcaption{}
\begin{threeparttable}
\begin{tabular}{lrrrrrrrrrrrr}
\hline
\hline
Pulsar name & Gl & Ref. & Epoch & ToA span & ToAs & $\Delta\nu/\nu$ & $\Delta\dot{\nu}/\dot{\nu}$ & $\Delta\ddot{\nu}$ & $\Delta\nu_{\mathrm{d}}$ & $\tau_{\mathrm{d}}$ & Q \\
PSR & No. &    No. & (MJD) & (MJD) & No. & ($10^{-9}$) & ($10^{-3}$) & ($10^{-23}\;\mathrm{s}^{-3}$) & ($10^{-8}\;\mathrm{s}^{-1}$) & (days) &  \\
\hline
B1823-13 & 1  & [2,10] & 49024.6(2) & 46612--53207 & 726  & 3110(30) & 50(20) &  &             80(30) &          31(4) &    0.025(8) \\
         &    &    &              &              &      &          &           &  &            200(90) &       600(200) &     0.06(3) \\
         & 4  & [2,4,7] & 53737.053(2) & 52744--56549 & 695  & 3607.9(6) & 44(2) &  &          20.3(8) &          $\lesssim11.5$ &   0.0057(2) \\
         &    &    &              &              &      &          &           &  &          22(2) &          60(5) &   0.0063(4) \\
         & 5  & [1] & 56555.955(3) & 53738--58863 & 850  & 2602.6(5) & 14.7(8) &  &          10.1(7) &          21(2) &   0.0039(3) \\
         &    &    &              &              &      &          &           &  &             40(50) &       300(200) &     0.02(2) \\
         & 6  & [JBO] & 58894.903(3) & 56562--59518 & 435  & 2493.4(6) & 28(2) &   &          14.1(9) &          11(1) &   0.0058(4) \\
         &    &    &              &              &      &          &           &   &          23(4) &            110(20) &    0.009(2) \\
B1830$-$08 & 2  & [2,10] & 48050.769(1) & 47154-54782 & 455  & 1865.7(1) & 1.6(2) &  &             2.9(3) &         310(80) &  0.0013(1) \\
J1837$-$0604 & 1  & [1,5] & 55821.11(2) & 51153-56776 & 214  & 1395(1) & 6.6(4) & +19(5) &                 -- &                 -- &         -- \\
J1841$-$0345 & 1  & [JBO] & 56829.35(3) & 51507-59241 & 431  & 3920(6) & 30(8) &  &              13(2) &              46(7) &   0.007(1) \\
         &         &    &              &             &      &          &       &  &             6.7(9) &         310(70) &  0.0035(5) \\
J1841$-$0524 & 3  & [2,6,7] & 54503.75(4) & 52607-56508 & 125  & 1031.8(6) & 1.1(1) &  &                 -- &                 -- &       -- \\
         & 4  & [1,5] & 55521.69(4) & 54544-56508 & 45   & 804.1(4) & 0.82(7) &            &                 -- &                 -- &       -- \\
         & 6  & [JBO] & 57101.53(3) & 55562-58557 & 52   & 1011.6(5) & 1.01(9) &            &                 -- &                 -- &       -- \\
         & 7  & [JBO] & 58738.6(6) & 57152-59248 & 42   & 130.4(7) & 0.6(2) &            &                 -- &                 -- &       -- \\
         & 8  & [JBO] & 59111.6(9) & 58119-59248 & 29   & 128(1) & 0.1(2) &            &                 -- &                 -- &       -- \\
J1842$+$0257 & 1  & [1] & 57113.411(4) & 52608-58528 & 188  & 12061.02(9) & 1.3(2) &  &                 -- &                 -- &         -- \\
J1850$-$0026 & 4  & [1] & 57708.632(7) & 56446-58462 & 73   & 1196.6(2) & 0.8(2) &  &                 -- &                 -- &       -- \\
B1853$+$01 & 1  & [2] & 54122.836(4) & 47576-57291 & 825  & 11636(8) & 100(10) &  &              36(3) &              17(2) &  0.0083(6) \\
         & 2  & [1] & 57340.78(2) & 54125-59236 & 468  & 3486(4) & 6(1) &         &                 -- &                 -- &         -- \\
J1856$+$0245 & 1  & [1] & 56388.13(1) & 53877-59170 & 443  & 2678(2) & 11(2) &  &              25(4) &           60(20) &   0.007(1) \\
B1859$+$01 & 3  & [1] & 58178.861(4) & 56266-59141 & 30   & 1053.01(5) & 2.0(4) &  &                 -- &                 -- &         -- \\
J1907$+$0631 & 1  & [1] & 56987.46(2) & 55729-57976 & 165  & 2107(1) & 5.3(3) &  &             90(60) &       500(200) &    0.14(8) \\
         & 2  & [1] & 57902.70(4) & 56910-58485 & 98   & 4985(6) & 5(1) &            &                 -- &                 -- &           -- \\
J1909$+$0749 & 1  & [1] & 57491.720(4) & 53692-58923 & 409  & 2904.1(7) & 3.7(2) &  &                 -- &                 -- &         -- \\
J1909$+$0912 & 2  & [1] & 57552.3(1) & 55037--59227 & 72   & 7530(60) & 400(300) &   &             70(30) &          30(7) &    0.019(8) \\
         &    &    &              &              &      &          &           &   &          16(1) &            280(60) &   0.0047(4) \\
J1921$+$0812 & 1  & [1] & 55349.269(4) & 53083-58495 & 279  & 3632.9(1) & 2.9(8) &  &                 -- &                 -- &         -- \\
B1930$+$22 & 1  & [2,11] & 50253.025(7) & 46860-58503 & 1150 & 4479(3) & 32(5) &  &              25(2) &              36(6) &  0.0080(6) \\
B1951$+$32 & 6  & [1] & 55327.556(3) & 54108-59191 & 642  & 1505(1) & 80(20) &  &              34(4) &              14(2) &  0.0088(9) \\
         &         &    &              &             &      &          &       &  &         110(40) &           600(200) &   0.028(9) \\
J2021$+$3651 & 1  & [18] & 52630.29(6) & 52589-54129 & 175  & 2599(5) & 3(2) &  &                 -- &                 -- &         -- \\
         & 2  & [2] & 54176.6(1) & 52704-55172 & 357  & 743(3) & 5.2(8) &         &                 -- &                 -- &         -- \\
         & 3  & [1] & 55110.149(5) & 54116-57192 & 370  & 2234(2) & 18(4) &         &              15(5) &           20(10) &   0.007(2) \\
         & 4  & [1] & 57205.007(8) & 55112-58228 & 235  & 3078(3) & 21(5) &         &              33(5) &           30(10) &   0.011(2) \\
         & 5  & [JBO] & 58245(5) & 57207-59243 & 72   & 1340(30) & 5(1) &         &                 -- &                 -- &         -- \\
J2229$+$6114 & 1  & [2] & 53063.720(7) & 51976-54109 & 416  & 1138.9(4) & 13.2(2) &  &                 -- &                 -- &           -- \\
         & 2  & [2] & 54111.16(1) & 53068-55108 & 445  & 328(1) & $-$0.1(20) &            &            $-$50(10) &          22(7) &   $-$0.07(2) \\
         & 4  & [1,19] & 55135.27(3) & 54112-56789 & 393  & 194(1) & 2(1) &            &            $-$20(10) &             20(40) &   $-$0.05(4) \\
         & 5  & [1,19] & 55601.001(4) & 54112-56789 & 393  & 1223.7(6) & 13.2(5) &            &                 -- &                 -- &           -- \\
         & 6*  & [1,19] & 56366.5(1) & 55604-58418 & 249  & 66.2(6) & 3.3(4) &            &                 -- &                 -- &           -- \\
         & 7  & [19] & 56788.59(4) & 55135-58418 & 328  & 928(2) & 11.9(5) &            &                 -- &                 -- &           -- \\
         & 8  & [1,19] & 58417.470(9) & 56829-59414 & 286  & 1047.0(8) & 12.7(5) &            &                 -- &                 -- &           -- \\
         & 9  & [19] & 59260.0(1) & 58304-59720 & 209  & 255(5) & $-$17(6) &            &         $-$76(9) &          13(2) &   $-$0.15(2) \\
B2334+61 & 1  & [2,20] & 53642.305(4) & 46717--58502 & 989  & 20480(2) & 36(4) &   &           4.5(6) &          43(8) &   0.0011(2) \\
         &    &    &              &              &      &          &           &   &           7.3(6) &            140(20) &   0.0018(2) \\
         &    &    &              &              &      &          &           &   &          15(2) &       900(100) &   0.0036(6) \\
\hline
\end{tabular}
\begin{tablenotes}
     \item[] \textit{Note.} Reference for discovery or measurement of glitches: 1. \citet{basu_jodrell_2022}; 2. \citet{espinoza_study_2011}; 3. \citet{lyne_massive_1987}; 4. \citet{yuan_29_2010}; 5. \citet{lower_impact_2021}; 6. \citet{weltevrede_pulsar_2010}; 7. \citet{yu_detection_2013}; 8. \citet{mckenna_psr1737-30_1990}; 9. \citet{zou_observations_2008}; 10. \citet{shemar_observations_1996}; 11. \citet{krawczyk_observations_2003}; 12. \citet{urama_glitch_2002}; 13. \citet{janssen_30_2006}; 14. \citet{jankowski_glitch_2016}; 15. \citet{basu_observed_2020}; 16. \citet{wang_glitches_2000}; 17. \citet{hobbs_very_2002}; 18. \citet{hessels_observations_2004}; 19. \citet{gugercinoglu_glitches_2022}; 20. \citet{yuan_very_2010}.
\end{tablenotes}
\end{threeparttable}
\end{center}
\end{table*}

Based on the work of \citet{basu_jodrell_2022}, pulsars with $0.03\mathrm{s}\leq P\leq1\mathrm{s}$ and $10^{-15}\leq\dot{P}\leq10^{-12}$ (characteristic age around $10^{1}\sim10^{3}$ Kyr) are more likely to undergo large glitches that satisfy our selection criterion of $\Delta\nu/\nu\geq10^{-6}$. Hence, the pulsars that form the sample of our work are located in a relatively small region in $P-\dot{P}$. 

Most pulsars in our sample (22 out of 35) have multiple glitches during our observations, whilst 13 pulsars have only one glitch. For instance, PSR B1737$-$30 has 36 reported glitches but, in a similar time interval, PSR~B1930$+$22 has just one.  

From the total of 157 glitches, 61 have $\Delta\nu/\nu\geq10^{-6}$ and hence meet our criteria for a `large' glitch, and 19 have an intermediate size ($10^{-7}\leq\Delta\nu/\nu<10^{-6}$). All glitches with $\Delta\nu \geq 10^{-7}$ are included in the model selection process. Small glitches, on the other hand, are typically treated just a steps in $\nu$ and $\dot{\nu}$ (calculated by a linear least square fit) because they do not commonly show detectable exponential recoveries. However, there are 5 small glitches for which preliminary analysis suggested the presence of additional recovery features, so these are included in the model selection process as well. These are the 1st glitch of PSR B0355$+$54, the 6th, 12th, 17th glitch of PSR J0631$+$1036, and the 6th glitch of PSR J2229$+$6114. We mark these glitches with asterisk in Table~\ref{tab: glitch_parameters}.

There are two pulsars (PSR J0631$+$1036 and PSR J1907$+$0631) where we did not strictly follow the model selection process in Section \ref{sub: selection}, specifically when comparing models with equivalent Bayes factors we did not choose the simplest model. In both these cases there is some ambiguity between glitch behaviour and timing noise, and sometimes equivalent models with additional terms improved the final timing solution in the full dataset. Specifically  PSR J0631$+$1036 has a large number of glitches close together and the overall timing solution is improved by including a single exponential term in the 4th and 17th glitches, and $\Delta\ddot{\nu}$ in the 6th and 15th glitches. For PSR J1907$+$0631 excluding the exponential recovery in the 1st glitch results in the same exponential recovery behaviour being absorbed into the timing noise model, which seems less likely. The logarithm of Bayes factor for all the models we tested are given in Table~\ref{tab: model_evidence}.

Among the 85 glitches for which we performed model selection, 37 include at least one exponential recovery term in the best model (see Equation \ref{eq: glitch_phase}). The preferred model for most of these includes only a single exponential, with a two exponential model favoured only for 15 glitches. In our dataset, only the 2nd glitch (MJD~46469.9) of PSR B0355$+$54 and the glitch at MJD~53642.3 of PSR B2334$+$61 present a Bayes factor indicating evidence for a third exponential. The timescales of the exponential recoveries range from around 10 days (our lower prior bound) to 879 days. We note that the lower prior bound is needed to prevent the timescale converging to extremely small values (which are impossible to resolve with current ToA cadence), and hence timescales consistent with 10 days should be taken to mean that the timescale might be smaller than what we can measure.

We list our solutions for the glitch parameters, including any exponential recovery components, in Table~\ref{tab: glitch_parameters}. In some cases, two large glitches were fitted simultaneously -- accordingly, the range of ToAs includes the previous or next large glitch, as shown in Table~\ref{tab: glitch_parameters}. This was done when the timing solution for one glitch could affect the derived parameters for the other due to their temporal proximity. For the small glitches, we adjust the glitch epoch until there is no perceivable phase discontinuity at the glitch.

\subsection{Spin Evolution}\label{sub: evolution}
It can be informative to visualise time-series of $\nu(t)$ and $\dot{\nu}(t)$. The glitch contribution can be computed by analytic derivatives of Equation~\ref{eq: glitch_phase}, however, we need to also consider the spin evolution attributed to timing noise in our model. A smooth model of the timing noise can be derived directly from the Gaussian process noise model (e.g. \citealp{keith_impact_2023}), however,  we can also directly measure $\nu$ and $\dot{\nu}$ over a given small time window from the residuals.
Therefore, after obtaining the best model, we use the striding boxcar method described by \citet{shaw_largest_2018, shaw_slow_2021} in order to visualise the evolution of the spin frequency $\nu$ and the spin frequency derivative $\dot{\nu}$. We fit for $\nu$, $\dot{\nu}$, and $\ddot{\nu}$ in small windows whose widths span 6 times the average cadence. We shift the fit window by half its width generating a series of spin parameter values over the entire data span. For any window near a glitch, we adjust the start or finish epoch, so that they are immediately after or before the glitch respectively, in order to avoid fitting the rotational parameters over the glitch. 

\begin{figure*}
    \centering
    \begin{subfigure}[b]{0.49\textwidth}
        \centering
        \includegraphics[scale=0.29]{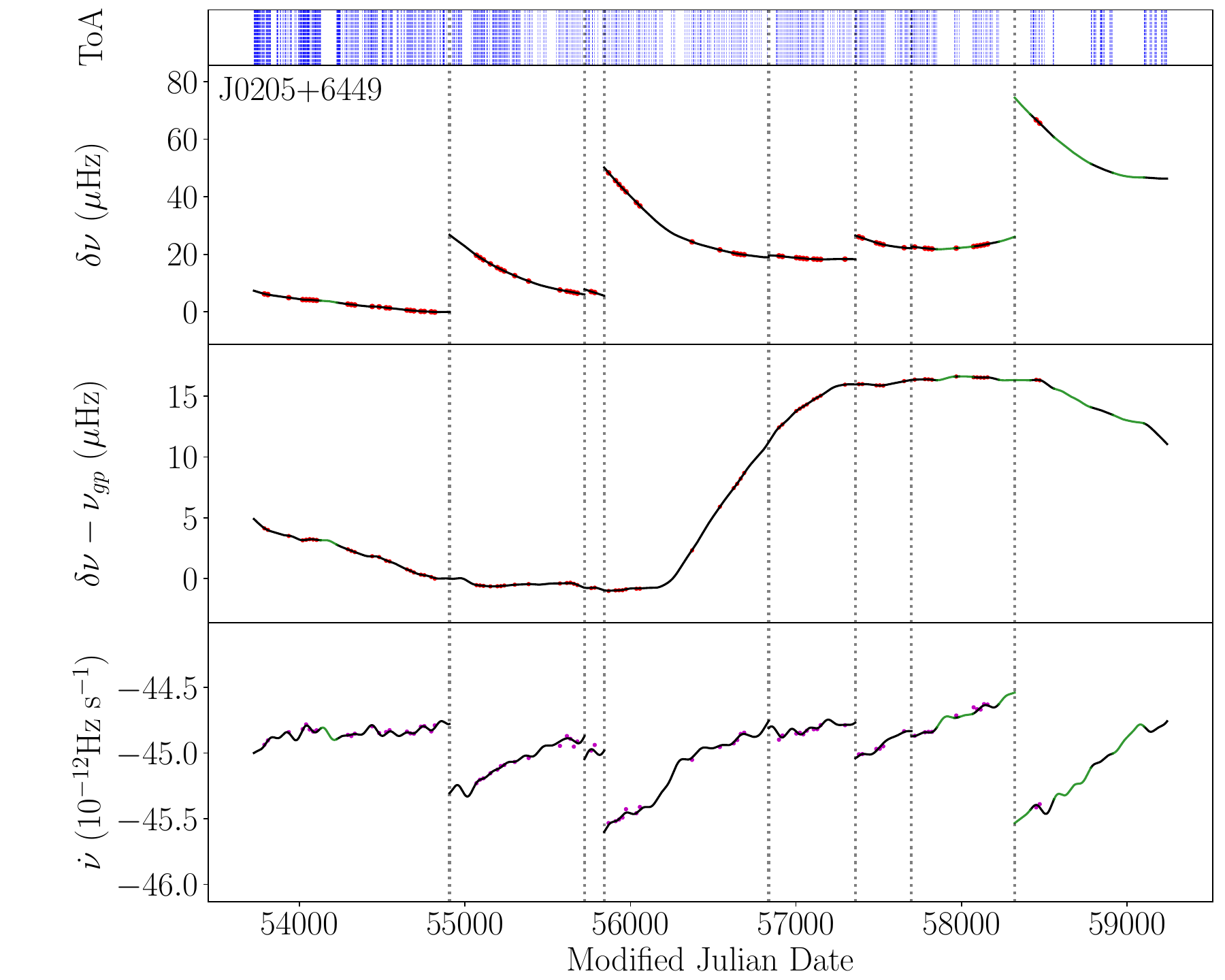}
    \end{subfigure}
    \begin{subfigure}[b]{0.49\textwidth}
        \centering
        \includegraphics[scale=0.29]{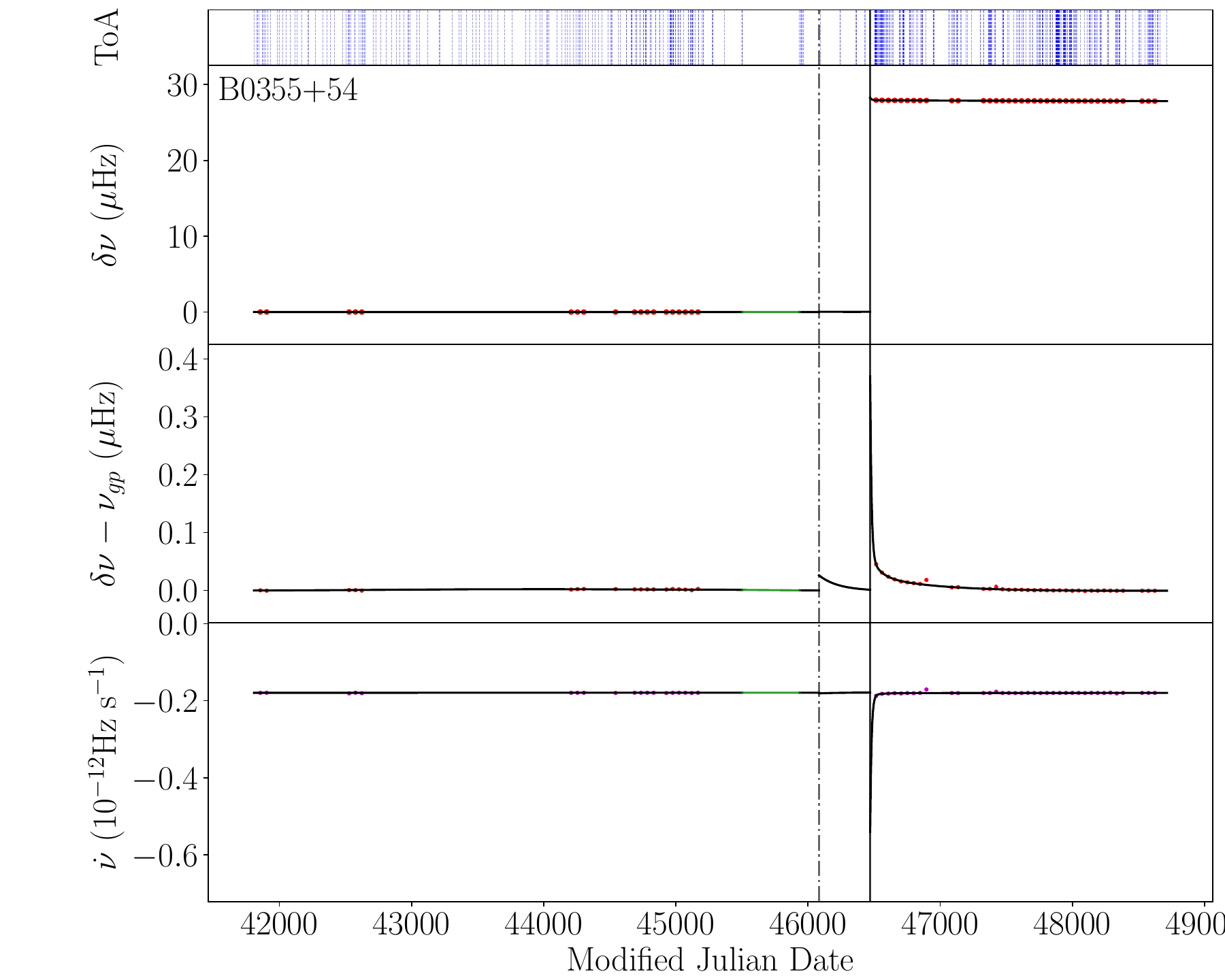}
    \end{subfigure}
    \begin{subfigure}[b]{0.49\textwidth}
        \centering
        \includegraphics[scale=0.29]{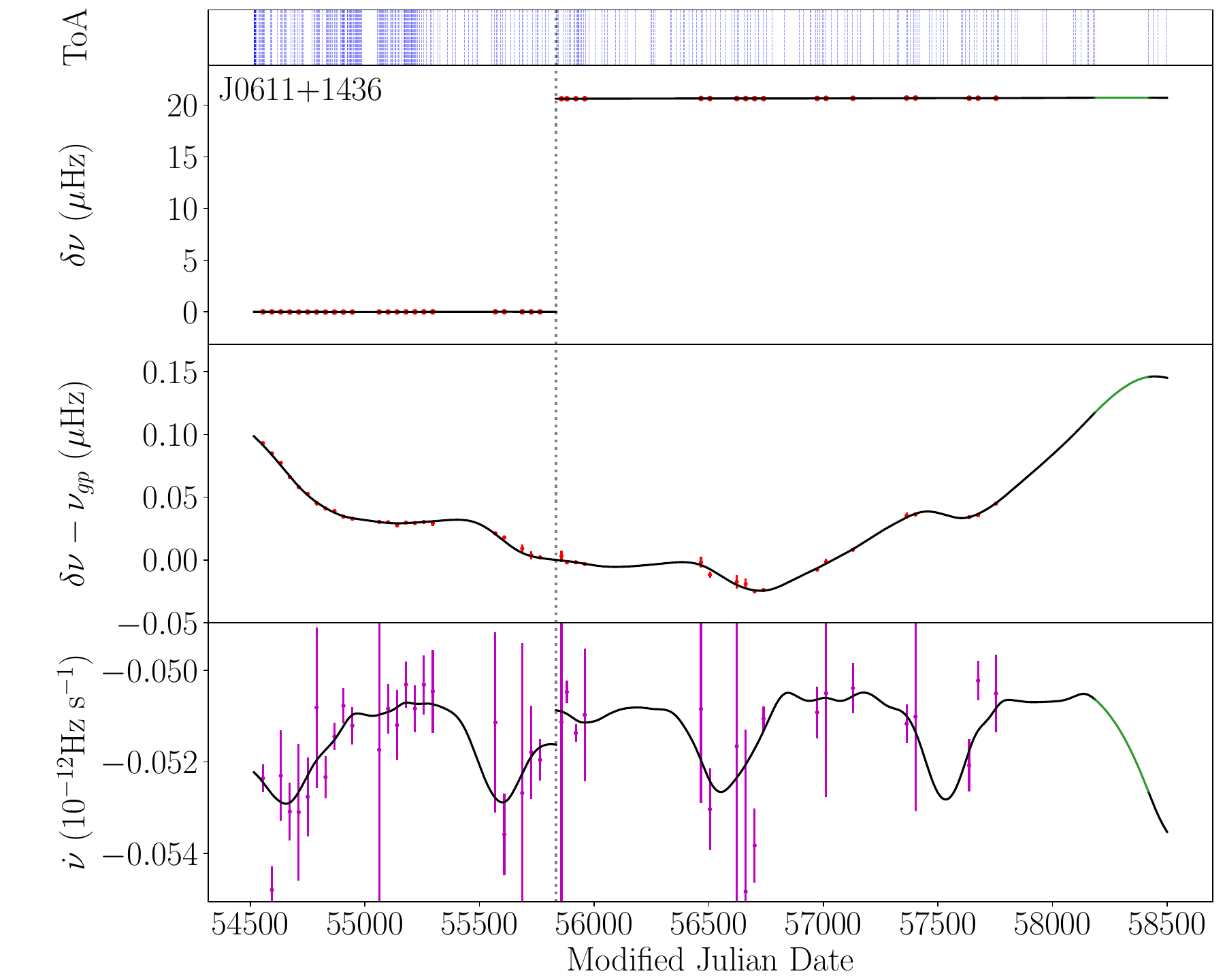}
    \end{subfigure}
    \begin{subfigure}[b]{0.49\textwidth}
        \centering
        \includegraphics[scale=0.29]{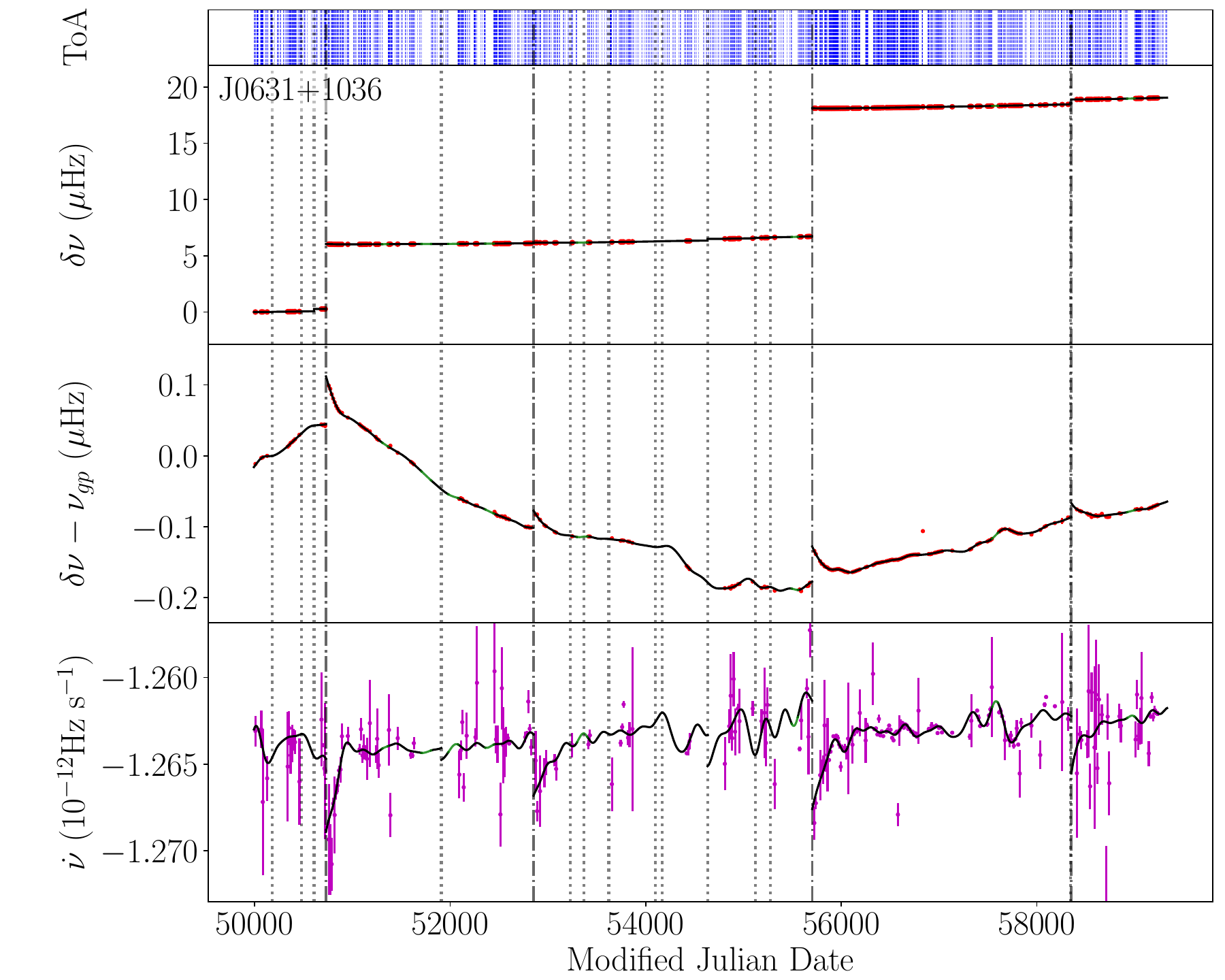}
    \end{subfigure}
    \begin{subfigure}[b]{0.49\textwidth}
        \centering
        \includegraphics[scale=0.29]{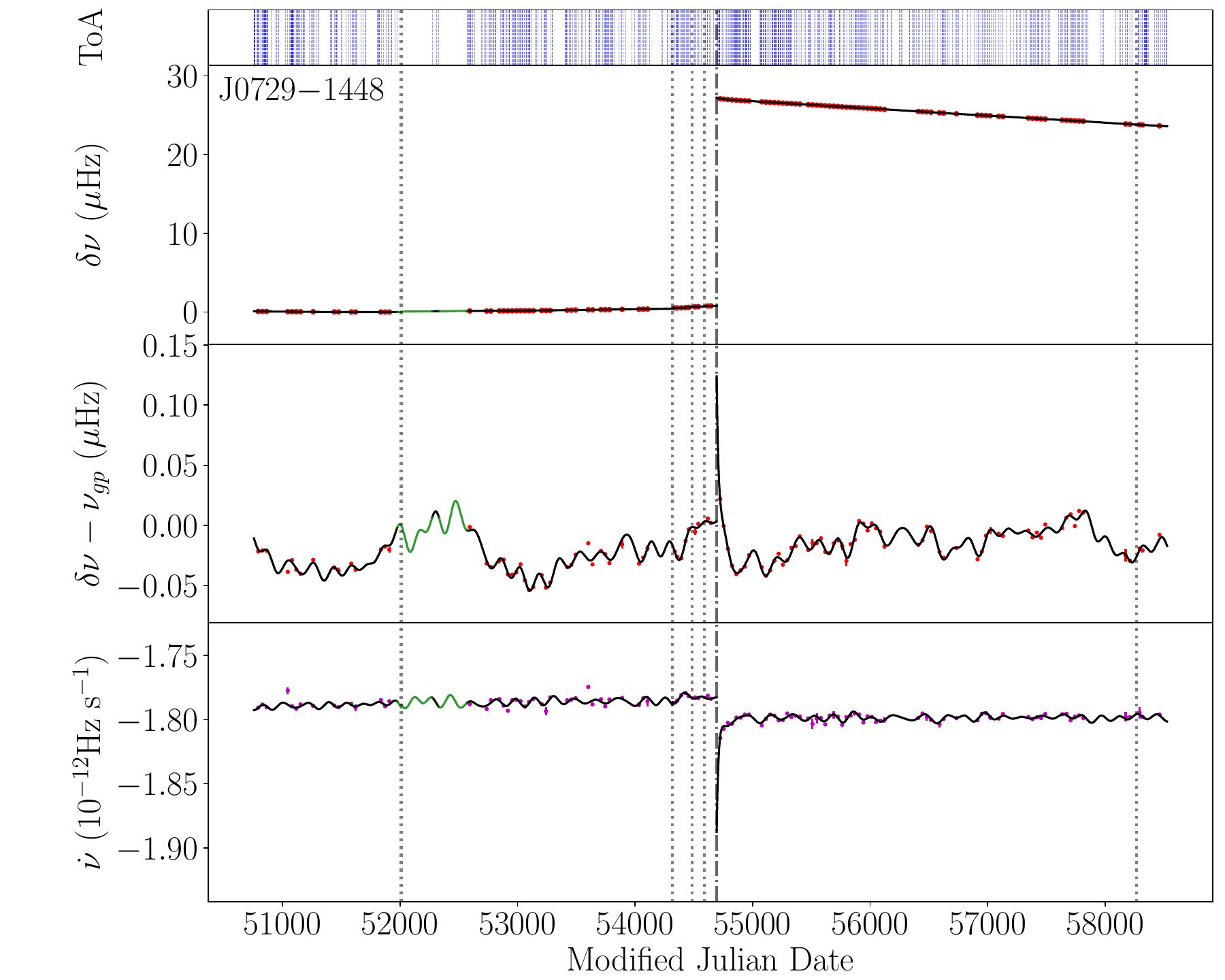}
    \end{subfigure}
    \begin{subfigure}[b]{0.49\textwidth}
        \centering
        \includegraphics[scale=0.29]{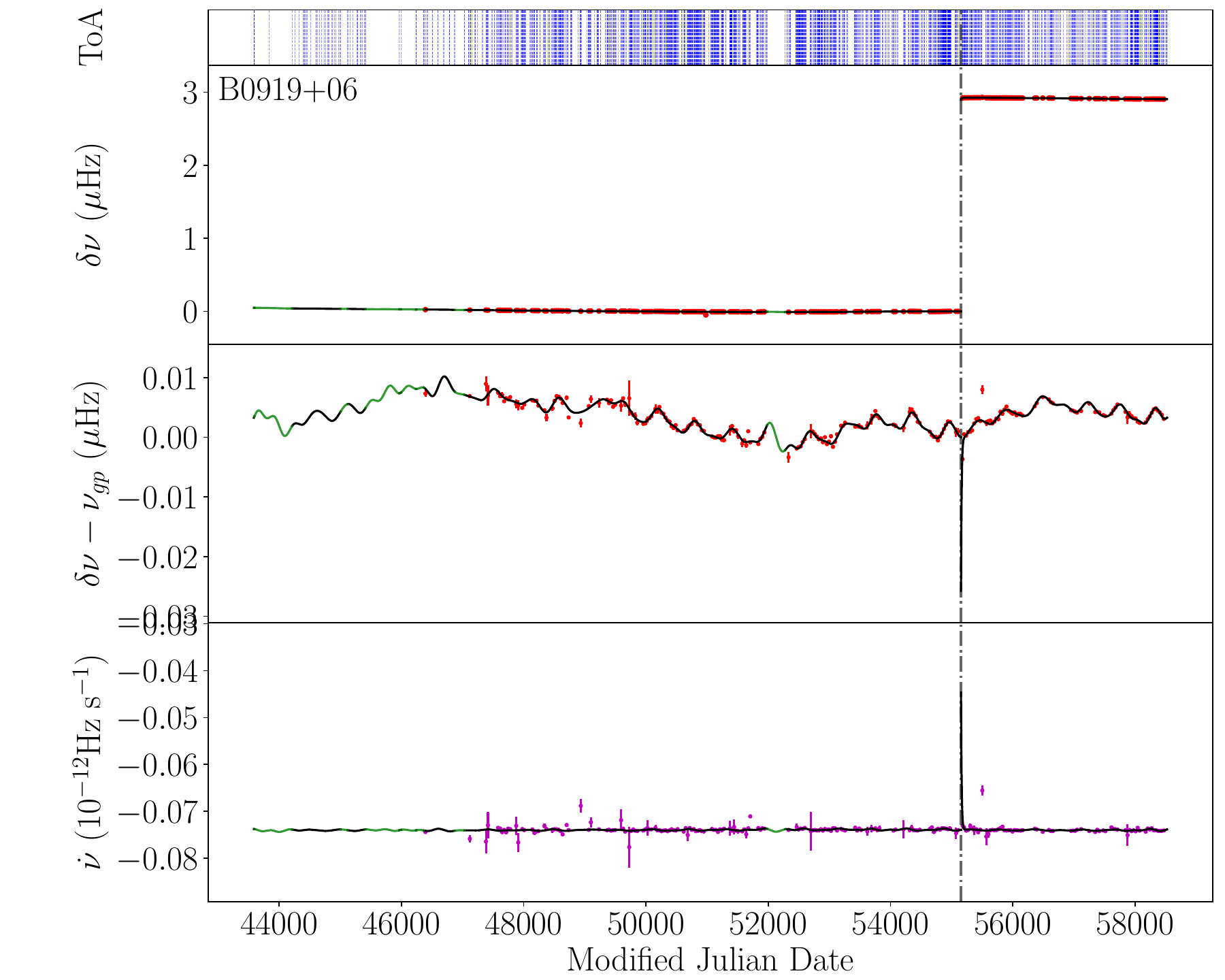}
    \end{subfigure}
    \caption{The evolution of $\delta\nu$ and $\dot{\nu}$ in PSRs J0205$+$6449, B0355$+$54, J0611$+$1436, J0631$+$1036, J0729$-$1448 and B0919$+$06. The top mini panel and other three panels show respectively: The epochs of ToAs by vertical blue lines; residual $\delta\nu$, which is $\nu$ after subtracting $\nu_{\mathrm{f}}$ and $\dot{\nu}_{\mathrm{f}}(t-t_{\mathrm{f}})$; $\delta\nu$ after further subtracting the permanent changes due to glitches $\nu_{\mathrm{gp}}=\Delta\nu_{\mathrm{p}}+\Delta\dot{\nu}_{\mathrm{p}}\Delta t+\frac{1}{2}\Delta\ddot{\nu}_{\mathrm{p}}\Delta t^{2}$ and $\frac{1}{2}\ddot{\nu}_{\mathrm{f}}(t-t_{\mathrm{f}})^{2}$, as well as some constant offset so that the point before the first glitch is zero; $\dot{\nu}$ as a function of time. The reference epoch $t_{\mathrm{f}}$, reference frequency $\nu_{\mathrm{f}}$, and its derivatives ($\dot{\nu}_{\mathrm{f}}$, $\ddot{\nu}_{\mathrm{f}}$) are some constants given in Appendix~\ref{sec: appendix} in Table~\ref{tab: plot_parameters}. Glitches best-modelled as steps in $\nu$ and $\dot{\nu}$ (without exponential recoveries) are indicated with dotted lines at their glitch epochs. Dash-dotted lines mark the epoch of glitches for which one exponential term was included in the best model, dashed lines for those with two exponentials, and solid lines for those with three exponentials. An analytical model derived from the best-fit glitch parameters is shown as a black-solid curve. We use green-solid lines to mark gaps longer than 10 times the pulsar's average cadence in ToAs. The stride fitting results for $\delta\nu$ and $\dot{\nu}$ are shown as the red and purple dots with errorbars respectively.}
    \label{fig: nu_nudot_J0205-B0919}
\end{figure*}

\begin{figure*}
    \centering
    \begin{subfigure}[b]{0.49\textwidth}
        \centering
        \includegraphics[scale=0.29]{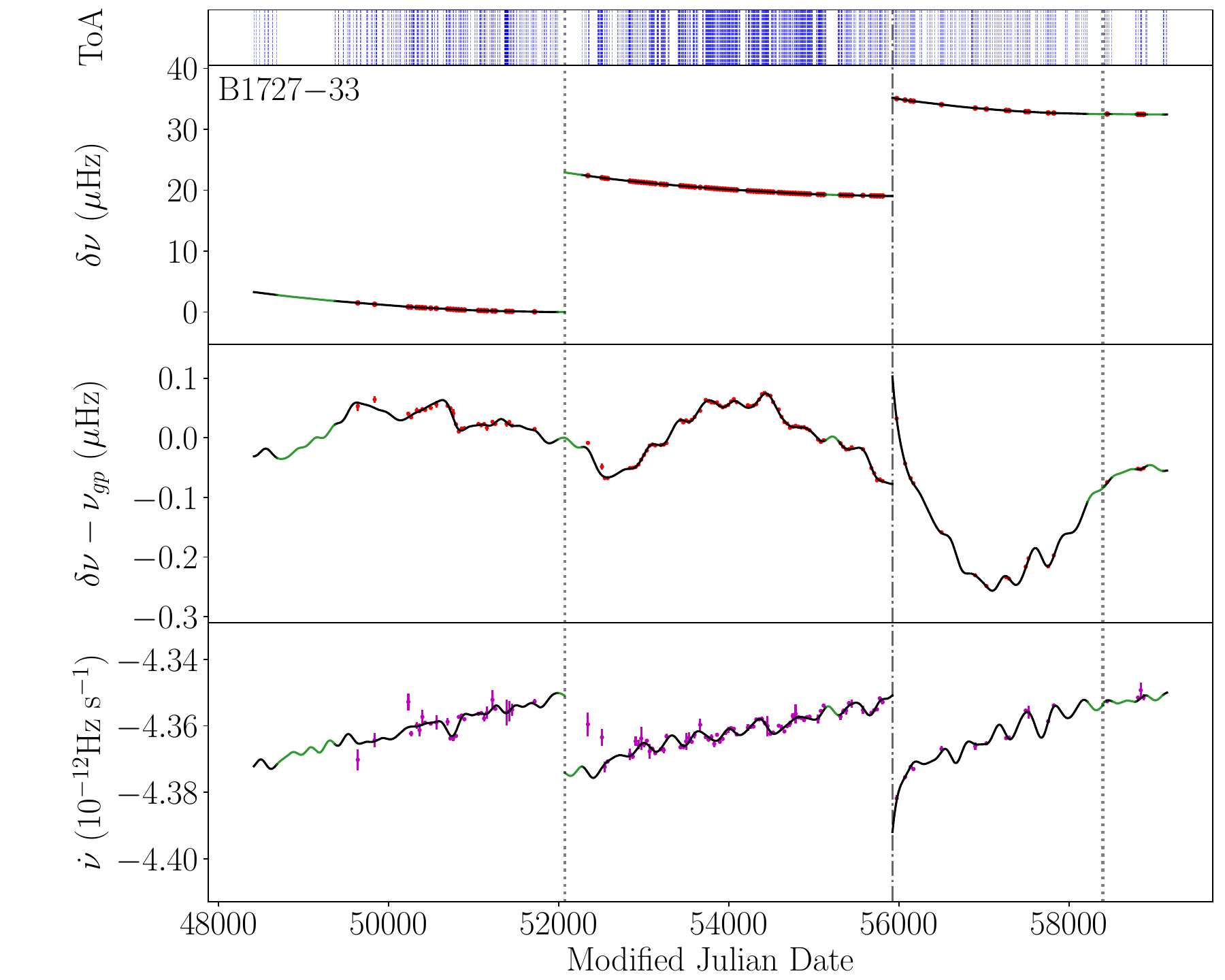}
    \end{subfigure}
    \begin{subfigure}[b]{0.49\textwidth}
        \centering
        \includegraphics[scale=0.29]{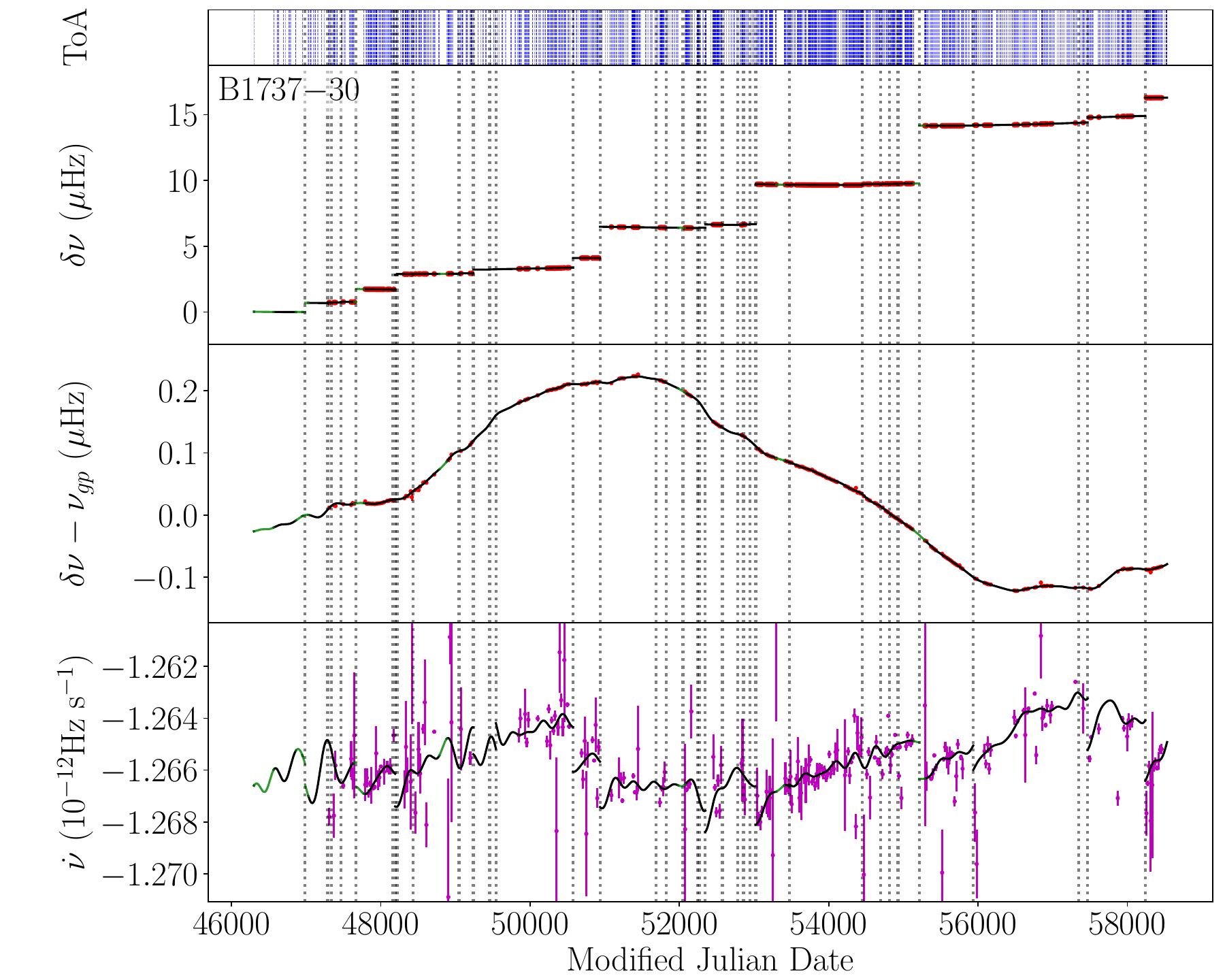}
    \end{subfigure}
    \begin{subfigure}[b]{0.49\textwidth}
        \centering
        \includegraphics[scale=0.29]{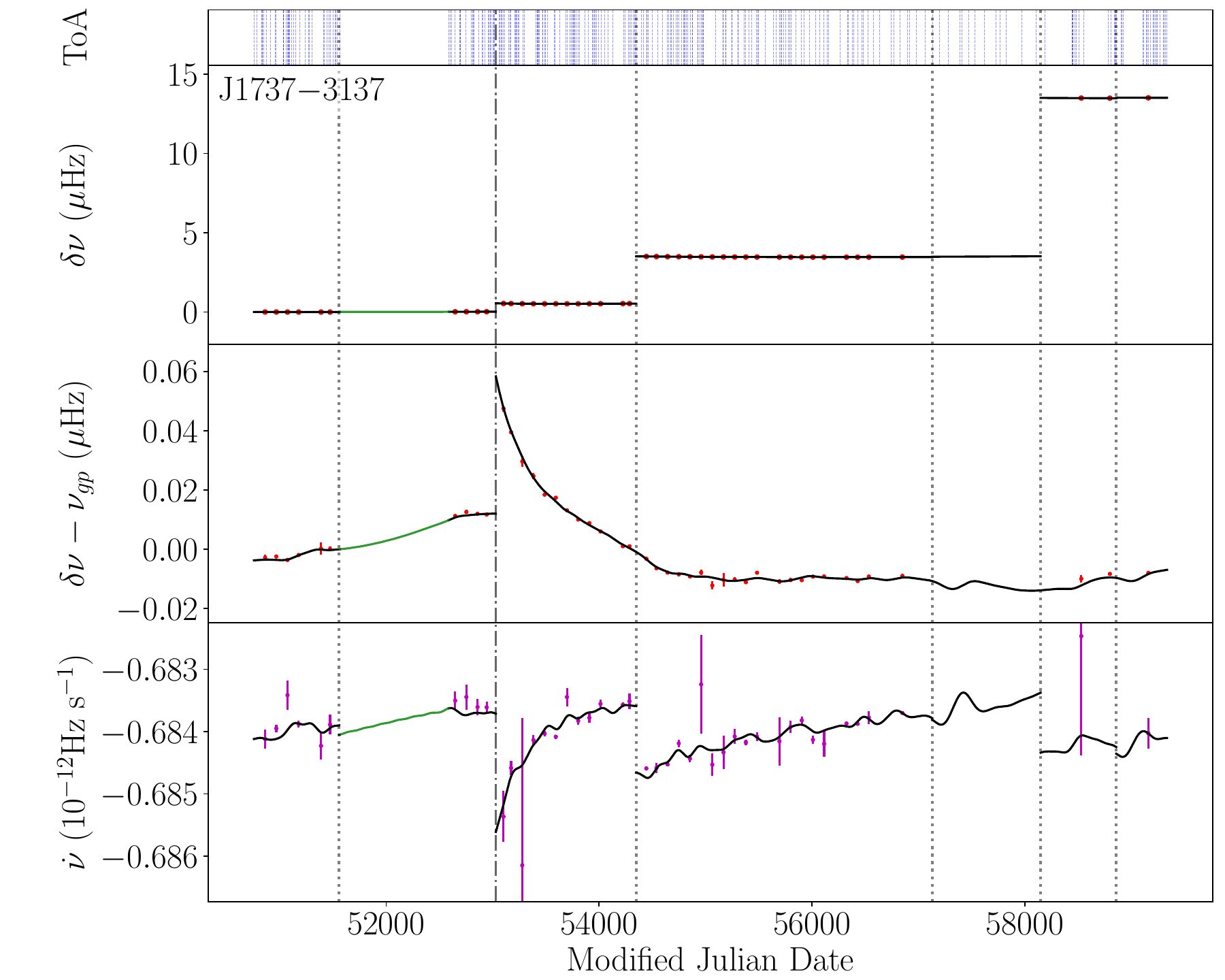}
    \end{subfigure}
    \begin{subfigure}[b]{0.49\textwidth}
        \centering
        \includegraphics[scale=0.29]{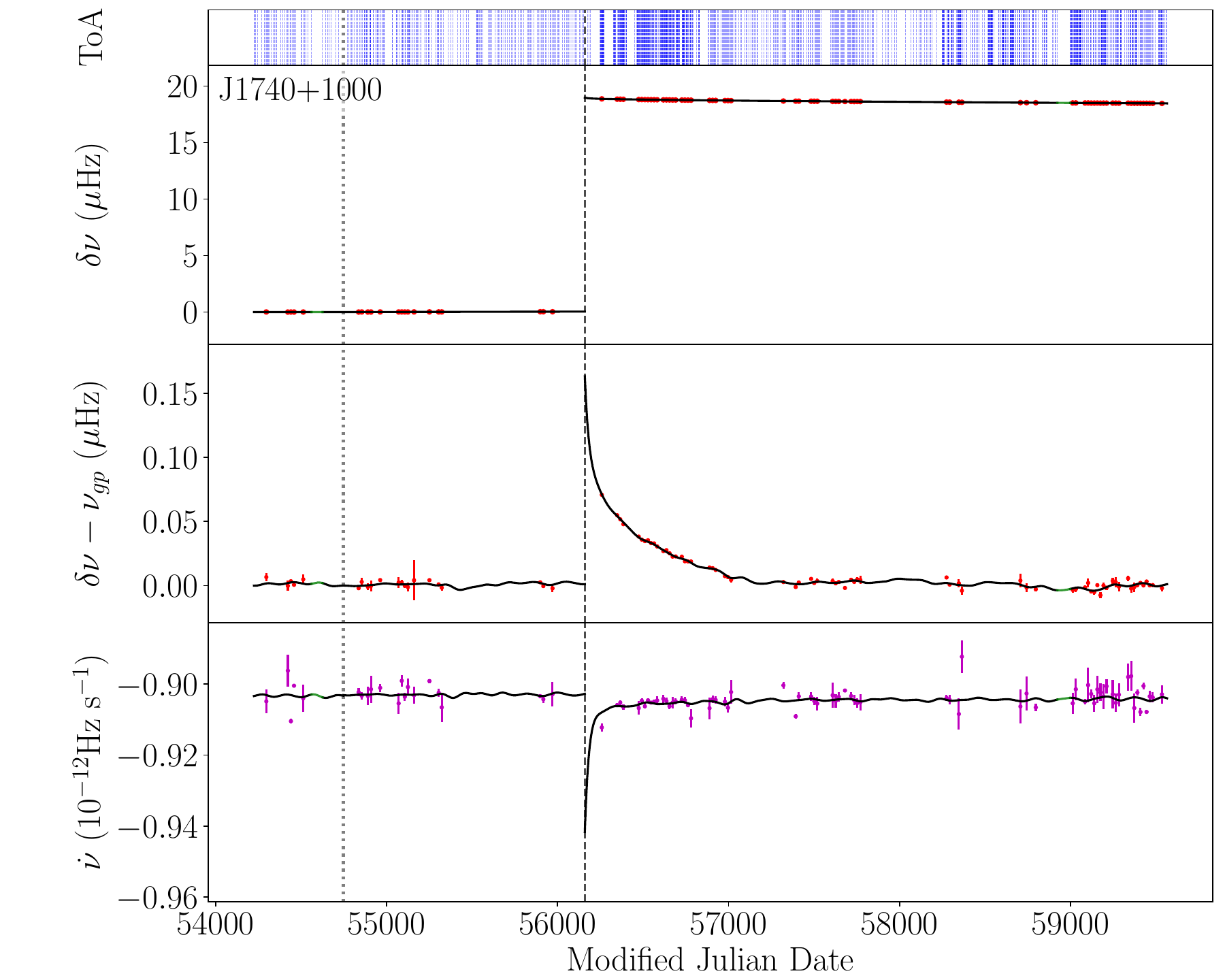}
    \end{subfigure}
    \begin{subfigure}[b]{0.49\textwidth}
        \centering
        \includegraphics[scale=0.29]{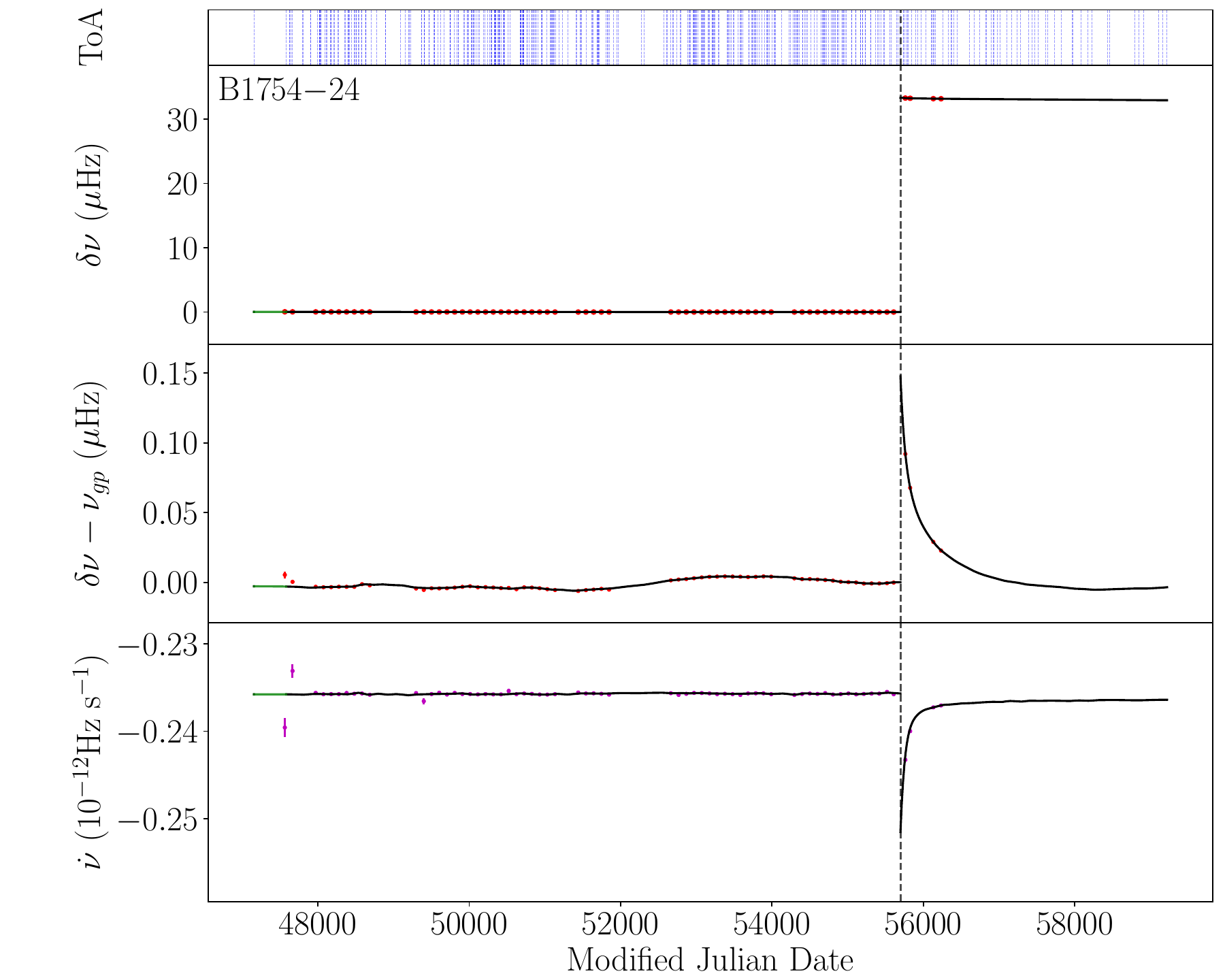}
    \end{subfigure}
    \begin{subfigure}[b]{0.49\textwidth}
        \centering
        \includegraphics[scale=0.29]{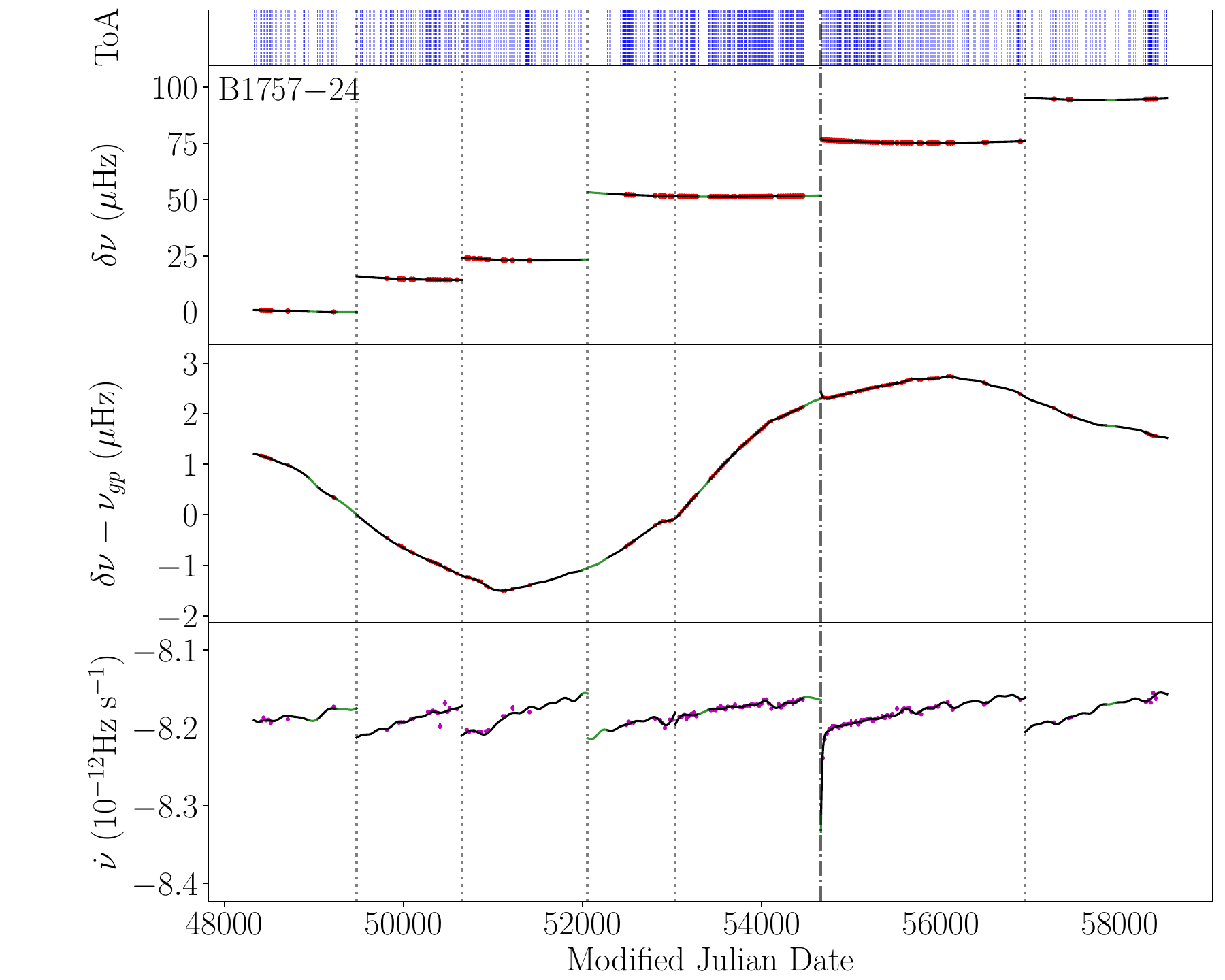}
    \end{subfigure}
    \caption{The evolution of $\delta\nu$ and $\dot{\nu}$ in PSRs B1727$-$33, B1737$-$30, J1737$-$3137, J1740$+$1000, B1754$-$24 and B1757$-$24. See the caption of Figure~\ref{fig: nu_nudot_J0205-B0919} for further details.}
    \label{fig: nu_nudot_B1727-B1757}
\end{figure*}

\begin{figure*}
    \centering
    \begin{subfigure}[b]{0.49\textwidth}
        \centering
        \includegraphics[scale=0.29]{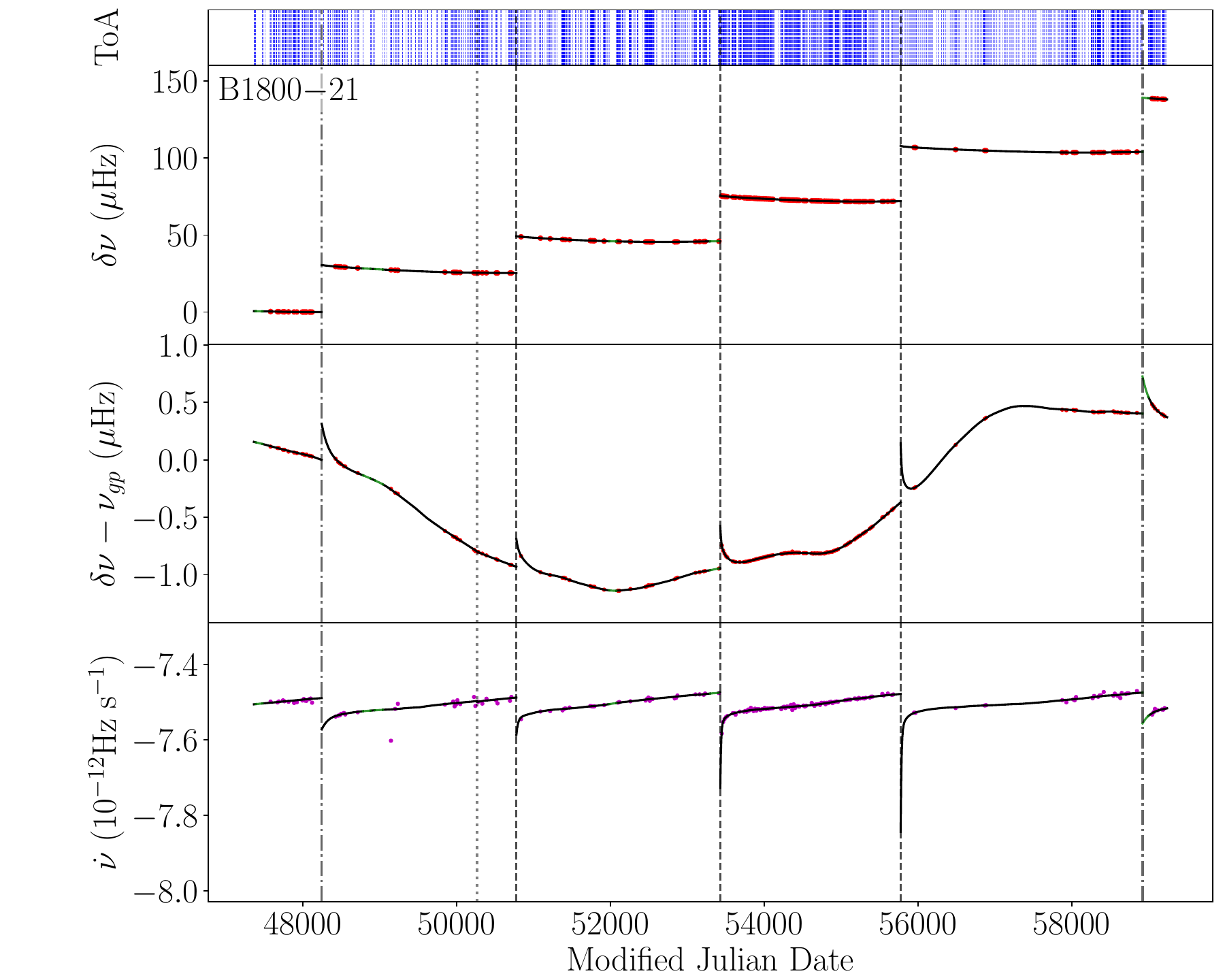}
    \end{subfigure}
    \begin{subfigure}[b]{0.49\textwidth}
        \centering
        \includegraphics[scale=0.29]{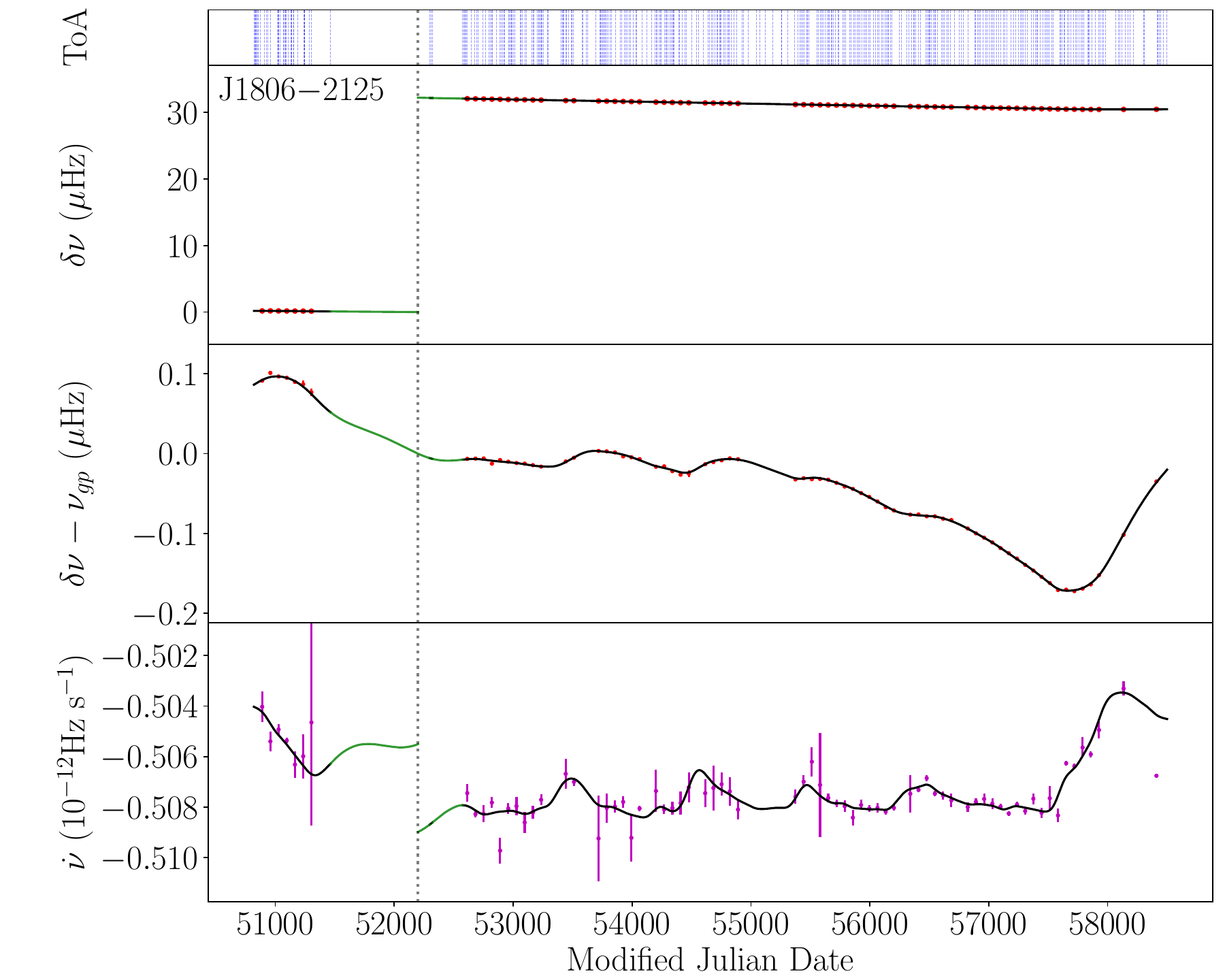}
    \end{subfigure}
    \begin{subfigure}[b]{0.49\textwidth}
        \centering
        \includegraphics[scale=0.29]{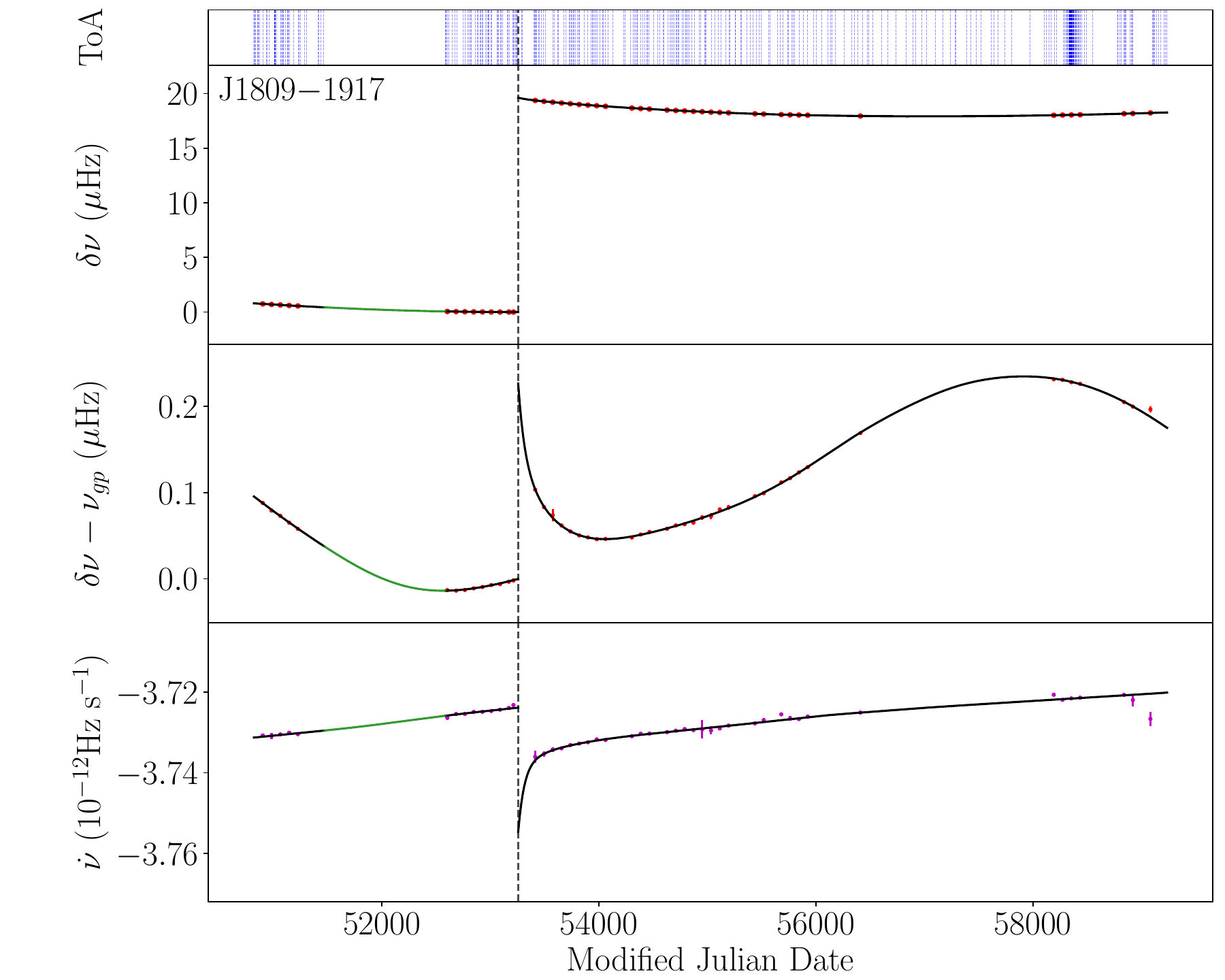}
    \end{subfigure}
    \begin{subfigure}[b]{0.49\textwidth}
        \centering
        \includegraphics[scale=0.29]{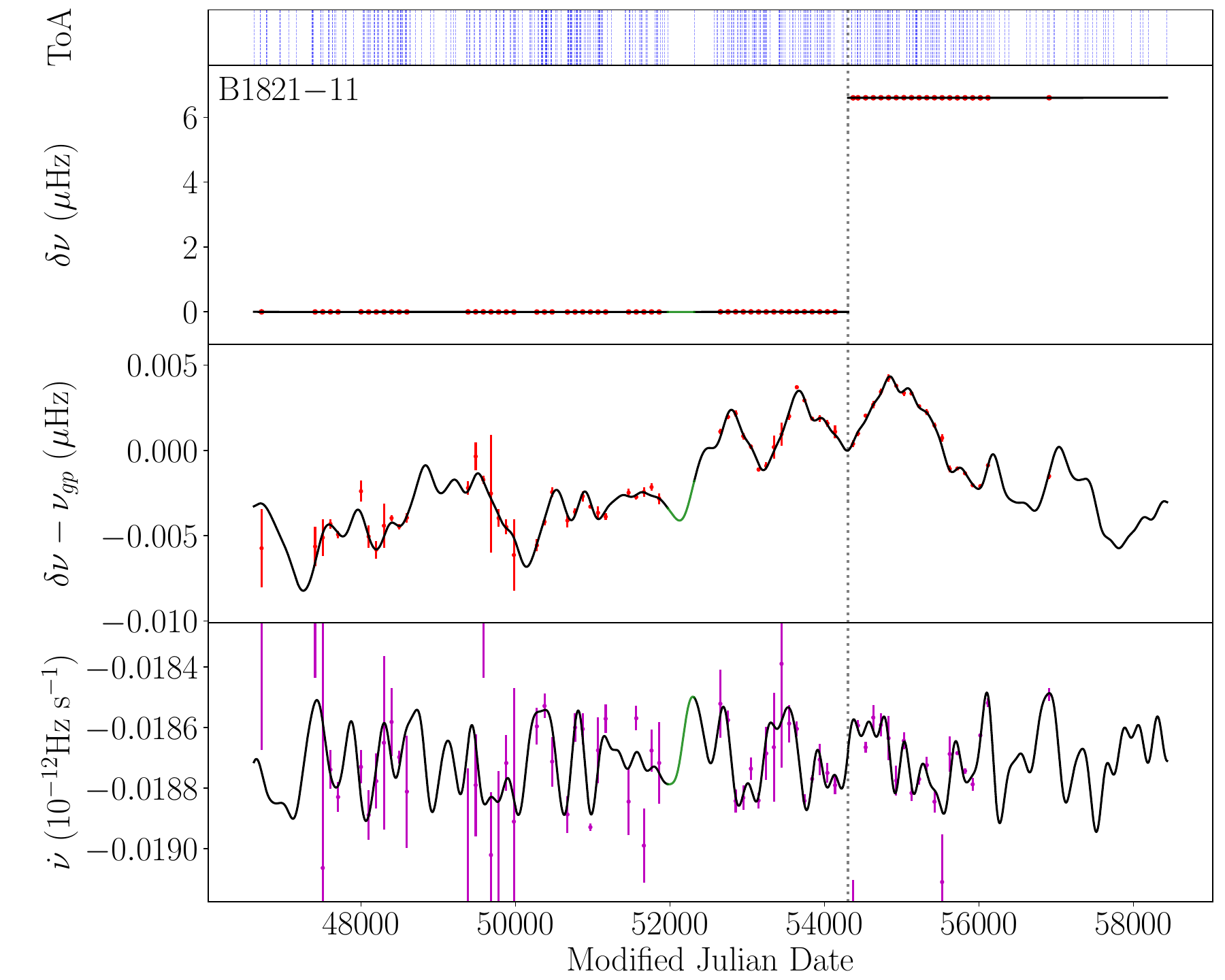}
    \end{subfigure}
    \begin{subfigure}[b]{0.49\textwidth}
        \centering
        \includegraphics[scale=0.29]{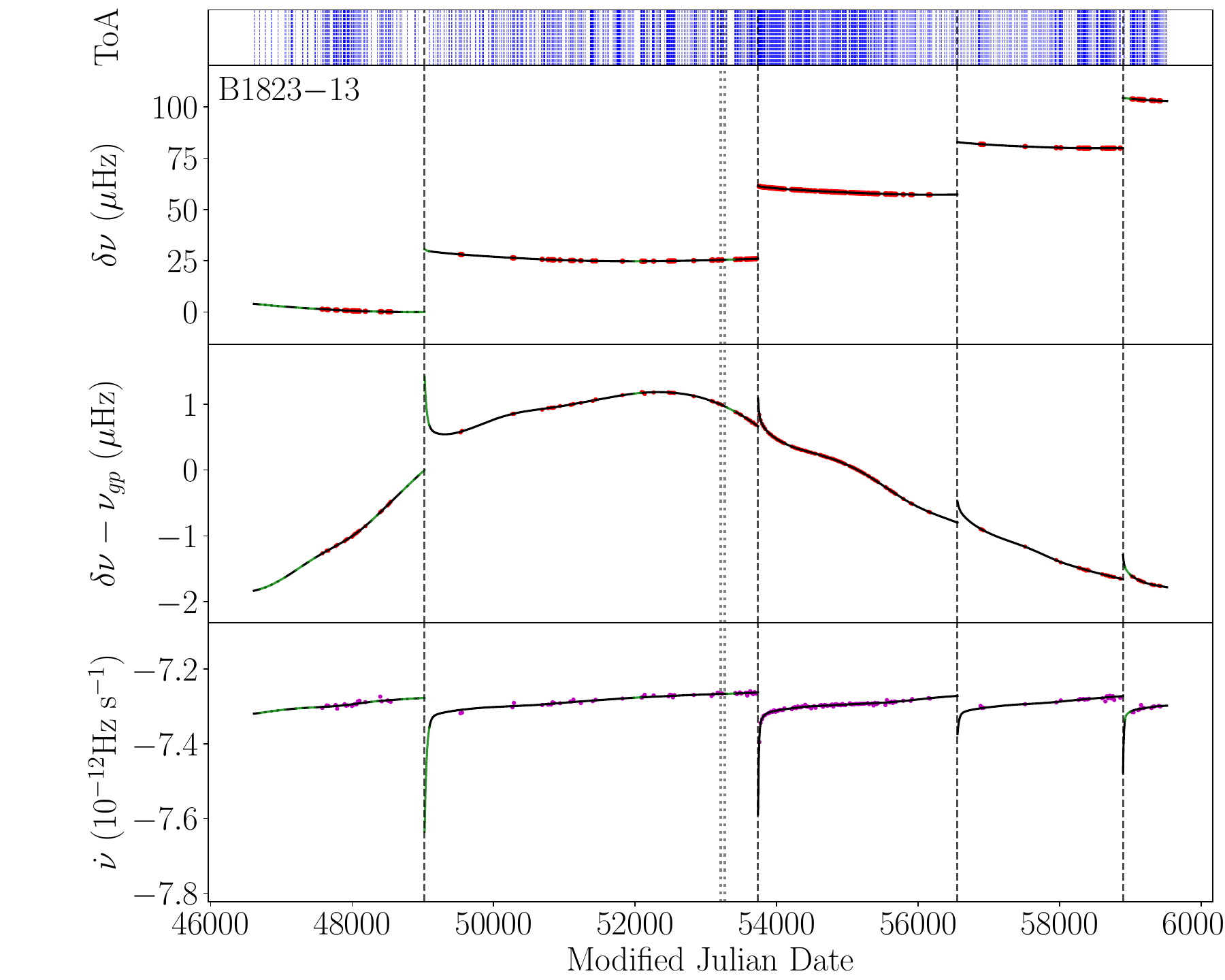}
    \end{subfigure}
    \begin{subfigure}[b]{0.49\textwidth}
        \centering
        \includegraphics[scale=0.29]{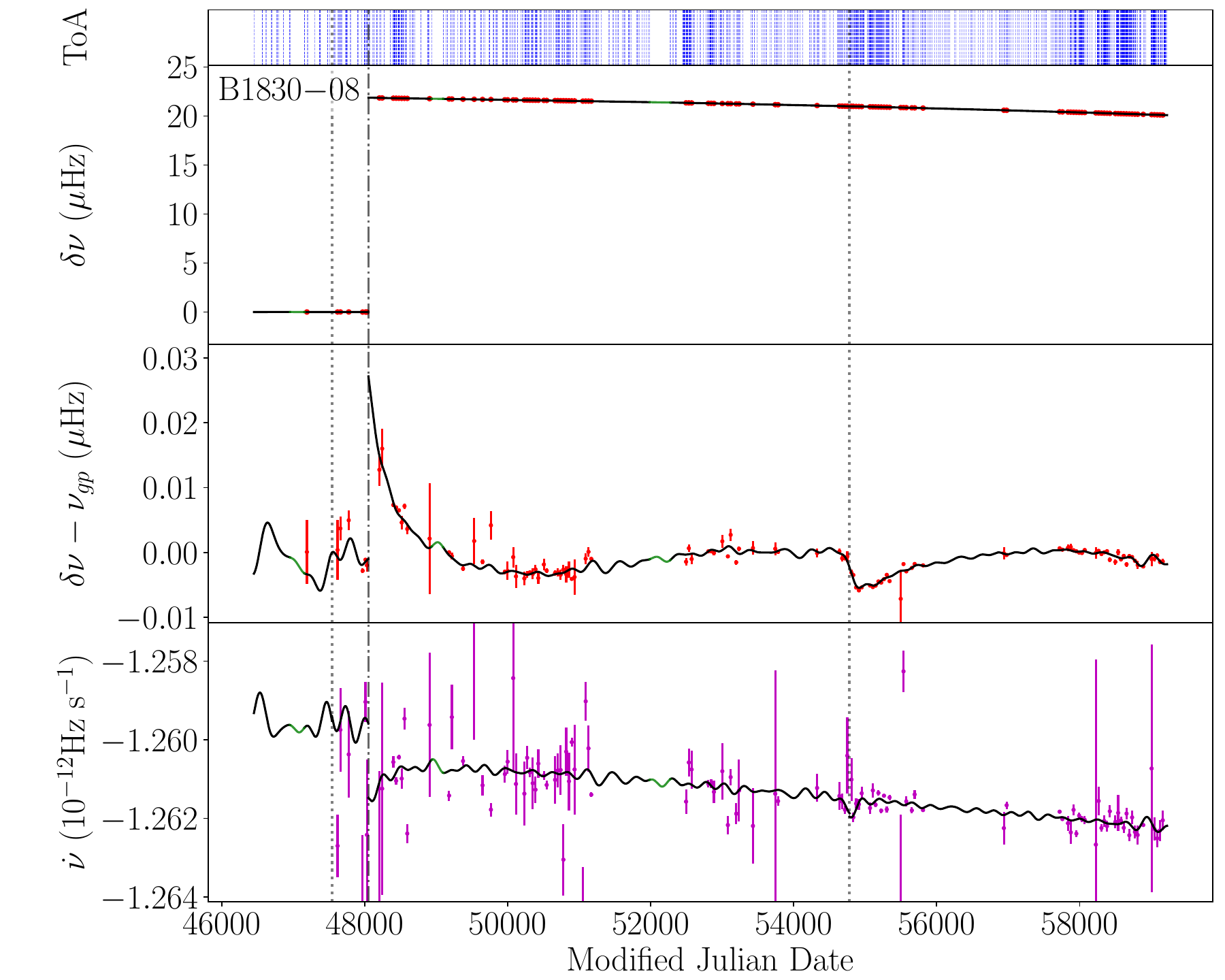}
    \end{subfigure}
    \caption{The evolution of $\delta\nu$ and $\dot{\nu}$ in PSRs B1800$-$21, J1806$-$2125, J1809$-$1917, B1821$-$11, B1823$-$13 and B1830$-$08. See the caption of Figure~\ref{fig: nu_nudot_J0205-B0919} for further details.}
    \label{fig: nu_nudot_B1800-B1830}
\end{figure*}

\begin{figure*}
    \centering
    \begin{subfigure}[b]{0.49\textwidth}
        \centering
        \includegraphics[scale=0.29]{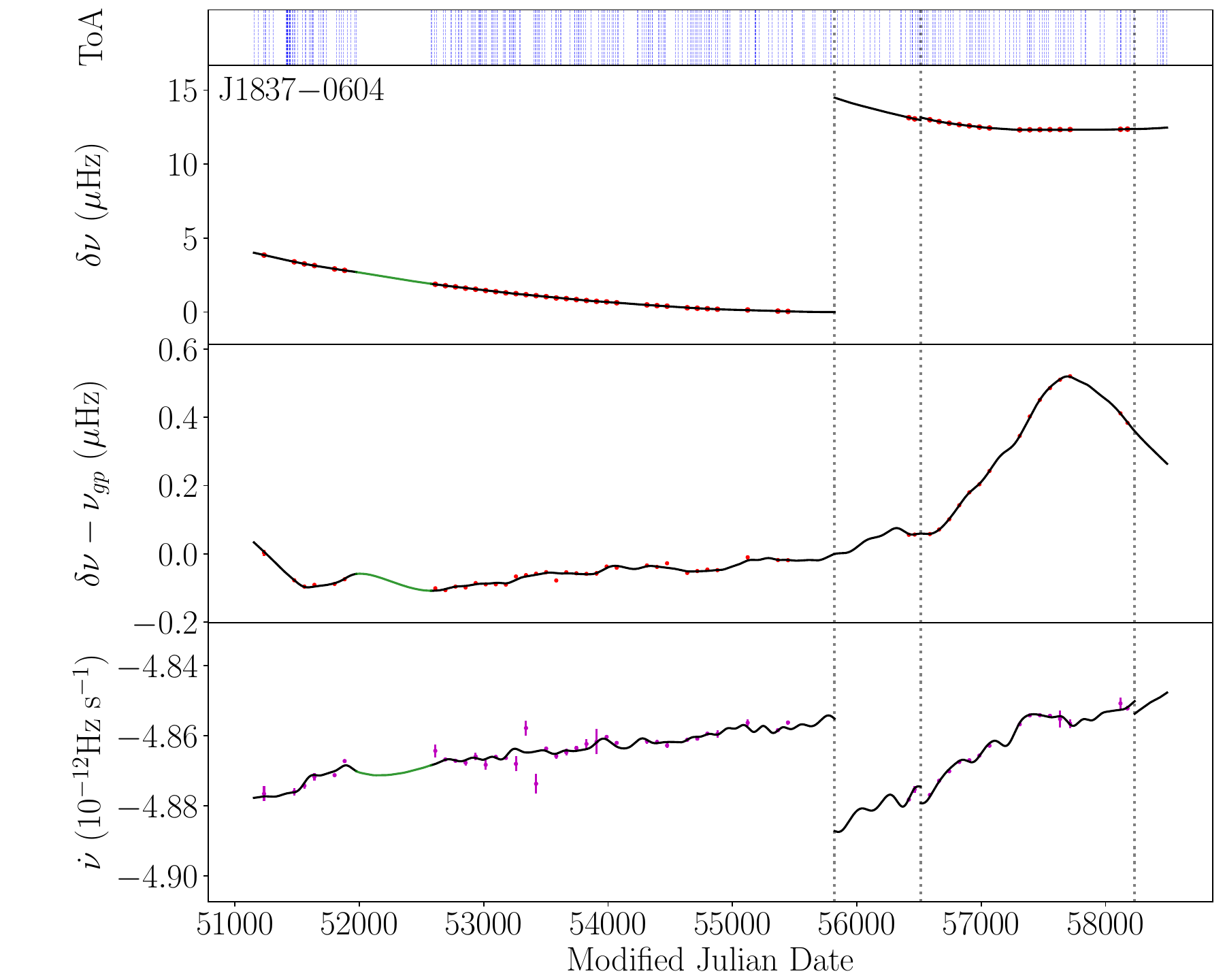}
    \end{subfigure}
    \begin{subfigure}[b]{0.49\textwidth}
        \centering
        \includegraphics[scale=0.29]{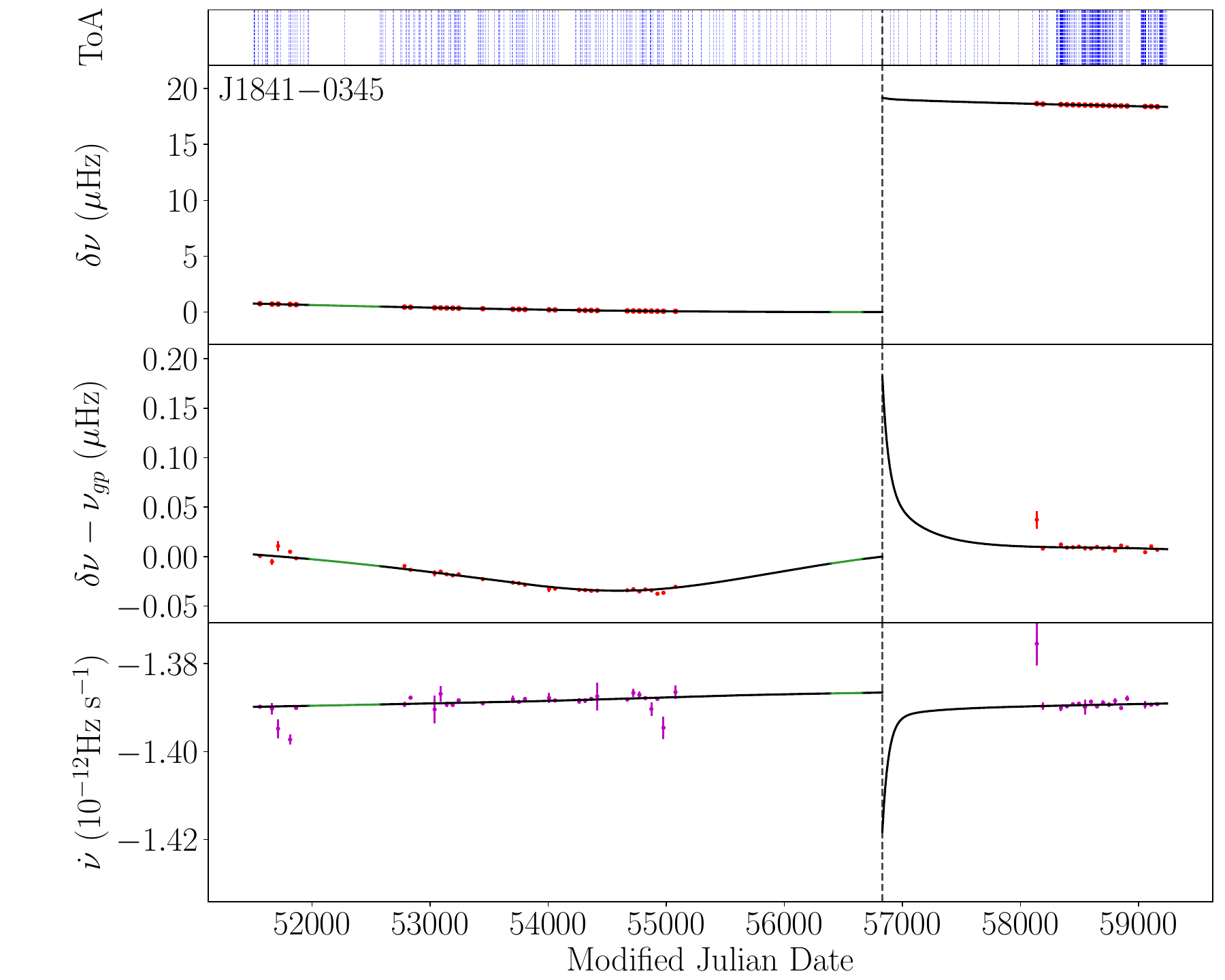}
    \end{subfigure}
    \begin{subfigure}[b]{0.49\textwidth}
        \centering
        \includegraphics[scale=0.29]{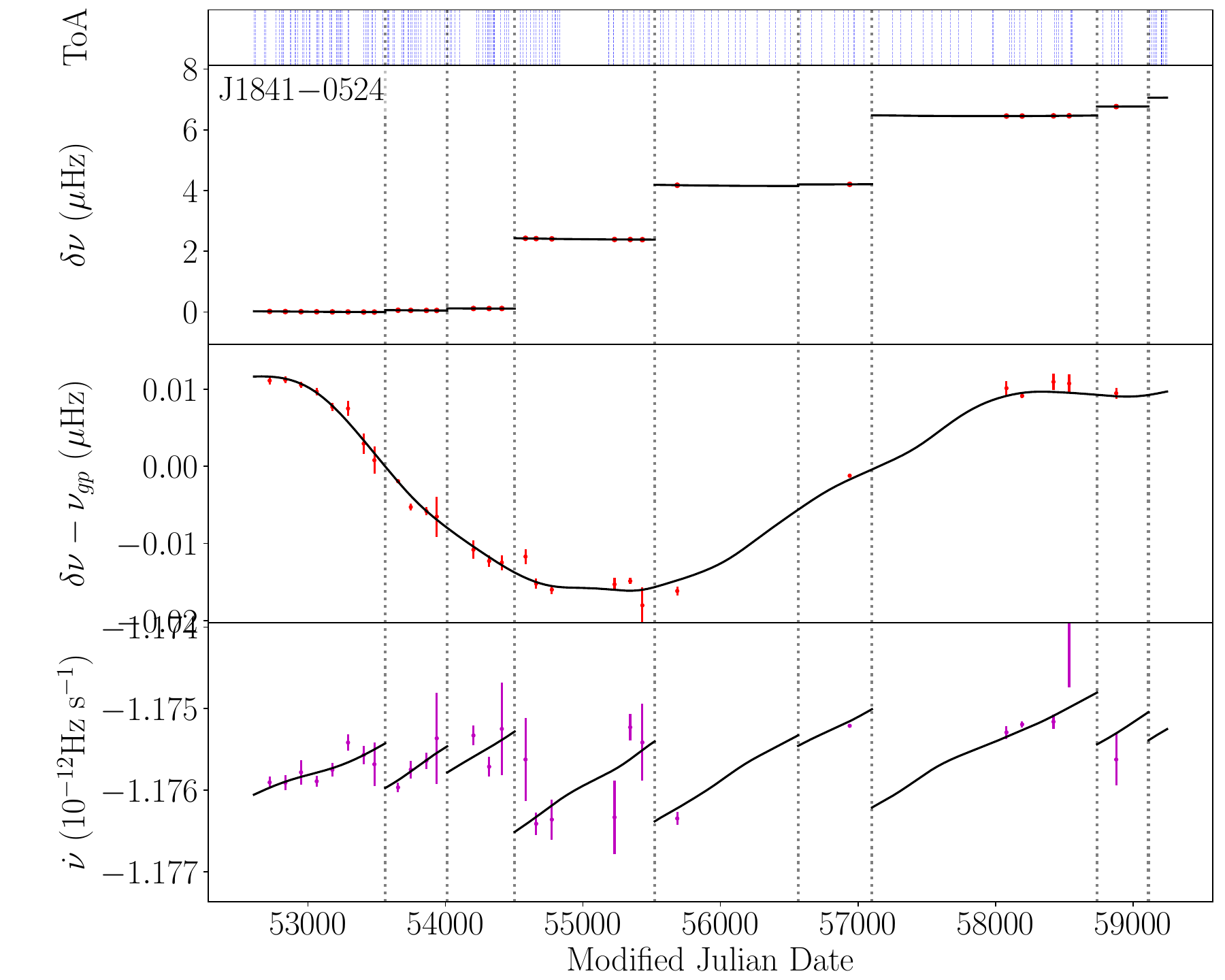}
    \end{subfigure}
    \begin{subfigure}[b]{0.49\textwidth}
        \centering
        \includegraphics[scale=0.29]{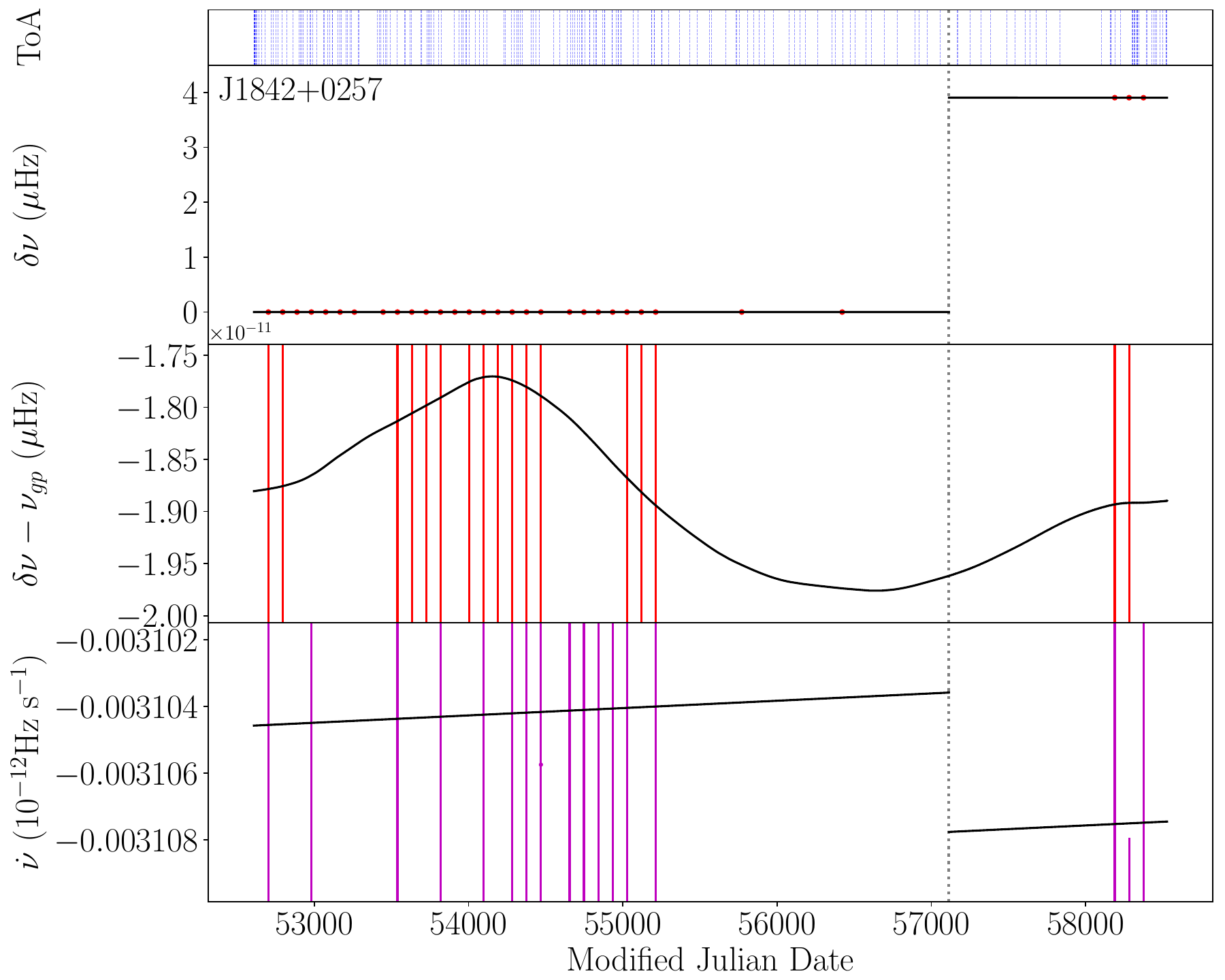}
    \end{subfigure}
    \begin{subfigure}[b]{0.49\textwidth}
        \centering
        \includegraphics[scale=0.29]{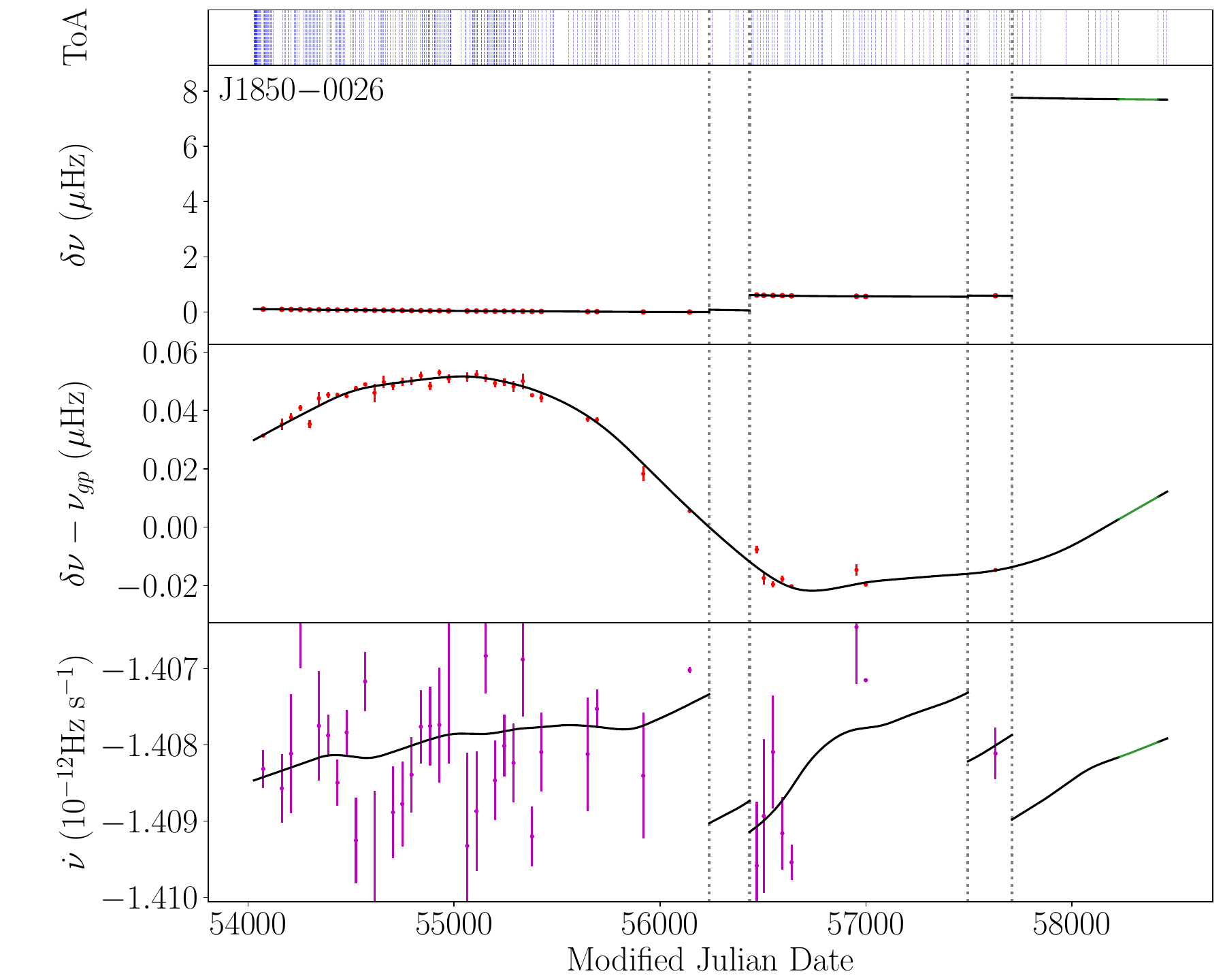}
    \end{subfigure}
    \begin{subfigure}[b]{0.49\textwidth}
        \centering
        \includegraphics[scale=0.29]{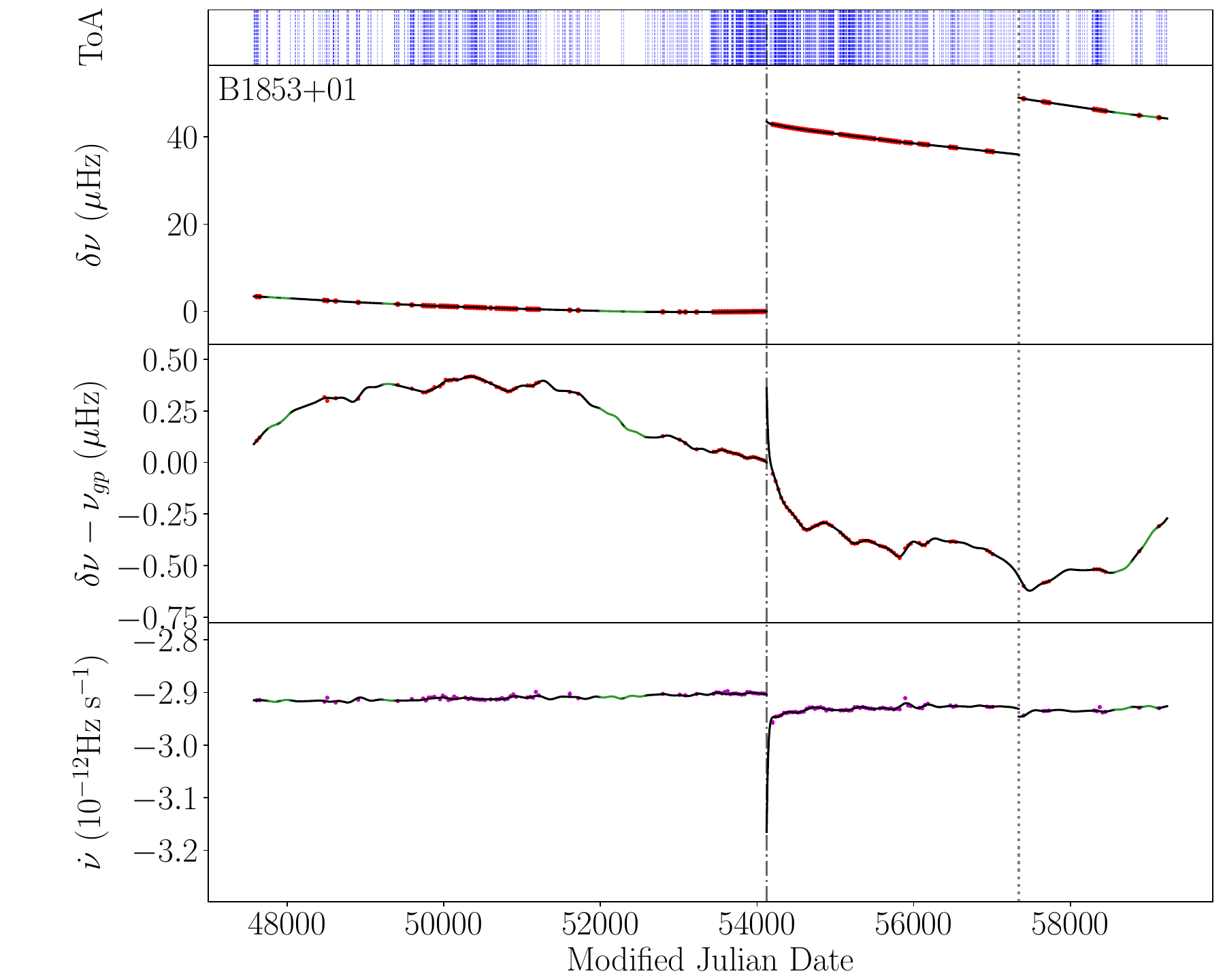}
    \end{subfigure}
    \caption{The evolution of $\delta\nu$ and $\dot{\nu}$ in PSRs J1837$-$0604, J1841$-$0345, J1841$-$0524, J1842$+$0257, J1850$-$0026 and B1853$+$01. See the caption of Figure~\ref{fig: nu_nudot_J0205-B0919} for further details.}
    \label{fig: nu_nudot_J1837-B1853}
\end{figure*}

\begin{figure*}
    \centering
    \begin{subfigure}[b]{0.49\textwidth}
        \centering
        \includegraphics[scale=0.29]{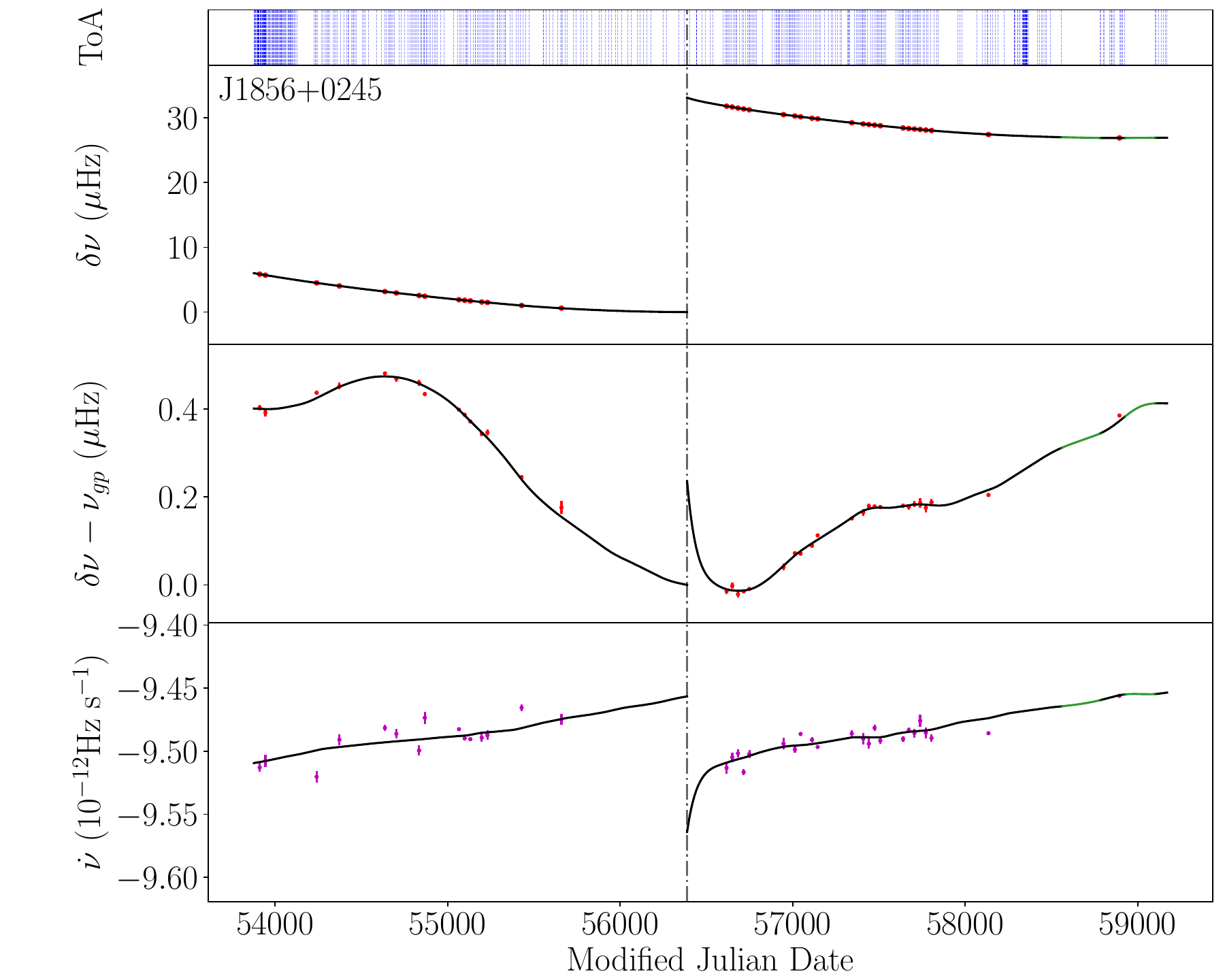}
    \end{subfigure}
    \begin{subfigure}[b]{0.49\textwidth}
        \centering
        \includegraphics[scale=0.29]{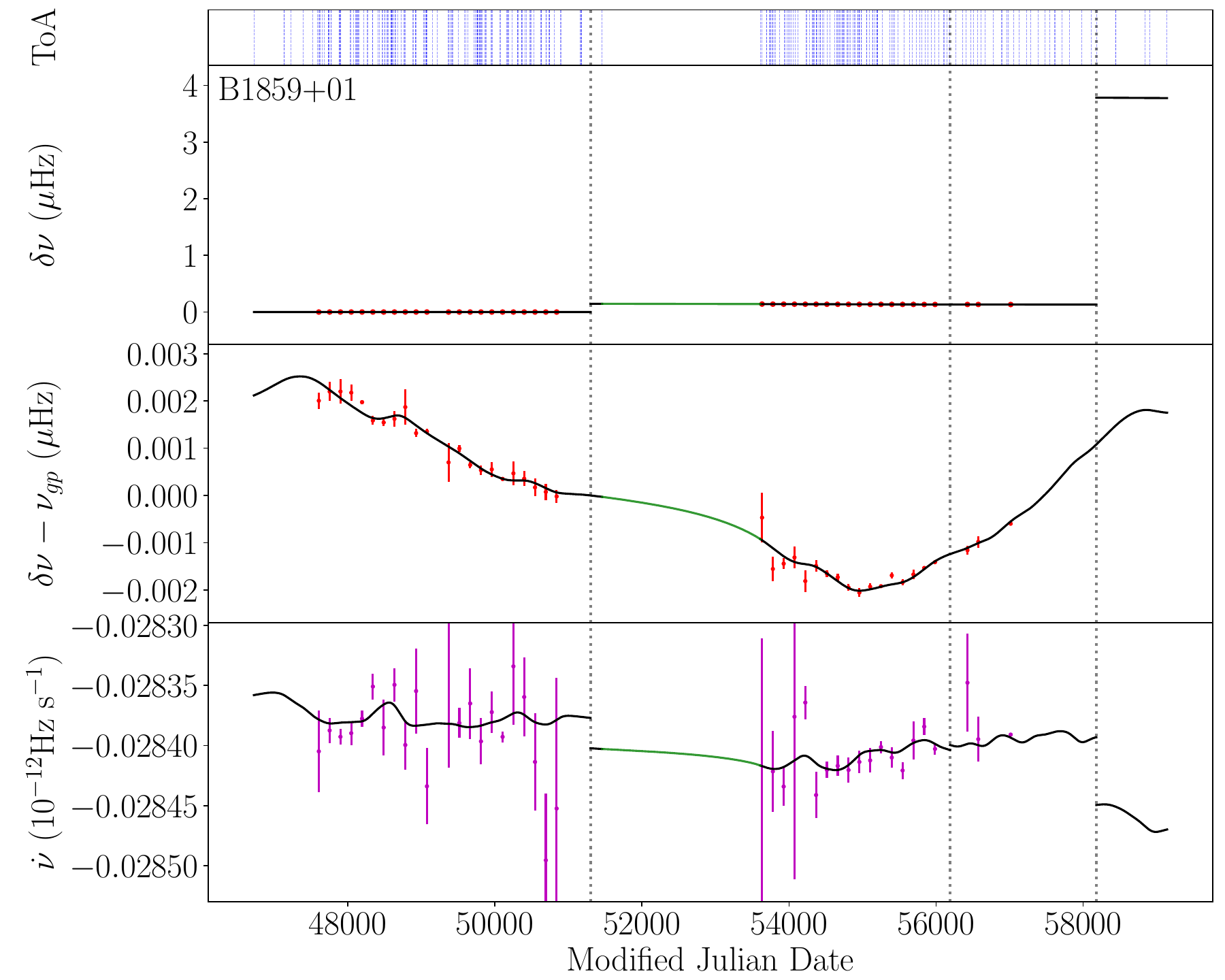}
    \end{subfigure}
    \begin{subfigure}[b]{0.49\textwidth}
        \centering
        \includegraphics[scale=0.29]{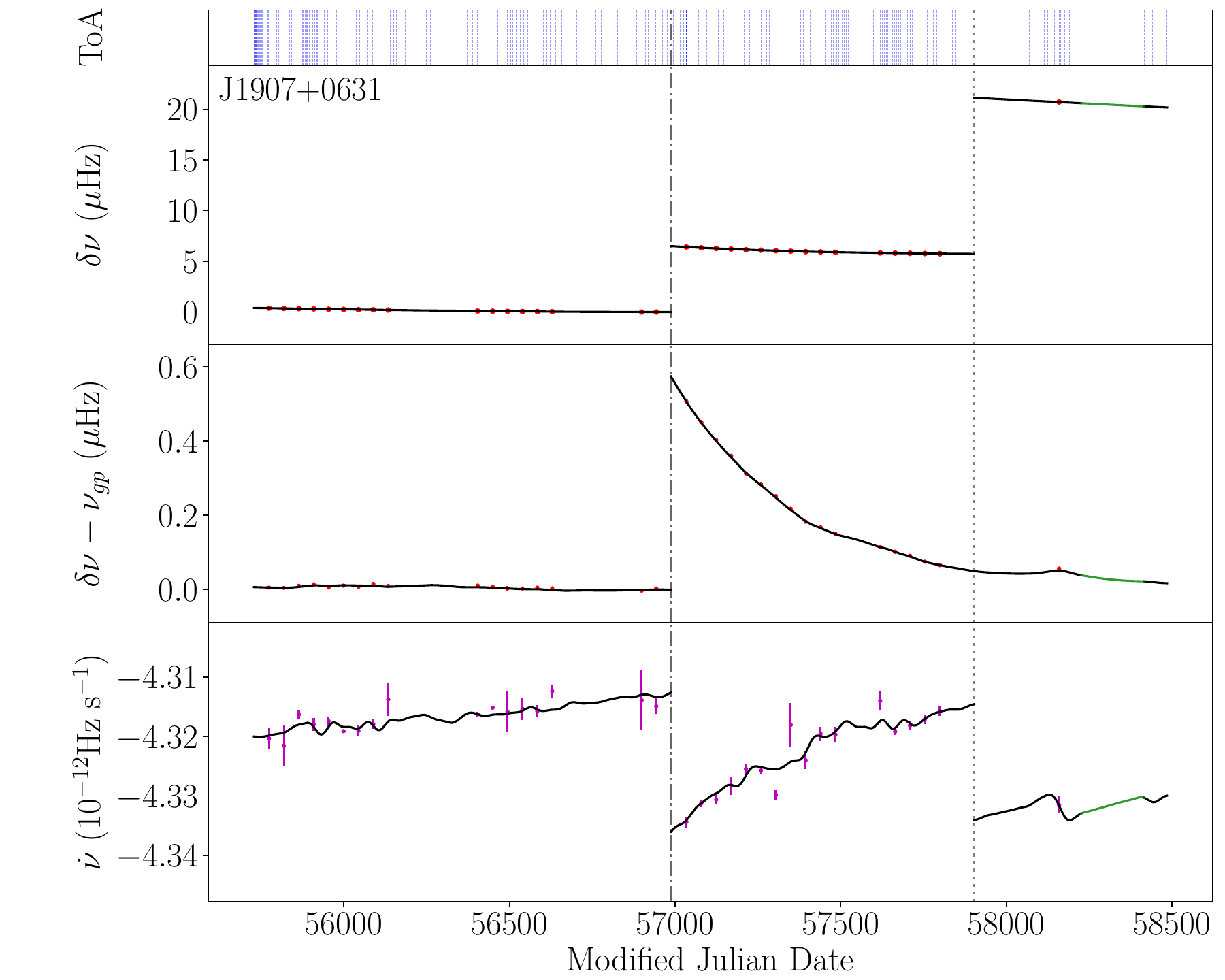}
    \end{subfigure}
    \begin{subfigure}[b]{0.49\textwidth}
        \centering
        \includegraphics[scale=0.29]{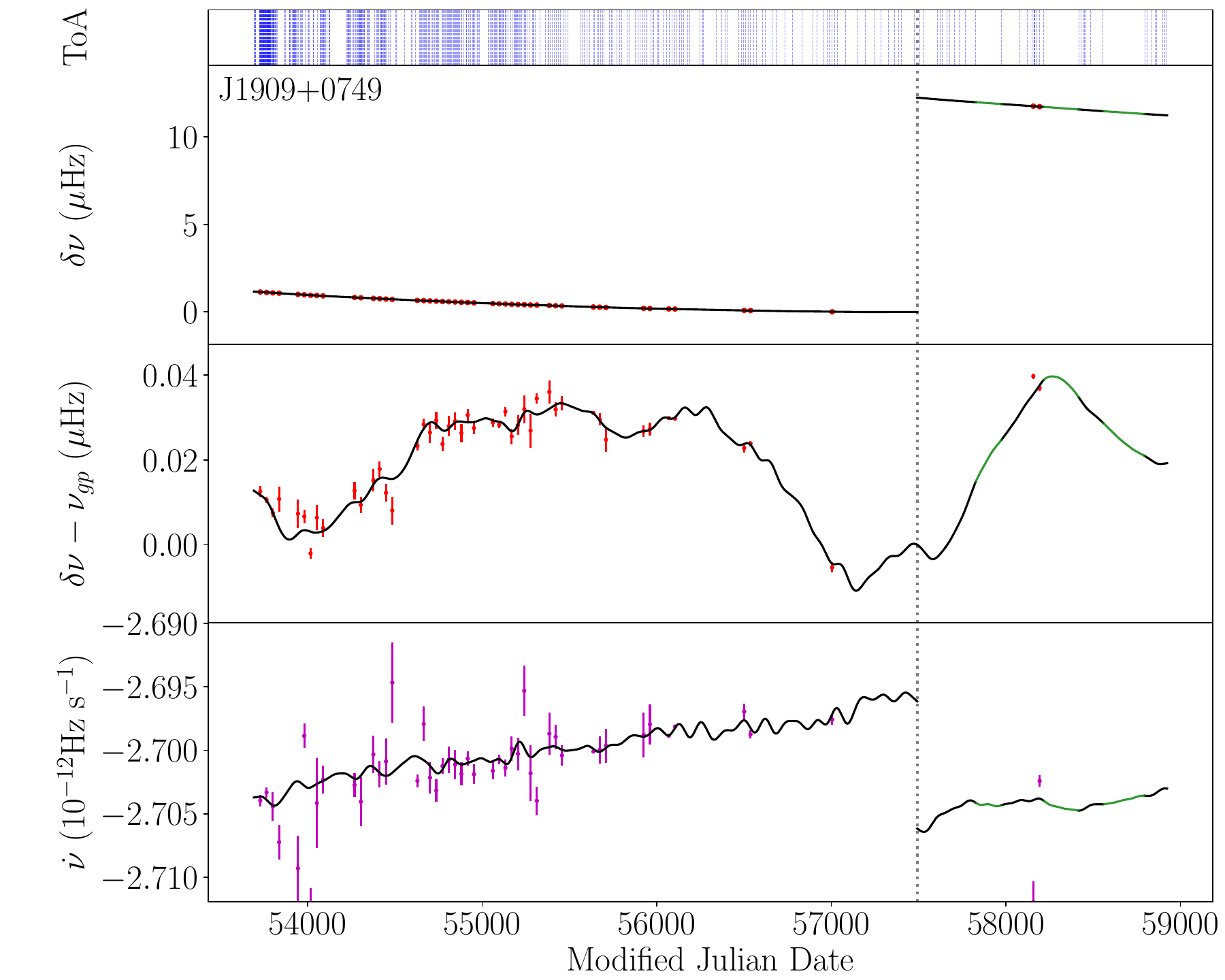}
    \end{subfigure}
    \begin{subfigure}[b]{0.49\textwidth}
        \centering
        \includegraphics[scale=0.29]{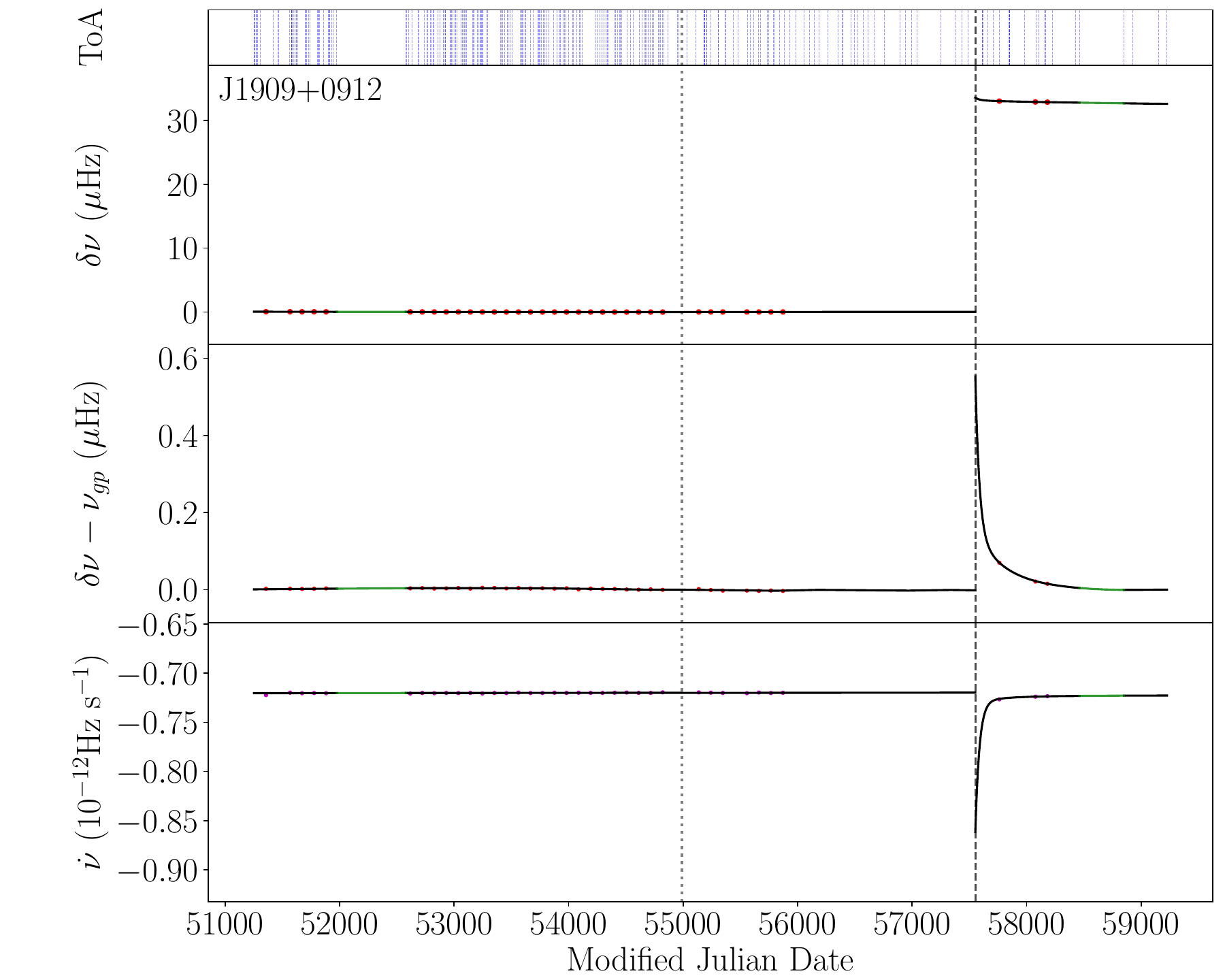}
    \end{subfigure}
    \begin{subfigure}[b]{0.49\textwidth}
        \centering
        \includegraphics[scale=0.29]{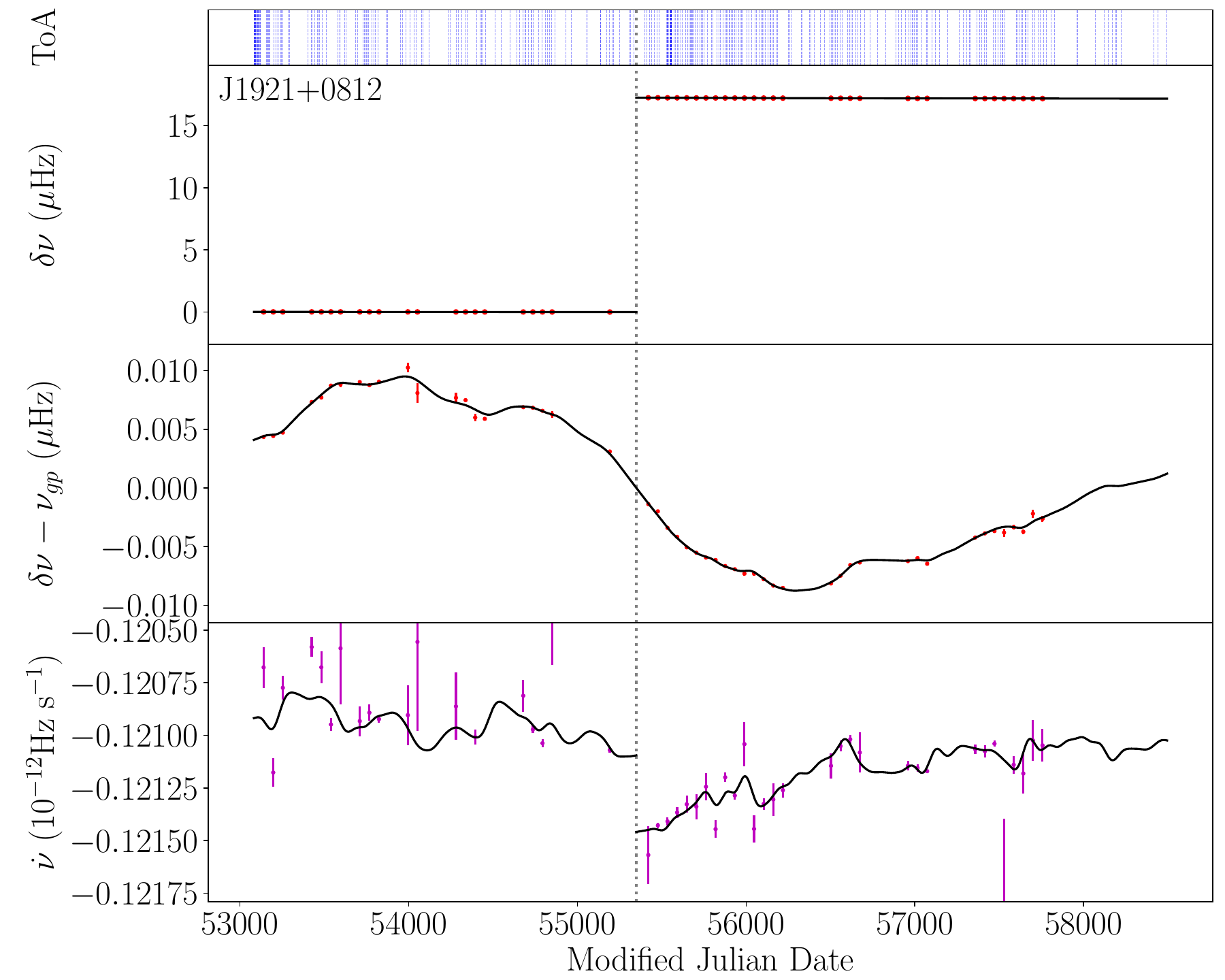}
    \end{subfigure}
    \caption{The evolution of $\delta\nu$ and $\dot{\nu}$ in PSRs J1856$+$0245, B1859$+$01, J1907$+$0631, J1909$+$0749, J1909$+$0912 and J1921$+$0812. See the caption of Figure~\ref{fig: nu_nudot_J0205-B0919} for further details.}
    \label{fig: nu_nudot_J1856-J1921}
\end{figure*}

\begin{figure*}
    \centering
    \begin{subfigure}[b]{0.49\textwidth}
        \centering
        \includegraphics[scale=0.29]{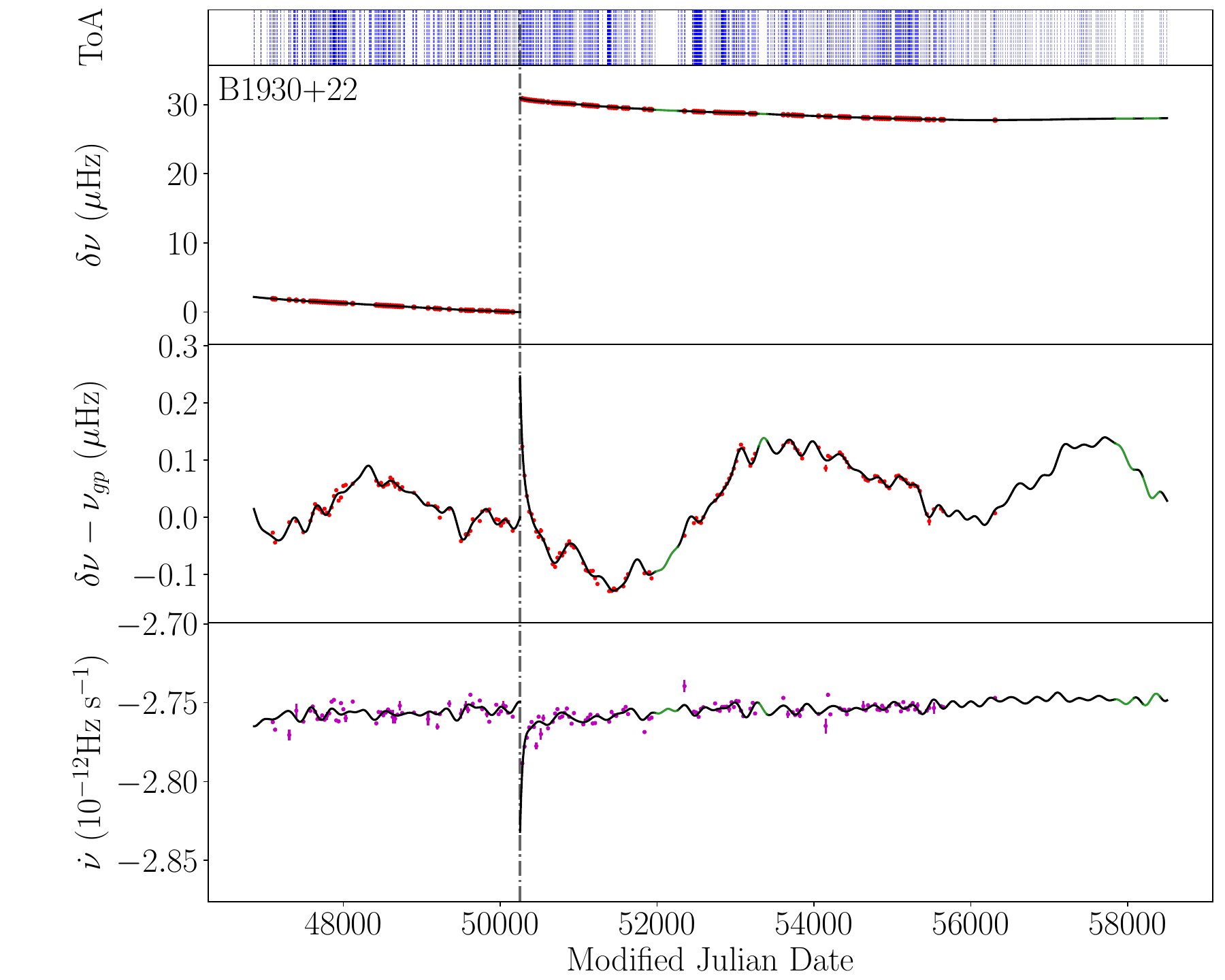}
    \end{subfigure}
    \begin{subfigure}[b]{0.49\textwidth}
        \centering
        \includegraphics[scale=0.29]{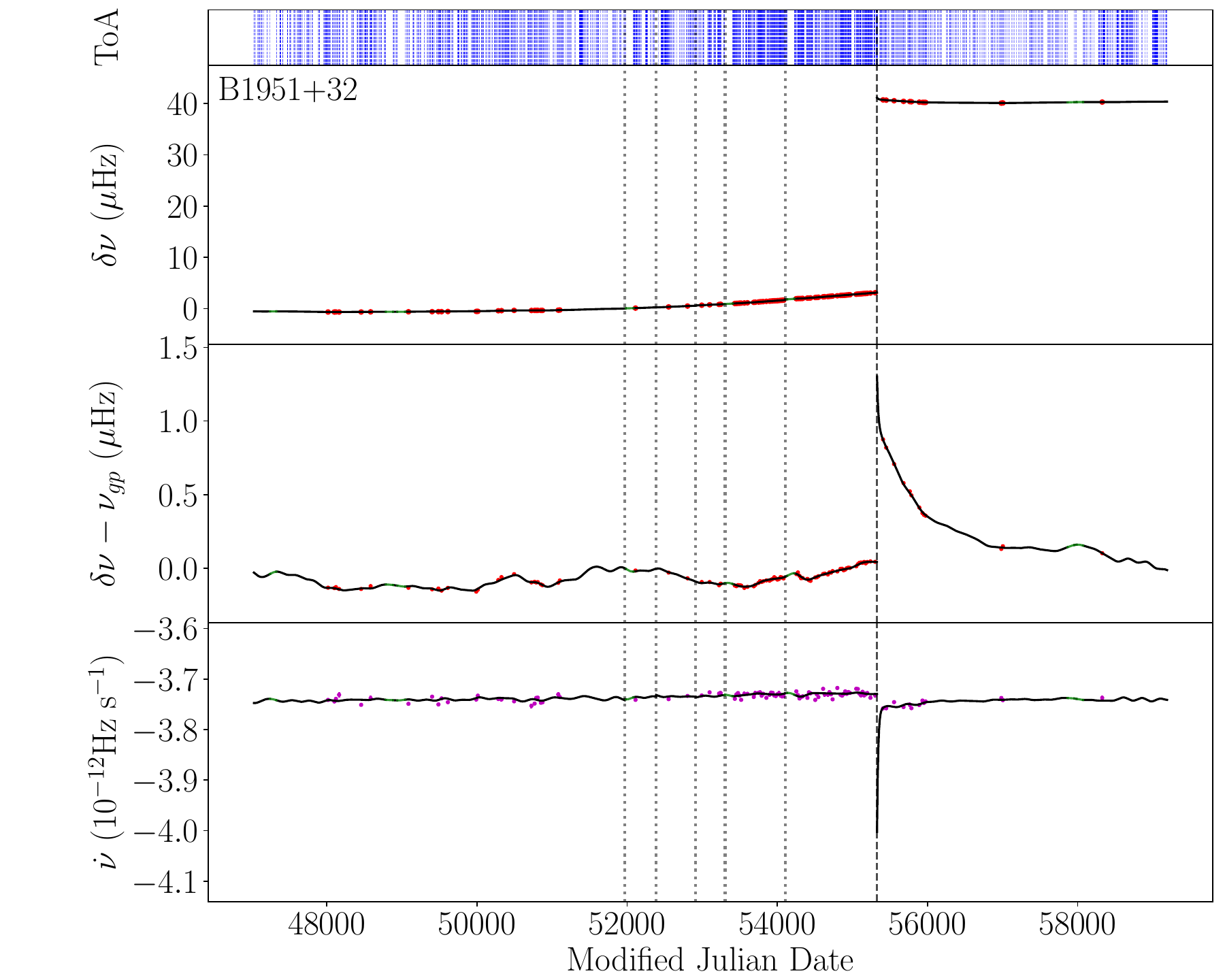}
    \end{subfigure}
    \begin{subfigure}[b]{0.49\textwidth}
        \centering
        \includegraphics[scale=0.29]{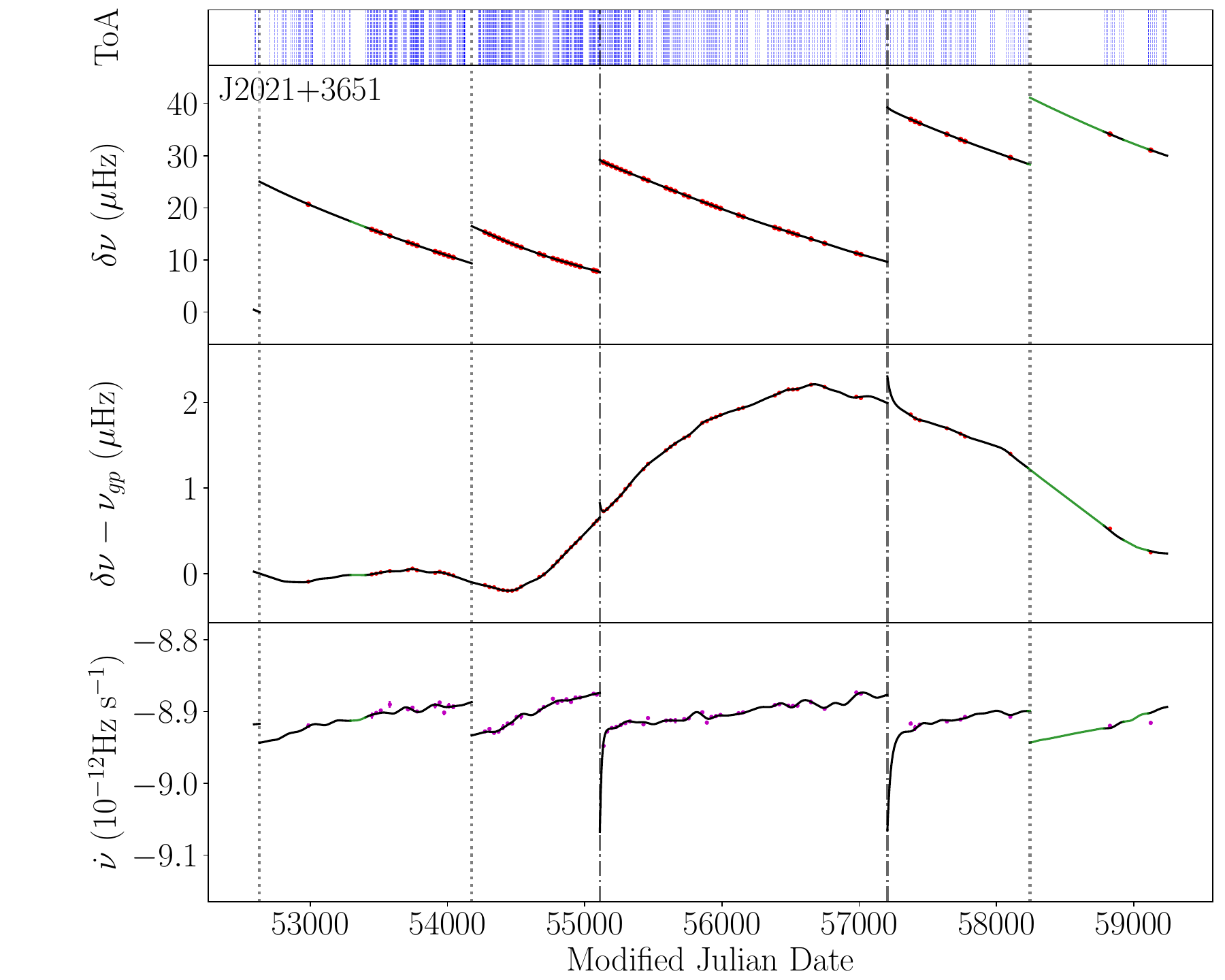}
    \end{subfigure}
    \begin{subfigure}[b]{0.49\textwidth}
        \centering
        \includegraphics[scale=0.29]{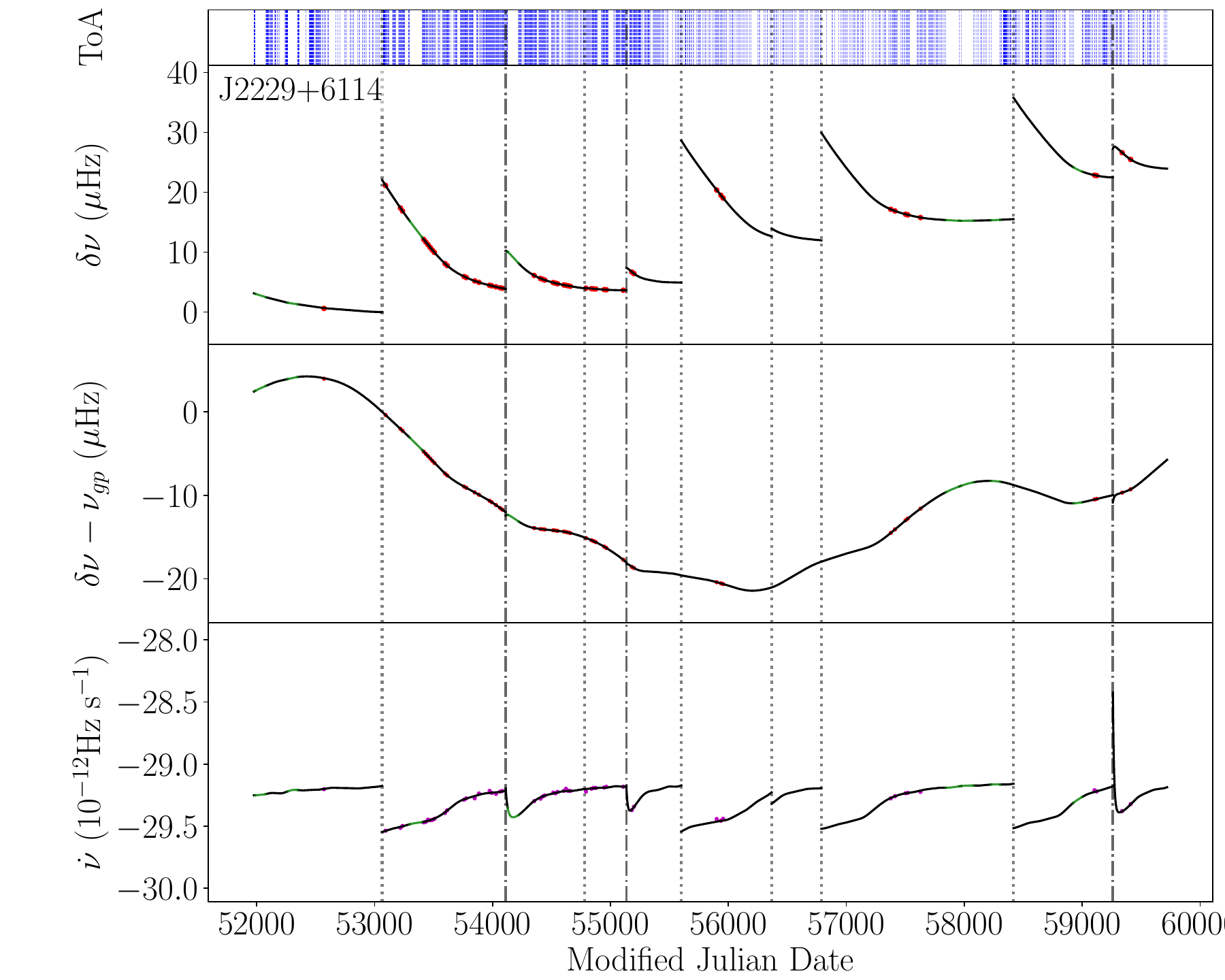}
    \end{subfigure}
    \begin{subfigure}[b]{0.49\textwidth}
        \centering
        \includegraphics[scale=0.29]{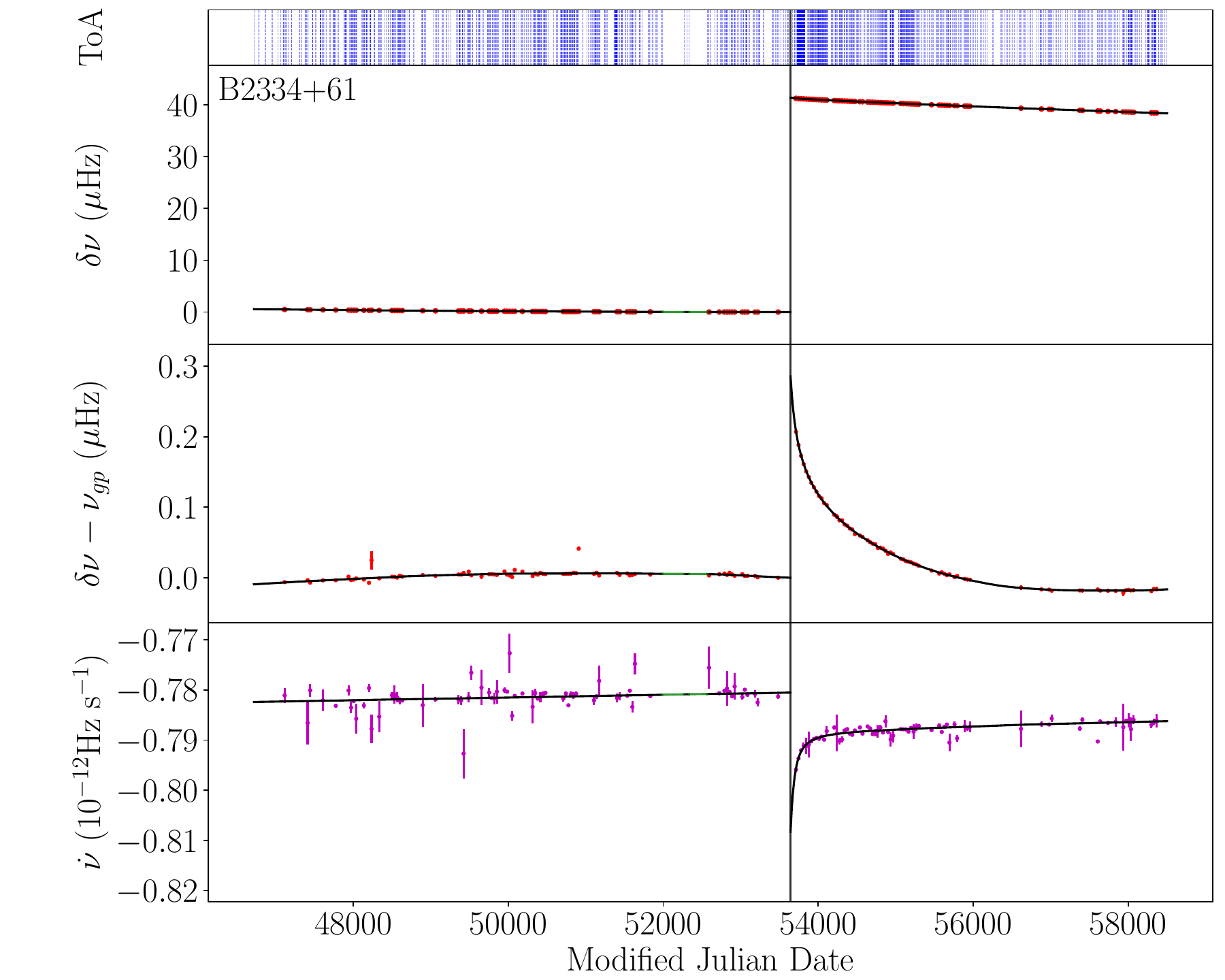}
    \end{subfigure}
    \caption{The evolution of $\delta\nu$ and $\dot{\nu}$ in PSRs B1930$+$22, B1951$+$32, J2021$+$3651, J2229$+$6114 and B2334$+$61. See the caption of Figure~\ref{fig: nu_nudot_J0205-B0919} for further details.}
    \label{fig: nu_nudot_B1930-B2334}
\end{figure*}

Figures~\ref{fig: nu_nudot_J0205-B0919} to \ref{fig: nu_nudot_B1930-B2334} present the rotational evolution of $\nu(t)$ and $\dot{\nu}(t)$ from the stride fitting method and as derived from the maximum likelihood timing solution, including the best glitch models and the timing noise model. All figures are organised in four panels as described in the caption of Figure~\ref{fig: nu_nudot_J0205-B0919}.

The $\delta{\nu}$ evolution for these pulsars is dominated by the large glitches that lead to their selection for the sample. Once the large permanent changes are subtracted, the evolution of the frequency is typically dominated by either the transient, exponential, recovery, or by timing noise. The $\dot{\nu}$ time series are typically dominated by glitches and their recoveries, though depending on the pulsar, this may be the `linear' recovery of $\dot{\nu}$ that is measured as $\ddot{\nu}$, or the exponential recoveries. For a handful of low $|\dot{\nu}|$ pulsars, the evolution of $\dot{\nu}$ mostly shows noise.

The scaling of vertical axes in the plots gives the impression that the amplitude of the exponential terms varies strongly over the sample. In fact, the decaying amplitude $\Delta\nu_{\mathrm{d}}$ only varies by about an order of magnitude ($10^{-8}-10^{-7}$Hz) over the sample, but this is seen against the background of the overall scale of $\Delta\dot{\nu}_{\mathrm{p}}$ and $\ddot{\nu}$, which both vary by seven orders of magnitude across our sample. The evolution of the glitch parameters over the $P-\dot{P}$ diagram will be discussed in section \ref{sec: analysis}.

Among our sample, we highlight PSRs J1740$+$1000 and B1757$-$24 as two pulsars which exhibit examples of some common glitch features. PSR J1740$+$1000 (middle right panel of Figure~\ref{fig: nu_nudot_B1727-B1757}) has a single large glitch over 13 years in our data span. Typically such pulsars have a relatively small $|\dot{\nu}|$ and $\ddot{\nu}$, and their post-glitch behaviour is often dominated by exponential relaxation. 
PSR B1757$-$24 (bottom right panel of Figure~\ref{fig: nu_nudot_B1727-B1757}), on the other hand, has several large glitches of similar amplitude and interglitch waiting times. Such a regular pattern of glitches resembles what is seen in the Vela pulsar, and hence such pulsars are often termed `Vela-like'. After the step change in $\nu$ and $\dot{\nu}$ at the glitch, the post-glitch recovery process resets most of the change in $\dot{\nu}$ before the next large glitch occurs (see Subsection \ref{sub: wait} for details on this). Whilst exponentially decaying components are detected in some of its glitches, the linear post-glitch recovery has the dominant effect on the long-term $\dot{\nu}$ evolution. 

The youngest pulsars in the sample, PSRs J0205$+$6449 and J2229$+$6114, show an interglitch $\dot{\nu}$ evolution that does not indicate a `linear' recovery nor can be described well by exponential terms. The post-glitch $\dot{\nu}(t)$ curve appears more like an ‘S-shaped’ curve with the gradient of $\dot{\nu}$ increasing smoothly a few hundred days after the glitch before flattening off again. As a consequence, our models describe glitches in PSR J0205$+$6449 (top left panel of Figure~\ref{fig: nu_nudot_J0205-B0919}) with permanent changes in $\nu$ and $\dot{\nu}$ only, while for PSR J2229$+$6114 (middle right panel of Figure~\ref{fig: nu_nudot_B1930-B2334}) our models favour exponential recoveries with negative changes in $\nu$ for the 2nd, 4th, and 9th glitch. The difference between the preferred model for these pulsars seems to come down largely to the timing noise model: the Bayesian code may attribute the post-glitch behaviour to a sum of exponentials or to the timing noise model. Such ‘S-shaped’ swings can also be seen in PSR~J1709$-$4429 \citep{lower_impact_2021} and are likely to be glitch related. Future improvements to glitch modelling may reduce the amount of this glitch activity currently attributed to timing noise.

PSR B0355$+$54 is a well studied (see e.g., \citet{janssen_30_2006}) pulsar which showed a very large glitch, preceded by a more moderate one by $\sim$387 days. Our model of the first glitch of PSR B0355$+$54 (top right panel of Figure~\ref{fig: nu_nudot_J0205-B0919}) finds an unusual positive value of $\Delta\dot{\nu}_{\mathrm{p}}$, which is not reported previously. However, there is a sparsity of data between the two glitches and the exponential recovery of the first glitch does not fully recover before the second glitch occurs. Therefore we suspect that the positive value of $\Delta\dot{\nu}_{\mathrm{p}}$ should be treated cautiously as it is not clear that there is sufficient data to unambiguously determine the `permanent' effect of the first glitch.

Another pulsar of interest is PSR B1737$-$30 (top right panel of Figure~\ref{fig: nu_nudot_B1727-B1757}), for which we have 36 glitches in our dataset. We do not find any exponential terms in the glitch recoveries of this pulsar, but we also note that the very high density of small glitches means that the fitting may struggle to distinguish between spin evolution or glitch recoveries and the parameters of the next glitch. \citet{urama_glitch_2002} reported an exponential recovery after the fifteenth (MJD~50936.8) glitch, with a timescale of 9 days. We have no ToAs within 10 days after this glitch and only 2 ToAs within 10-20 days, so it not possible to verify this exponential with our data. \citet{zou_observations_2008} found a big ($Q=0.103$) exponential recovery ($\tau_{\mathrm{d}}\sim50$ days) at the twenty-first (MJD~52347.7) glitch, and a smaller ($Q=0.0302$) exponential recovery of 100 days at the twenty-sixth (MJD~53036) glitch. Our one-exponential model for the twenty-first glitch ($\tau_{\mathrm{d}}\sim40$ days, $Q=0.04(4)$) did agree with their results, but the Bayes factor for this model is not large enough ($\ln{B_{21}}=2$) and the exponential timescale is poorly constrained so we did not accept it. In addition, \citet{liu_one_2019} reported an exponential of $\Delta\nu_{\mathrm{d}}=9.5(6)\times10^{-9}\mathrm{Hz}$ with a timescale of 71(6) days for the thirty-sixth (MJD~58232.4) glitch, but our Bayes factor does not favour models with exponential recovery for this glitch.

PSR J0631$+$1036 (middle right panel of Figure~\ref{fig: nu_nudot_J0205-B0919}) also has very frequent glitches, and has a $\dot{\nu}=-1.265\times10^{-12}\;\mathrm{Hz/s}$ similar to PSR B1737$-$30 ($\dot{\nu}=-1.266\times10^{-12}\;\mathrm{Hz/s}$). The 17 glitches of PSR J0631$+$1036 in our data span are described by a range of different glitch models: four of them have exponential recoveries, two of them additionally have a noticeable change in $\ddot{\nu}$. The frequent small glitches and noisy $\dot\nu(t)$ variations in PSR J0631$+$1036 and PSR B1737$-$30 make it hard to be certain that all the glitch recovery is fully captured or discernible from other noise-like processes. Other pulsars with similar frequency derivatives are PSR J1841$-$0524 (middle left panel of Figure~\ref{fig: nu_nudot_J1837-B1853}), PSR J1850$-$0026 (bottom left panel of Figure~\ref{fig: nu_nudot_J1837-B1853}) and PSR B1830$-$08 (bottom right panel of Figure~\ref{fig: nu_nudot_B1800-B1830}). They do not glitch as frequently as PSR J0631$+$1036 and PSR B1737$-$30, but they also show some modulations in their $\dot{\nu}$ evolution.

PSR J1921+0812 (bottom right panel of Figure~\ref{fig: nu_nudot_J1856-J1921}) has an increased $\ddot{\nu}$ for $\sim1000$ days following the glitch around MJD~55349 but then $\ddot{\nu}$ returns to its pre-glitch level. The particular Bayes factors do not support exponential recovery models nor a permanent change in $\ddot{\nu}$. Instead, our algorithm prefers to describe the change in $\ddot{\nu}$ as timing noise.

All pulsars with a measurable $\ddot{\nu}$ show a positive value, except PSR B1830$-$08 (bottom right panel of Figure~\ref{fig: nu_nudot_B1800-B1830}), which has a clear systematic linear increase in $|\dot{\nu}|$ since its large 2nd glitch. Unfortunately there is insufficient data to unambiguously determine if $\ddot{\nu}$ before this large glitch is also negative. Unlike other pulsars, this negative $\ddot{\nu}$ and the effect of glitches work together to increase $|\dot{\nu}|$. Understanding the origins of a negative $\ddot{\nu}$ will be important for any theory that hopes to explain the long-term evolution of pulsars.

Whilst large glitches are typically accompanied by a negative $\Delta\dot{\nu}$, the glitch of PSR B0919$+$06 (bottom right panel of Figure~\ref{fig: nu_nudot_J0205-B0919}) at MJD~55151.7 is an unusual one. Our best model suggests a negative decaying component $\Delta\nu_{\mathrm{d}}=-2.2\times10^{-8}\;\mathrm{s}^{-1}$ and a positive $\Delta\dot{\nu}=2.4\times10^{-14}\;\mathrm{Hz/s}$. The negative exponential recovers on a very short timescale of about 10 days as shown in the bottom right panel of Figure~\ref{fig: nu_nudot_J0205-B0919}. We note that this model is favoured mostly due to the two ToAs right after the glitch. So our model for this glitch is highly dependent on the accuracy of these ToAs. Apart from this negative exponential and the three in PSR J2229$+$6114 mentioned earlier in this section, all other exponential recoveries we measure are associated with positive changes in $\nu$.

Many large glitches have significant exponential recoveries, however not every large glitch is necessarily associated with an exponential recovery. The exponential timescale is sometimes shorter than we can reliably measure, and in these cases we caution that the exponential model may not be correct as typically only one or two ToAs close to the glitch are giving rise to the requirement for a transient recovery term. In these cases it is worth noting that the uncertainty on the instantaneous change in $\nu$  strongly depends on the exponential model.

Finally, we note that several pulsars (PSRs J0729$-$1448, B1727$-$33, B1821$-$11, B1830$-$08, and B1930$+$22) exhibit quasi-periodic oscillations in $\dot{\nu}$, which can be seen by eye in the $\dot{\nu}$ time series. Such behaviour is not uncommon across the pulsar population, though not fully understood \citep{hobbs_analysis_2010, lyne_switched_2010, nitu_search_2022}. The quasi-periodic oscillations are thought to originate from changes in the magnetosphere, though there have been hints of a connection to glitch activity \citep{keith_connection_2013}. Further investigation of quasi-periodic oscillations are left for future work.

\section{Statistical Analysis}\label{sec: analysis}

\subsection{Evolution of $\ddot{\nu}$ and braking indices across the pulsar population}\label{sub: ppdot}
The spin-down of a pulsar under a braking torque can be assumed to have the form $\dot{\nu}\propto-\nu^{n}$, with $n$ called the braking index. The $\ddot\nu$ of a pulsar can then be predicted to be
\begin{equation}\label{eq: F2_n_brake}
\ddot{\nu}=n \nu^{-1}{\dot{\nu}}^{2}.
\end{equation}
For the case where the braking is purely due to electromagnetic dipole radiation, we expect $n_{\mathrm{b}}=3$. However, as seen in the results of Table~\ref{tab: pulsars_parameters} we typically find that $n$ for the linear interglitch phase is of order 10 to 100. Note that the fitted parameter $\ddot{\nu}$ in these glitching pulsars does not reflect the long-term evolution of the pulsar spin down, which -- as can be seen in Figures~\ref{fig: nu_nudot_J0205-B0919} to \ref{fig: nu_nudot_B1930-B2334} -- is in general slower (smaller effective $\ddot\nu$). It has been postulated that the high interglitch $\ddot{\nu}$ observed in several glitching pulsars is the result of internal, superfluid, torques \citep{antonopoulou_pulsar_2022}. 

\begin{figure}
	\includegraphics[scale=0.31]{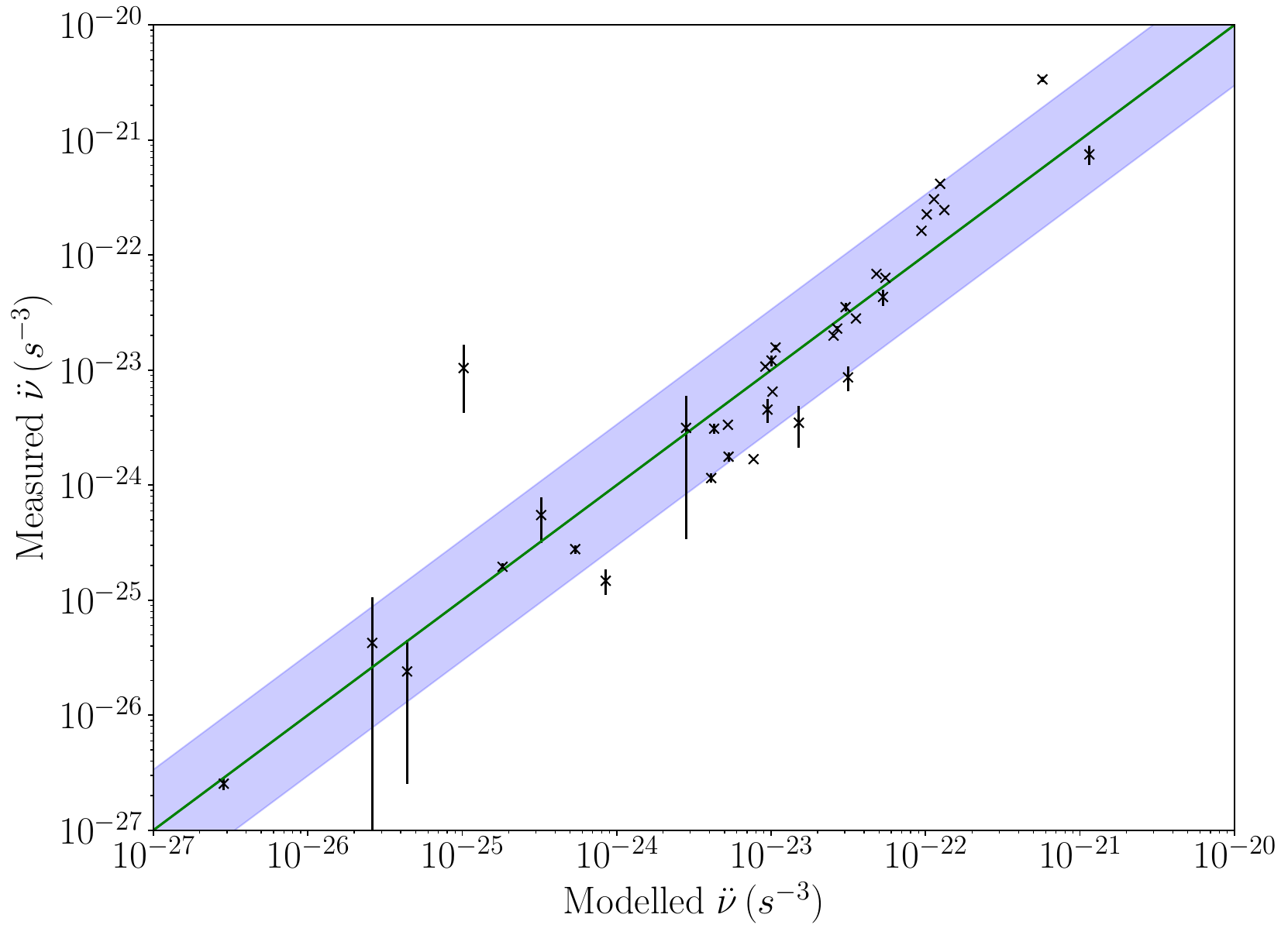}
    \caption{Comparison between measured $\ddot{\nu}$ and modelled $\ddot{\nu}$, the result of fitting Equation~\ref{eq: nuddot_int_nunudot_result} for pulsars in our sample. The modelled $\ddot{\nu}$ are calculated with the mean values of $a$, $b$, and $c$ in Equation~\ref{eq: nuddot_int_nunudot_result}. The green-solid line shows the one to one relation. The shaded blue region is the area between $\pm1\sigma$ (variance of the Gaussian noise term in Equation~\ref{eq: nuddot_int_nunudot_result}) lines off the green line.}
    \label{fig: F2_F2fit}
\end{figure}

Our results indicate that the interglitch $\ddot{\nu}$ does not substantially change between glitches of the same pulsar. Conversely, $\ddot{\nu}$ can vary considerably between different glitching neutron stars.
To explore this, we investigate the evolution of $\ddot{\nu}$ across the $P-\dot{P}$ diagram. We do this by fitting $\ddot{\nu}$ as a function of powers of $\nu$ and $\dot{\nu}$ using a Bayesian solver \textsc{emcee} with MCMC (Monte Carlo Markov Chain) \citep{foreman-mackey_emcee_2013}. Specifically we model
\begin{equation}\label{eq: nuddot_int_nunudot_result}
    \log{\left(\frac{\ddot{\nu}}{\mathrm{s}^{-3}}\right)} = a \log {\left(\frac{\nu}{\mathrm{s}^{-1}}\right)} + b \log{\left(\frac{-{\dot{\nu}}}{\mathrm{s}^{-2}}\right)} + c + \epsilon,
\end{equation}
where $a$, $b$ and $c$ are fit parameters and $\epsilon$ is a Gaussian noise term with variance $\sigma^2$. The result is shown in Figure~\ref{fig: F2_F2fit} where $a=-0.2\pm0.4$, $b=1.4\pm0.1$, $c=-6.1\pm1.9$, and $\sigma=0.53\pm0.07$. Comparing the exponents of $\nu$ and $\dot{\nu}$ with Equations~\ref{eq: E_dot}-\ref{eq: B_sur}, we see that $\ddot{\nu}$ does not strictly track any of $\dot{E}$, $\tau_{\mathrm{c}}$ or $B_{\mathrm{S}}$. We find that although $\ddot{\nu}$ generally tends to decrease with $\tau_{\mathrm{c}}$, with Spearman rank correlation coefficient $\rho_{s}=-0.90$ (see e.g. \citealp{spearman_proof_2010}), it most strongly correlates with $\dot{\nu}$ directly, $\rho_{s}=-0.93$ (here we only consider measurements with $\ddot{\nu}>0$ and at least 3$\sigma$ confidence.).

\begin{figure}
	\includegraphics[scale=0.31]{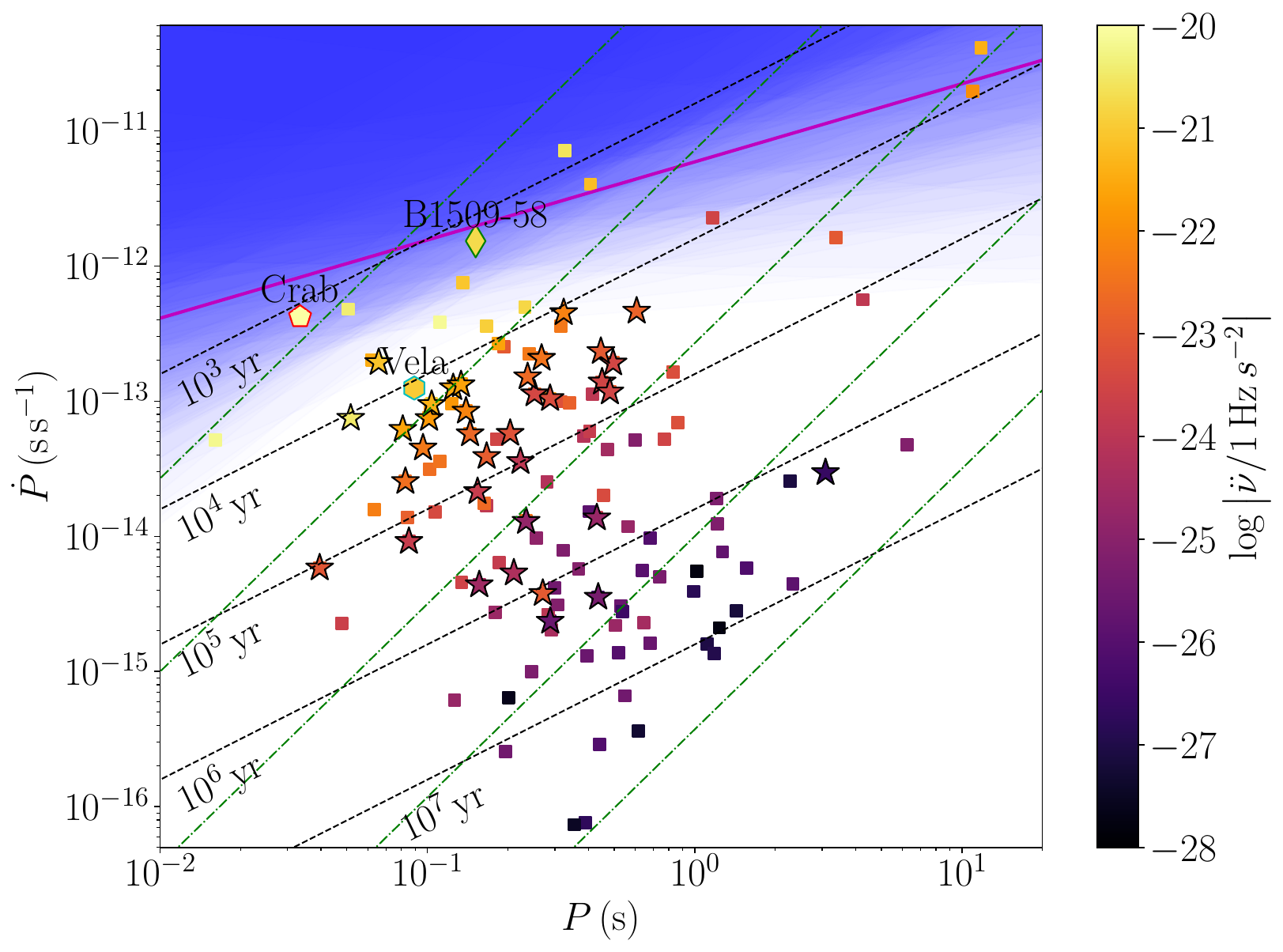}
    \caption{$P-\dot{P}$ diagram showing the $\ddot{\nu}$ of pulsars in our sample (stars) and glitching pulsars from the pulsar catalogue (squares). Two young pulsars PSRs J0534$+$2200 (the Crab pulsar, in red pentagon) and B1509$-$58 (in green diamond) with $n\sim3$ are highlighted. The Vela pulsar is also highlighted in cyan hexagon for comparison with the `Vela-like' pulsars. The color-bar represents the long-term $\ddot{\nu}$. We draw 100 samples from the posterior of $a$, $b$, $c$ and shade the region in blue where the modelled $\ddot{\nu}$ is less than that expected from $n_{\mathrm{b}}=3$ braking. The magenta line shows this $\ddot{\nu}\nu/\dot{\nu}^2=3$ threshold for the mean solution. The green dot-dashed lines correspond to constant $\ddot{\nu}$ values according to the best fit of Equation.~\ref{eq: nuddot_int_nunudot_result}.}
    \label{fig: PPdot_F2}
\end{figure}

The evolution of $\ddot{\nu}$ across the $P-\dot{P}$ diagram is shown in Figure~\ref{fig: PPdot_F2}, which includes our sample of pulsar as well as all other glitching pulsars found in the pulsar  catalogue\footnote{https://www.atnf.csiro.au/research/pulsar/psrcat/} \citep{manchester_australia_2005}. If we assume that the observed $\ddot{\nu}$ (described by Equation~\ref{eq: nuddot_int_nunudot_result}) is the result of internal and external torques acting on the star, and that the external torque has an $n_{\mathrm{b}}\simeq3$, then it is clear that for most pulsars with large glitches the internal torque is dominant between glitches. However, using Equation~\ref{eq: nuddot_int_nunudot_result} we find that there is a region at the upper-left corner of the the $P-\dot{P}$ diagram for which we might expect the observed $\ddot{\nu}$ to be dominated by the external torque instead. The blue area in Figure~\ref{fig: PPdot_F2} represents the region where the expected $\ddot{\nu}$ (calculated from 100 draws from the posterior of Equation~\ref{eq: nuddot_int_nunudot_result}) is smaller than what is expected from dipole braking ($n_{\mathrm{b}}=3$). The cross-over region roughly coincides with $\tau_{\mathrm{c}}\approx10^{2}\sim10^{4}\mathrm{yr}$, so only the youngest glitching pulsars may be expected to show $n_{\mathrm{b}}\simeq3$. 

In Figure~\ref{fig: PPdot_F2} we highlight two young pulsars with well measured braking indices that reside close to the boundary of external torque versus internal torque domination on $\ddot\nu$: the Crab pulsar (PSR J0534$+$2200), and PSR B1509$-$58. The Crab pulsar has several observed glitches and its interglitch braking index has been found to be $n\sim2.5$ \citep{lyne_45_2015}. This value is consistent with our aforementioned expectation. On the other hand, PSR B1509$-$58 has never glitched in its 25 years observing history and has a braking index of $n=2.832\pm0.003$ \citep{livingstone_long-term_2011}, very similar to that of the Crab. This could indicate that $n\simeq3$ is typical of very young pulsars, irrespectively of whether they glitch or not. 

Most pulsars with $\tau_{\mathrm{c}}\approx10^{4}\sim10^{5}\mathrm{yr}$ have interglitch braking indices which are an order of magnitude larger than expected for long-term dipole braking. For older pulsars, towards the bottom-right corner in $P-\dot{P}$ diagram, we find that in practice the observed spin-down evolution becomes dominated by stochastic timing noise, for which the power on the longest timescales is indistinguishable from $\ddot{\nu}$. To differentiate a long-term $\ddot{\nu}$ from timing noise effects and measure $\ddot{\nu}$ accurately, extremely long observational spans are needed \citep{parthasarathy_timing_2020}. Consequently, there is a rapid drop-off in the number of pulsars with accurate $\ddot{\nu}$ measurements above a characteristic age of $10^{5} \mathrm{yr}$ \citep{lower_impact_2021}. As the interglitch $\ddot{\nu}$ becomes hard to detect for pulsars of large characteristic age, it is difficult to know if there is a transition back towards $n_{\mathrm{b}}\simeq3$, or if superfluid effects continue to play a significant role in the evolution of these old pulsars.

\subsection{Correlations between waiting times and glitch parameters}\label{sub: wait}
Previous works such as \citet{melatos_size-waiting-time_2018} have focused on the relation between glitch size and glitch waiting time, and many of them suggested some correlation between them \citep{carlin_autocorrelations_2019, fuentes_glitch_2019}. Such a relation has implications on the physical mechanism behind glitches. Moreover, establishing a well-defined statistical procedure for predicting a glitch in a given pulsar will be greatly beneficial for long-term planning of observations. 

The most active glitching pulsar that we know of is PSR~J0537$-$6910, which has over 3 glitches per year on average. This is the only source for which a clear, strong, correlation between glitch amplitude $\Delta\nu$ and the time to the next glitch $\Delta t^{+}$ has been established \citep{middleditch_predicting_2006, antonopoulou_pulsar_2018}. The same studies also showed a weaker correlation between the change $\Delta\dot{\nu}$ and the time preceding a glitch $\Delta t^{-}$. 

Another reported correlation is between the instantaneous change of frequency derivative at a glitch $|\Delta\dot{\nu}|$ and the subsequent $\ddot{\nu}^+\Delta t^{+}$, where $\ddot{\nu}^+$ is measured over the post-glitch interval $\Delta t^{+}$. \citet{lower_impact_2021} modelled this relation as a power-law
\begin{equation}\label{eq: lower}
\ddot{\nu}_{i}^{+}=10^{\alpha}\left(\left|\Delta \dot{\nu}_{i}\right| / \Delta t^{+}_{i}\right)^{\beta} \;\text{.}
\end{equation}
and used collectively the glitch parameters for 16 pulsars, finding $ \alpha=-4.3_{\scriptscriptstyle-2.6}^{\scriptscriptstyle+2.5}$ and $\beta=0.80\pm0.12$ and a Spearman's rank correlation coefficient of $0.74$. The glitches of PSR~J0537$-$6910 also show this correlation, and follow the general trend suggested by the other pulsars \citep{ho_timing_2022}.

\begin{table*}
\begin{center}
\centering
\caption{The values of parameters (at the 95 percent confidence interval) in Equation~\ref{eq: lower} and the Spearman coefficients $\rho_{s}$ with corresponding $p$-value. The data are from \citet{lower_impact_2021} (L21), \citet{ho_timing_2022} (H22),  and our data (JBO). The third column is the cut-off in $\Delta{\nu}/\nu$, the glitches below this threshold are not included, and the waiting time are therefore recalculated between the remaining large or intermediate size glitches. The forth column is the glitch parameters we are fitting in the right hand side of Equation~\ref{eq: lower}. In addition, we test the correlation for 7 `Vela-like' pulsars (PSRs B1727$-$33, B1757$-$24, B1800$-$21, B1823$-$13, J0205$+$6449, J2021$+$3651, and J2229$+$6114) in our sample as well. $\Delta t^{-}_{i}$ represents the backward waiting time, while $\Delta t^{+}_{i}$ is the forward waiting time.}
\label{tab: powerlaw}
\renewcommand{\arraystretch}{1.5}
\begin{tabular}{lrrrrrrr}
\hline
\hline
Dataset & Pulsars & $\Delta{\nu}/\nu$ cut-off & Glitch parameters & $\alpha$ &  $\beta$ &  $\rho_{s}$ & $p$-value \\
\hline
L21 & All & - & $\Delta\dot{\nu}/\Delta t^{+}_{i}$ & $-4.3_{-2.6}^{+2.5}$ & $0.80\pm0.12$ & $0.74$ & - \\
JBO & All & - & $\Delta\dot{\nu}/\Delta t^{+}_{i}$ & $-0.6_{-3.3}^{+4.0}$ & $0.99_{-0.15}^{+0.18}$ & $0.69$ & $4.7\times10^{-13}$ \\
JBO & All & - & $\Delta\dot{\nu}/\Delta t^{-}_{i}$ & $-2.7_{-5.9}^{+8.0}$ & $0.90_{-0.27}^{+0.37}$ & $0.55$ & $5.1\times10^{-8}$ \\
JBO & All & - & $\Delta\dot{\nu}_{\mathrm{p}}/\Delta t^{+}_{i}$ & $0.1_{-3.1}^{+3.5}$ & $1.02_{-0.14}^{+0.16}$ & $0.74$ & $2.0\times10^{-15}$ \\
JBO & All & - & $\Delta\dot{\nu}_{\mathrm{p}}/\Delta t^{-}_{i}$ & $-1.0_{-3.7}^{+4.5}$ & $1.06_{-0.17}^{+0.20}$ & $0.72$ & $1.0\times10^{-14}$ \\
JBO & All & $\geq10^{-7}$ & $\Delta\dot{\nu}_{\mathrm{p}}/\Delta t^{+}_{i}$ & $-0.3\pm1.6$ & $0.98 \pm 0.07$ & $0.96$ & $4.1\times10^{-25}$ \\
JBO & All & $\geq10^{-7}$ & $\Delta\dot{\nu}_{\mathrm{p}}/\Delta t^{-}_{i}$ & $-1.0_{-2.0}^{+2.1}$ & $0.95 \pm 0.09$ & $0.94$ & $2.3\times10^{-22}$ \\
JBO & `Vela-like' & $\geq10^{-7}$ & $\Delta\dot{\nu}_{\mathrm{p}}/\Delta t^{+}_{i}$ & $-1.2_{-2.0}^{+2.1}$ & $0.94 \pm 0.10$ & $0.93$ & $4.6\times10^{-12}$ \\
JBO & `Vela-like' & $\geq10^{-7}$ & $\Delta\dot{\nu}_{\mathrm{p}}/\Delta t^{-}_{i}$ & $-3.0_{-3.0}^{+3.4}$ & $0.86_{-0.14}^{+0.16}$ & $0.92$ & $2.9\times10^{-11}$ \\
H22 & J0537$-$6910 & $\geq10^{-7}$ & $\Delta\dot{\nu}_{\mathrm{p}}/\Delta t^{+}_{i}$ & $-5.2_{-4.8}^{+7.5}$ & $0.74_{-0.24}^{+0.38}$ & $0.57$ & $9.4\times10^{-5}$ \\
H22 & J0537$-$6910 & $\geq10^{-7}$ & $\Delta\dot{\nu}_{\mathrm{p}}/\Delta t^{-}_{i}$ & $-1.8_{-6.3}^{+9.4}$ & $0.91_{-0.32}^{+0.47}$ & $0.58$ & $4.8\times10^{-5}$ \\
\hline
\end{tabular}
\end{center}
\end{table*}

\begin{figure}
    \centering
    \begin{subfigure}[b]{0.49\textwidth}
        \centering
        \includegraphics[scale=0.31]{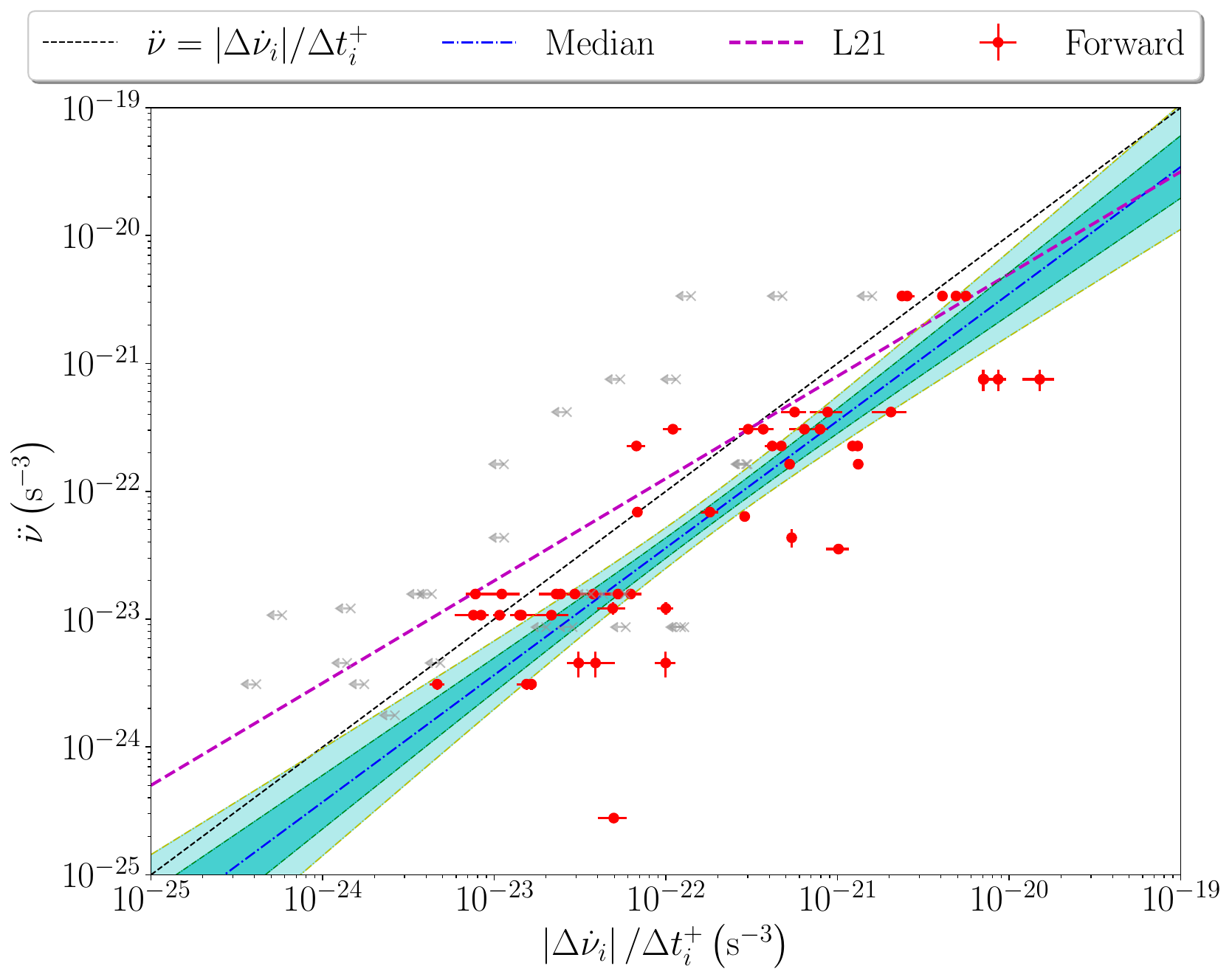}
        \label{F2nuddot_GLF1ins_Tga}
    \end{subfigure}
    \begin{subfigure}[b]{0.49\textwidth}
        \centering
        \includegraphics[scale=0.31]{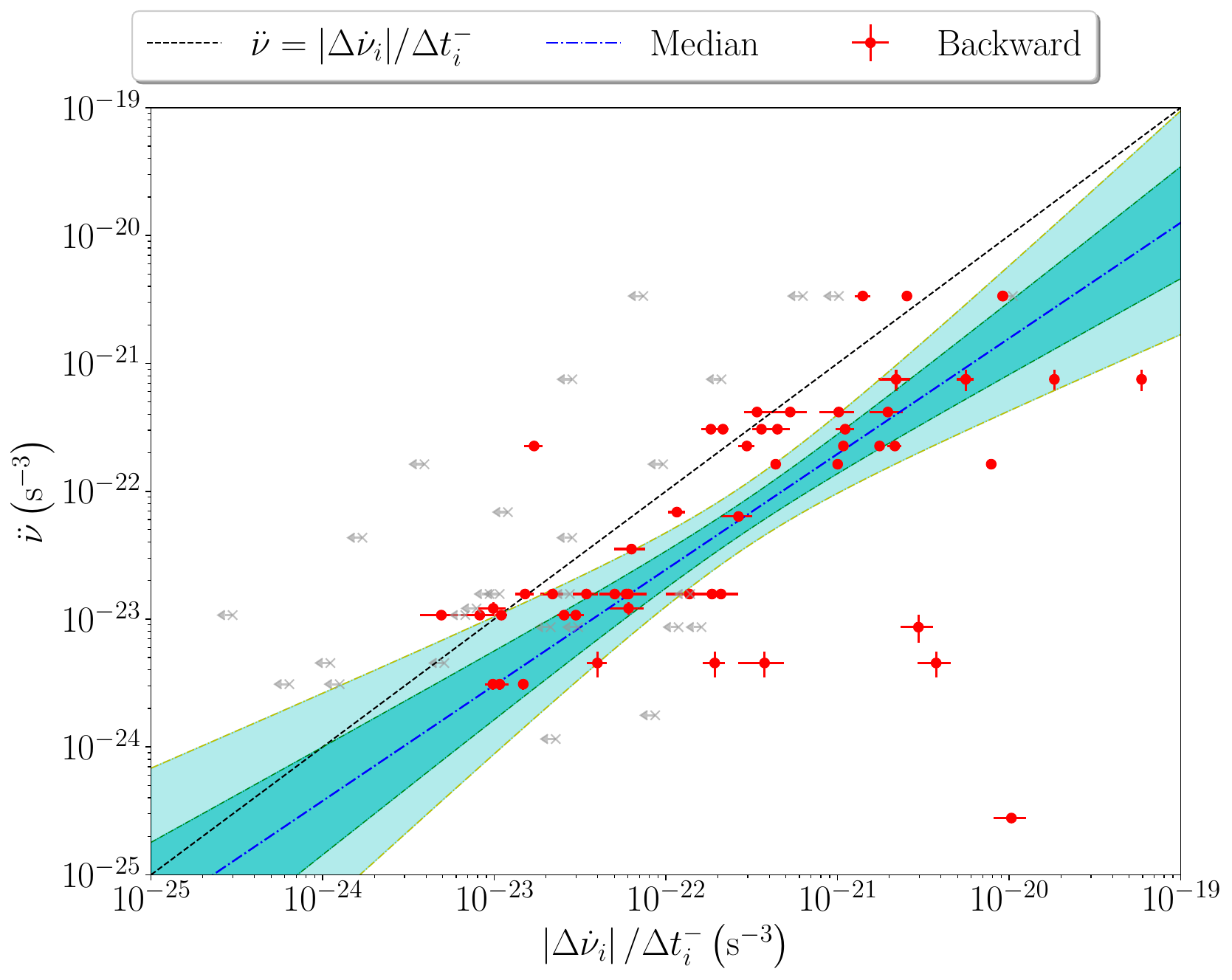}
        \label{F2nuddot_GLF1ins_Tgb}
    \end{subfigure}
    \caption{Top panel: Comparison between $\ddot\nu$ and the $|\Delta\dot{\nu}_{i}|$ of the $i$ th glitch divided by the forward waiting time $\Delta t^{+}_{i}$ for 157 glitches in 35 pulsars. Bottom panel: Comparison between $\ddot\nu$ and the $|\Delta\dot{\nu}_{i}|$ of the $i$ th glitch divided by the backward waiting time $\Delta t^{-}_{i}$ for the same glitches. The black-dashed line correspond to $\ddot\nu$ being equal to $|\Delta\dot{\nu}_{i}|/\Delta t^{\pm}_{i}$. The magenta-dashed line shows the fitted relation obtained by \citet{lower_impact_2021} (Equation~\ref{eq: lower}). Blue dot-dashed lines stands for our median power-law fit, and the shaded areas are the $68\%$ (dark cyan) and $95\%$ (light cyan) confidence regions. Note that the measurements below $3-\sigma$ confidence are not included in this fit, we marked the upper limits of these measurements by grey points and arrows. The red points show measurements with at least $3\sigma$ confidence and their $1\sigma$ errorbars.}
    \label{fig: F2nuddot_GLF1ins_Tg}
\end{figure}

We assume a functional form as in Equation~\ref{eq: lower} for a possible $\Delta\dot{\nu}$--interglitch time interval relation, and investigate the presence of a correlation with either forward ($\Delta t^{+}_{i}$) or backward waiting time ($\Delta t^{-}_{i}$), since indications of both have been previously found  \citep{middleditch_predicting_2006, antonopoulou_pulsar_2018,lower_impact_2021}. Using our sample, we compute the Spearman's rank correlation coefficient $\rho_{s}$, as well as the exponents $\alpha$ and $\beta$, using a bi-variate Gaussian likelihood \citep{hogg_data_2010} to account for the Gaussian uncertainties in both $\Delta \dot{\nu}_{i}$ and $\ddot{\nu}_{\mathrm{i}}$. The results are presented in Table~\ref{tab: powerlaw}. Figure~\ref{fig: F2nuddot_GLF1ins_Tg} shows the data and the median power-law model for both $\Delta t^{+}_{i}$ and $\Delta t^{-}_{i}$. We note that in our timing solutions $\ddot{\nu}_i$ is typically the same for each interglitch interval ($\ddot{\nu}_i=\ddot{\nu}$), except in the very few cases for which a change $\Delta\ddot{\nu}$ was included in the glitch model. On the other hand, \citet{lower_impact_2021} used $\ddot{\nu}_{i}$ measurements for each interval. Nonetheless our results largely agree with theirs and we confirm the correlation between $\ddot{\nu}_{i}$ and $|\Delta\dot{\nu}_{i}|/\Delta t^{+}_{i}$. The difference between our result and \citet{lower_impact_2021} is due to the different sample of pulsars; for the glitches that are in both samples we get consistent measurements. Our sample has a larger number of glitches in pulsars with $\ddot{\nu}<2\times{10^{-23}}$, whereas \citet{lower_impact_2021} have only one, and we once again note the inherent bias in our sample as pulsars were selected on the basis of having at least one large glitch. 

The $\Delta\dot{\nu}_{i}$ values used in the above investigation stand for the total change in $\dot{\nu}$ at a glitch. Of this, part recovers exponentially whilst the rest ($\Delta\dot{\nu}_{\mathrm{p},i}$) is treated as `permanent' in our timing solutions and recovers linearly. The  $\ddot{\nu}$ values represent the linear evolution of $\dot{\nu}$ in between glitches. In most cases, the exponential decay timescales are shorter than the interglitch waiting times, hence the sum of any $\Delta\dot{\nu}_{\mathrm{p},i}$ terms has decayed to negligible levels, compared to $\Delta\dot{\nu}_{\mathrm{p},i}$ at the time of the next glitch. Therefore, $|\Delta\dot{\nu}_{i}|/\ddot{\nu}$ can be viewed as the time required for $\dot{\nu}$ to return to the value it had right before glitch $i$. The situation where a following glitch happens when $|\Delta\dot{\nu}|$ of the previous glitch fully recovers ($\ddot{\nu}=|\Delta\dot{\nu}_{i}|/\Delta t^{+}_{i}$) is represented by the black line in the top panel of Figure~\ref{fig: F2nuddot_GLF1ins_Tg}. As we can see, this scenario is not inconsistent with our best-fit parameters $\alpha$ and $\beta$, given their uncertainties, and in fact the median $\beta$ of our power-law fit is very close to one.

From the bottom panel of Figure~\ref{fig: F2nuddot_GLF1ins_Tg} a trend is also apparent for the case where the backwards interglitch time intervals have been used, though this correlation is weaker (as reflected in the smaller correlation coefficient in Table~\ref{tab: powerlaw}). Such a relation would not be unexpected within the superfluid glitch model: a longer waiting time means that more of the superfluid that decoupled at glitch $(i-1)$ had time to recouple before glitch $i$, and can thus decouple anew leading to a larger $|\Delta\dot{\nu}_{i}|$. 

Since $\ddot{\nu}$ describes the approximately linear evolution of $\dot{\nu}$, we find it informative to examine how the correlations change if we consider the permanent change, $\Delta\dot{\nu}_{\mathrm{p}}$, rather than the instantaneous change, $\Delta\dot{\nu}$ (which is the sum of the permanent change and any exponentially-relaxing changes). If the superfluid components which relax exponentially have time to recouple almost fully before a subsequent glitch, then $\Delta t^{-}_{i}$ will mostly determine $\Delta\dot{\nu}_{{\mathrm{p}},i}$ instead of the total $\Delta\dot{\nu}_{i}$. We therefore include in Table~\ref{tab: powerlaw} the results for the correlations $\ddot{\nu}$--$|\Delta\dot{\nu}_{{\mathrm{p}},i}|/\Delta t^{+}_{i}$ and $\ddot{\nu}$--$|\Delta\dot{\nu}_{{\mathrm{p}},i}|/\Delta t^{-}_{i}$. In accordance to our earlier argument, including or not the $\Delta\dot{\nu}_{\mathrm{d}}$ components has minimal impact on the correlation with the forward waiting time, whilst the correlation with the backwards waiting time gets significantly stronger when only the permanent changes $|\Delta\dot{\nu}_{{\mathrm{p}},i}|$ are considered. 

Following \citet{antonopoulou_pulsar_2018} and \citet{fuentes_glitch_2019} we also repeat the analysis focusing only on intermediate and large glitches, as mostly these affect $\dot{\nu}$ (whilst, for example, the small glitches in PSRs B1800$-$21 and B1823$-$13 have no noticeable effect on $\dot{\nu}$). We therefore ignore glitches with $\Delta\nu/\nu<10^{-7}$ and find that both correlations improve considerably and their correlation coefficients are very close to unity.

Ideally, when exploring the above correlations, $\ddot{\nu}^+_{i}$ (post-glitch $i$) and $\ddot{\nu}^-_{i}$ (pre-glitch $i$) should be used with $\Delta t^{+}_{i}$ and $\Delta t^{-}_{i}$ respectively. Nonetheless, our use of the average $\ddot{\nu}$ is justified because $\ddot{\nu}$ is often not well defined over a single interglitch interval, and we do not find strong evidence for a $\Delta\ddot{\nu}$ in most glitches. Such small changes in $\ddot{\nu}$ of a given pulsar can introduce some (vertical) scatter in Figure~\ref{fig: F2nuddot_GLF1ins_Tg} but the overall correlation still emerges as $\ddot{\nu}$ for the collection of pulsars varies by $\sim4$ orders of magnitude. Using the overall $\ddot{\nu}$ means, however, that we cannot check whether the correlations hold for each pulsar of our sample individually, as is the case for PSR~J0537$-$6910 \citep{ho_timing_2022}.

It is possible, given the large scatter in Figure~\ref{fig: F2nuddot_GLF1ins_Tg}, that one correlation is the artefact of the other. The reason is that -- for many of the pulsars included -- the waiting times distribution is rather narrow (see for example PSR~B1757$-$24 with a characteristic waiting time around 1000 days, or PSR~B1800-21 with $\Delta t$ about 2000 days). Therefore the interchange of $\Delta t^{+}_{i}$ with $\Delta t^{-}_{i}$ results in a relatively small shift  compared to the overall range of $|\Delta\dot{\nu}_{i}|/\Delta t_{i}$ values. 

\begin{figure}
    \centering
	\includegraphics[scale=0.31]{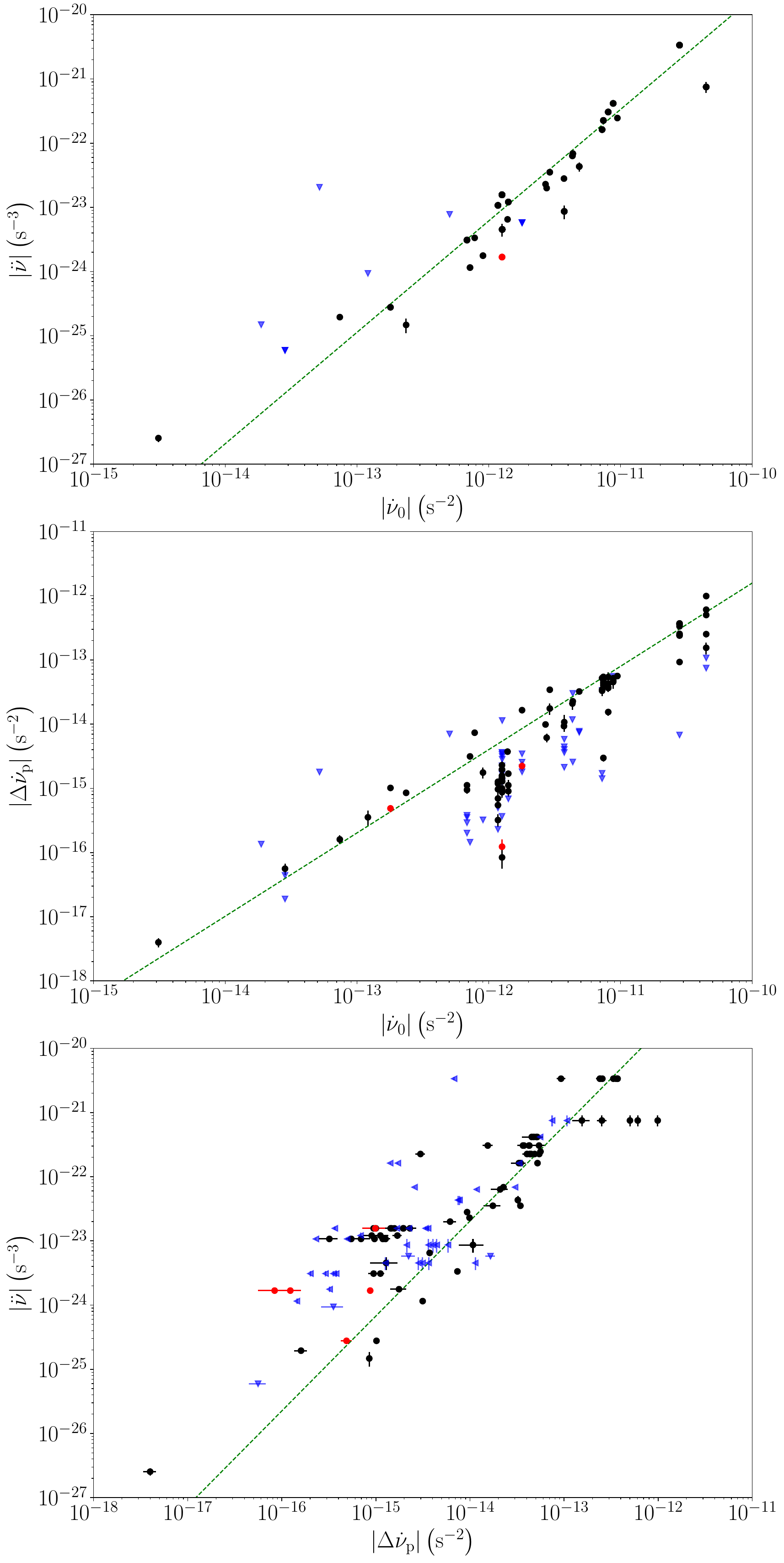}
    \caption{Top panel and middle panel: Comparison between the correlation of $|\ddot{\nu}|$ and $|\Delta\dot{\nu}_{\mathrm{p}}|$ with $|\dot{\nu}_{0}|$. Bottom panel: Correlation between $|\ddot{\nu}|$ and $|\Delta\dot{\nu}_{\mathrm{p}}|$. Black circles are the $|\ddot{\nu}|$ and $|\dot{\nu}_{0}|$ of the pulsars, and the $|\Delta\dot{\nu}_{\mathrm{p}}|$ of individual glitches, while green-dashed lines stand for the best linear fit. Red points correspond to negative $\ddot{\nu}$ (in top and bottom panel) or positive $\Delta\dot{\nu}_{\mathrm{p}}$ (in middle and bottom panel). The blue arrows are non-detections (measurements below $3-\sigma$) and represent the 95\% confidence upper limit (arrows towards left are upper limits for $\Delta\dot{\nu}_{\mathrm{p}}$, and arrows towards bottom are for $\ddot{\nu}$). The fitted lines (at the 95 percent confidence interval) are $\ddot{\nu}=10^{-2.38\pm0.07}\left|\dot{\nu}_{0}\right|^{1.74\pm0.01}$, $\left|\Delta\dot{\nu}_{\mathrm{p}}\right|=10^{1.17\pm0.10}\left|\dot{\nu}_{0}\right|^{1.30\pm0.01}$, and $\ddot{\nu}=10^{-1.98\pm0.21}\left|\Delta\dot{\nu}_{\mathrm{p}}\right|^{1.48\pm0.02}$. For all three panels, only the black points are used in fitting and the calculation for correlation coefficients.}
    \label{fig: F2_GLF1_nudot}
\end{figure}

Most importantly though, we notice two strong correlations, as shown in Figure~\ref{fig: F2_GLF1_nudot}. First, $\ddot\nu$ correlates with $|\dot{\nu}_{0}|$ ($\rho_{s}=0.96$). Secondly, $|\Delta\dot{\nu}_{\mathrm{p}}|$ also correlates with $|\dot{\nu}_{0}|$ ($\rho_{s}=0.93$). Here we only consider measurements with $\ddot{\nu}>0\,,\Delta\dot{\nu}_{\mathrm{p}}<0$ and at least 3$\sigma$ confidence. As a result, pulsars with greater $\ddot{\nu}$ will generally display larger changes of their spin-down rate at glitches. Indeed, $\ddot{\nu}$ presents an even stronger correlation directly with $|\Delta\dot{\nu}_{\mathrm{p}}|$ (Spearman's rank coefficient of $\rho_{s}=0.91$, here we only consider measurements with $\ddot{\nu}>0\,,\Delta\dot{\nu}_{\mathrm{p}}<0$ and at least 3$\sigma$ confidence.) than the one shown in Figure~\ref{fig: F2nuddot_GLF1ins_Tg}. These correlations should be accounted for when interpreting our results because the trend observed in Figure \ref{fig: F2nuddot_GLF1ins_Tg} could be a consequence of the separate correlations with $|\dot{\nu}_{0}|$. We alleviate the above effects by looking instead at the relation between $\left|\Delta \dot{\nu}_{\mathrm{p}}/\ddot{\nu}\right|$ and waiting times.

\begin{figure}
	\includegraphics[scale=0.31]{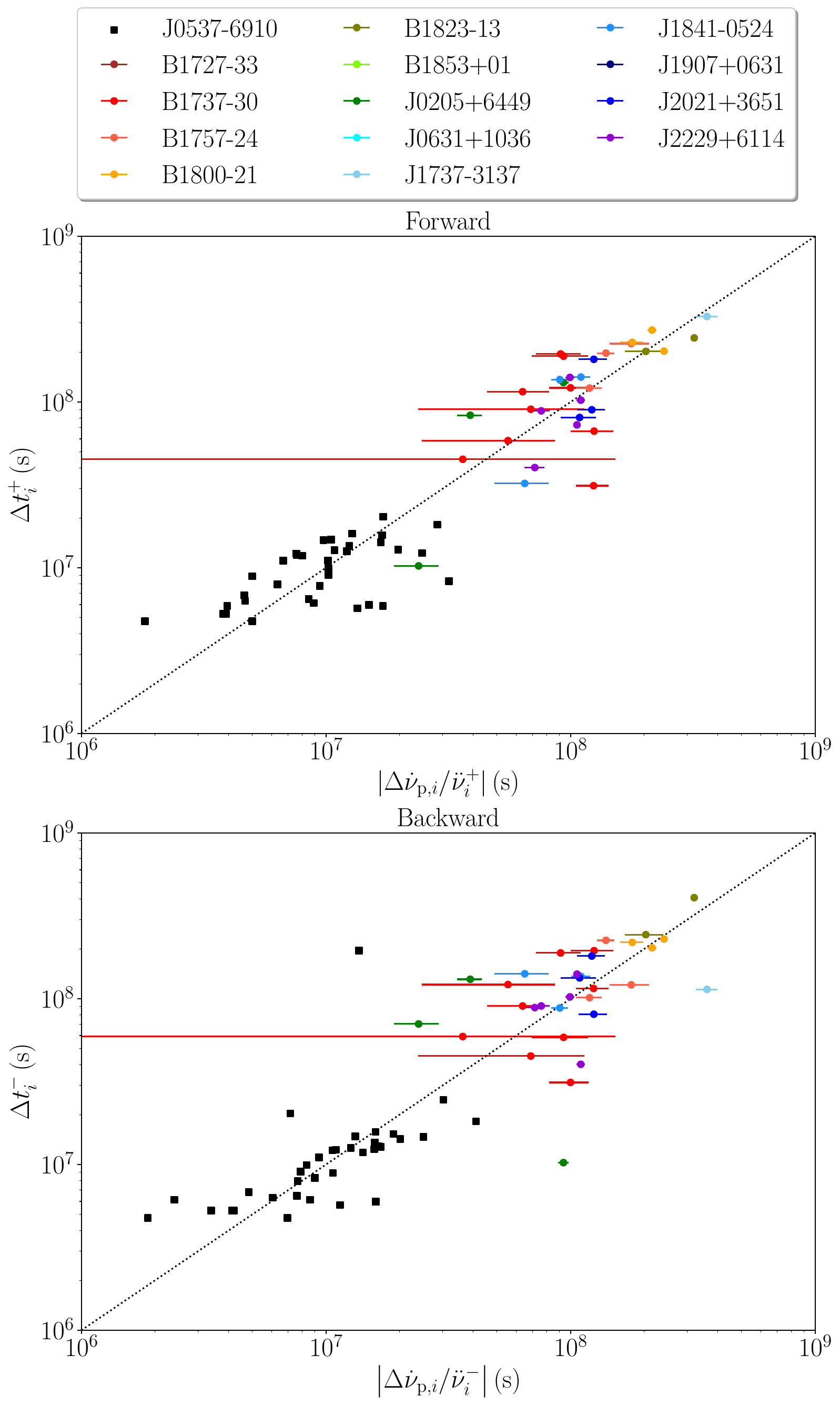}
    \caption{Comparison between the actual waiting time and the waiting time predicted by $\left|\Delta \dot{\nu}_{\mathrm{p},i}/\ddot{\nu}^{\pm}_{i}\right|$. The top panel shows the forward waiting time $\Delta t^{+}$ between target glitch and next glitch, and the $\ddot{\nu}^{+}_{i}$ measured in this range. The bottom panel shows the backward waiting time $\Delta t^{-}_{i}$ between previous glitch and target glitch, and the $\ddot{\nu}^{-}_{i}$ measured in this range. The black dotted line is $\Delta t^{\pm}=\left|\Delta \dot{\nu}_{\mathrm{p}}/\ddot{\nu}^{\pm}_{i}\right|$, and each pulsars are marked with different colors. PSR~J0537$-$6910 is a X-ray pulsar with frequent large glitches, and its data is taken from \citet{antonopoulou_pulsar_2018} and \citet{ho_return_2020, ho_timing_2022}.}
    \label{fig: L_Tg_predict}
\end{figure}

In doing so, the stronger correlation observed in our sample of pulsars is in the case where we ignore glitches with $\Delta\nu/\nu<10^{-7}$ (denoted as the JBO with $\Delta\nu/\nu$ cut-off in Tables \ref{tab: powerlaw} and \ref{tab: waiting}, which includes 13 pulsars). We also took into account changes in $\ddot{\nu}$ that might have occurred after each glitch. The Spearman correlation coefficient between $\left|\Delta \dot{\nu}_{{\mathrm{p}},i}/\ddot{\nu}^{-}_{i}\right|$ and $\Delta t^{-}_{i}$ is 0.54 whilst between $\left|\Delta \dot{\nu}_{{\mathrm{p}},i}/\ddot{\nu}^{+}_{i}\right|$ and $\Delta t^{+}_{i}$ it is 0.70 (Table~\ref{tab: waiting}). We present the data together with the identity line in Figure \ref{fig: L_Tg_predict}. We also showcase data from the X-ray pulsar PSR J0537$-$6910, for which the correlation with backward waiting times is stronger ($\rho_{s}=0.75$, as opposed to $\rho_{s}=0.64$ for forward waiting times). Restricting the sample to `Vela-like' pulsars strengthens the correlations. 

Based on the second panel of Figure \ref{fig: L_Tg_predict}, a (very) rough estimate of the epoch of the next glitch can be made for a given pulsar, as long as we can measure $\Delta\dot{\nu}_{\mathrm{p}}$ at its latest large glitch, and $\ddot{\nu}$ after that. On average, the difference between predicted waiting time and real $\Delta t^+$ is of the order of $\sim100$ days for the `Vela-like' pulsars, and can be up to $\sim1000$ days for others.  However, the standard deviation of this average is around 300 days for the `Vela-like' pulsars, and can be an order of magnitude larger for other pulsars. 

\begin{table}
\begin{center}
\centering
\caption{The Spearman coefficients $\rho_{s}$ and $p$-values for the correlation between waiting time and $|\Delta\dot{\nu}_{\mathrm{p},{i}}/\ddot{\nu}^{\pm}_{i}|$ with cut-off $\Delta\nu/\nu\geq10^{-7}$. See the caption of Table~\ref{tab: powerlaw} for further details.}
\label{tab: waiting}
\begin{tabular}{lrrrr}
\hline
\hline
Dataset & Pulsars & Waiting time &  $\rho_{s}$ & p-value \\
\hline
JBO & All & $\Delta t^{+}_{i}$ & $0.70$ & $9.1\times10^{-8}$ \\
JBO & All & $\Delta t^{-}_{i}$ & $0.54$ & $1.3\times10^{-4}$ \\
JBO & `Vela-like' & $\Delta t^{+}_{i}$ & $0.79$ & $1.5\times10^{-6}$ \\
JBO & `Vela-like' & $\Delta t^{-}_{i}$ & $0.77$ & $4.8\times10^{-6}$ \\
H22 & J0537$-$6910 & $\Delta t^{+}_{i}$ & $0.64$ & $4.8\times10^{-6}$ \\
H22 & J0537$-$6910 & $\Delta t^{-}_{i}$ & $0.75$ & $1.5\times10^{-8}$ \\
\hline
\end{tabular}
\end{center}
\end{table}

\subsection{Exponential recovery}\label{sub: exp}

\begin{figure}
    \centering
    \begin{subfigure}[b]{0.49\textwidth}
        \centering
        \includegraphics[scale=0.31]{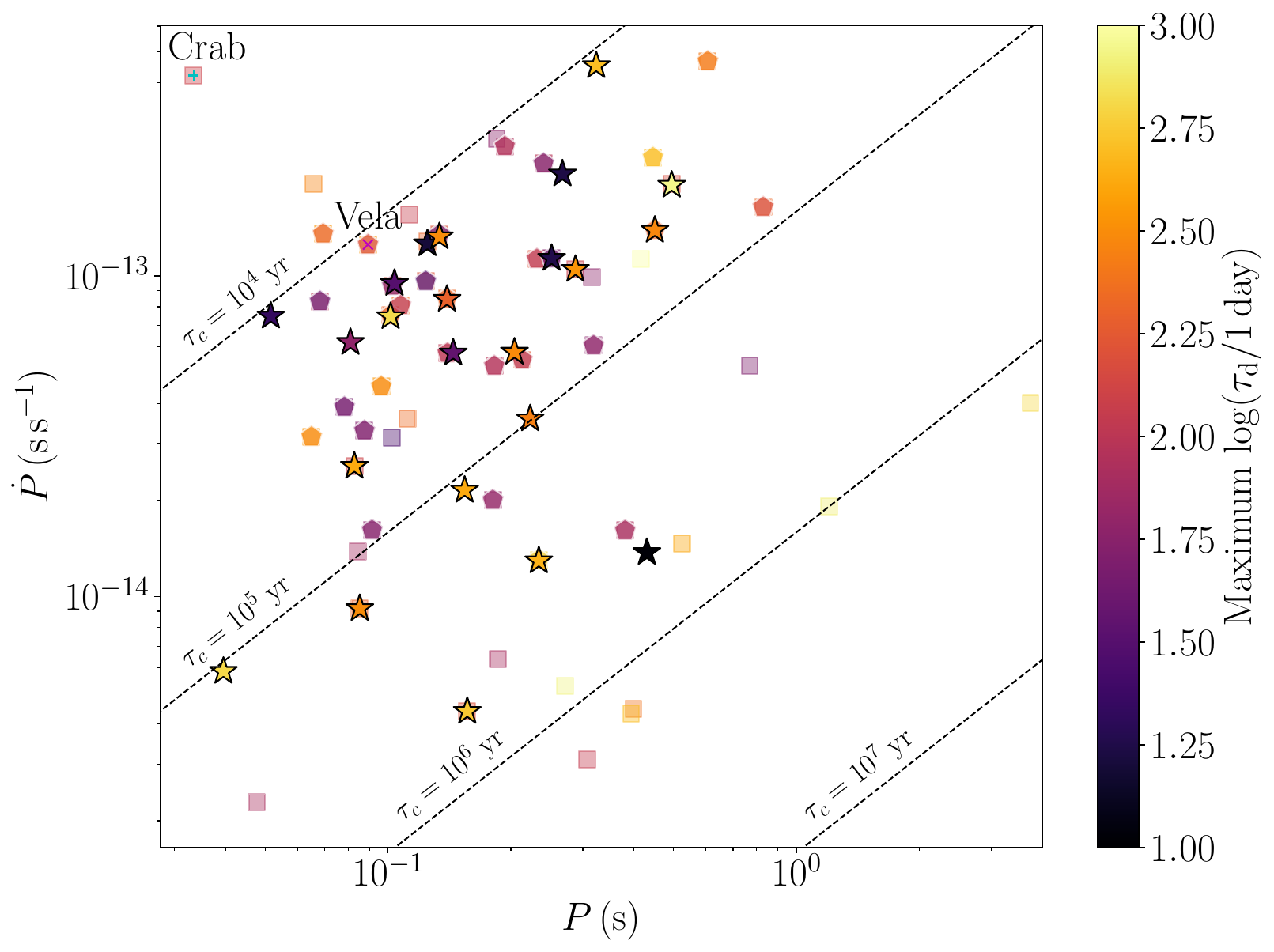}
    \end{subfigure}
    \begin{subfigure}[b]{0.49\textwidth}
        \centering
        \includegraphics[scale=0.31]{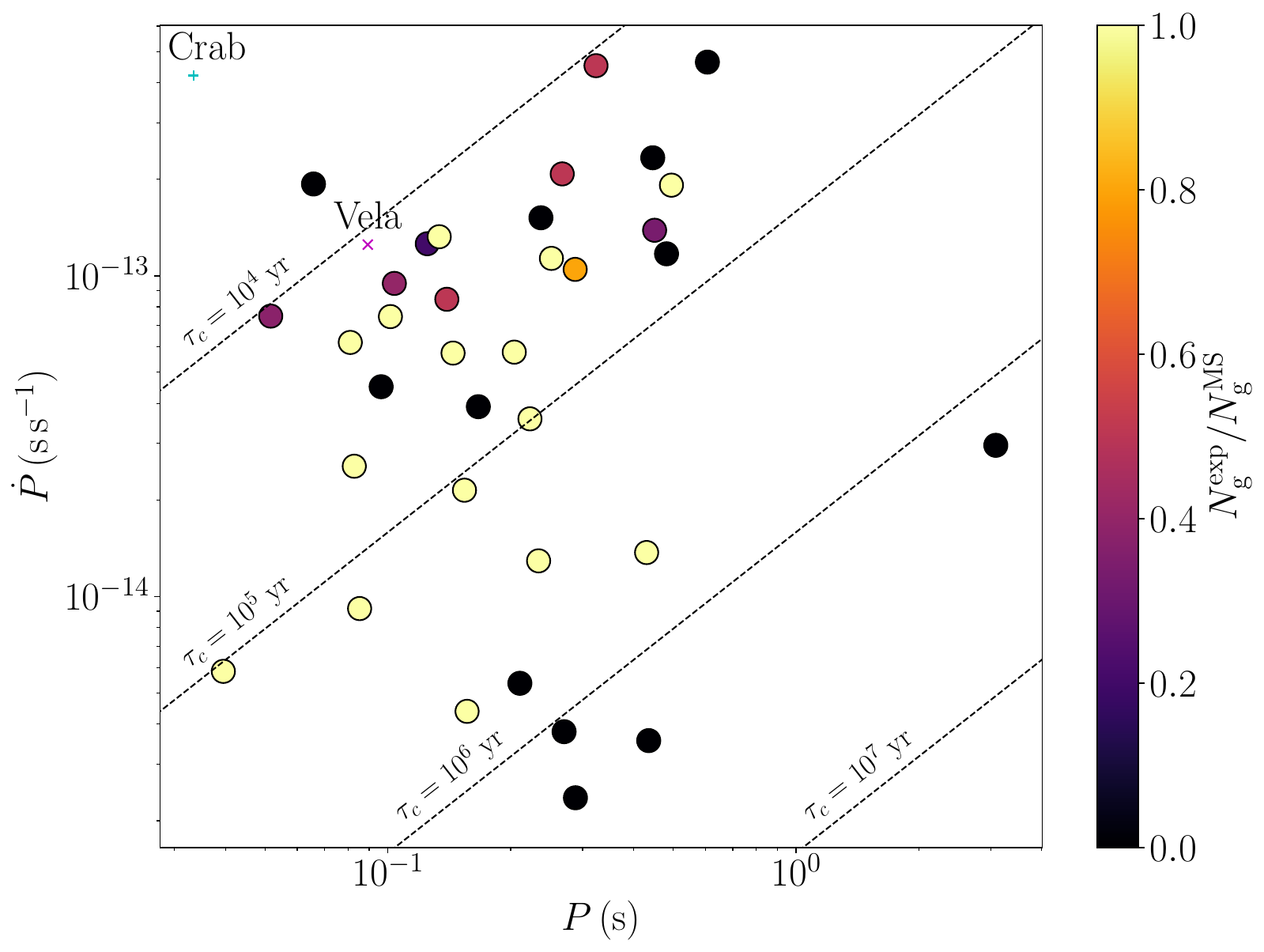}
    \end{subfigure}
    \caption{Top panel: $P-\dot{P}$ diagram showing the largest detected exponential glitch recovery timescale (colour scale) for pulsars in our sample (in stars), those in \citet{lower_impact_2021} (in pentagons), and those from ATNF Pulsar Catalogue (in squares). The black-dashed lines represent lines of constant characteristic age. Bottom panel: $P-\dot{P}$ diagram showing the fraction of glitches for which exponential recoveries are preferred by our model selection for pulsars in our sample.}
    \label{fig: PPdot_maxgltd_exp}
\end{figure}

Our results of glitch model selection for the intermediate to large glitches confirm that exponential relaxation is a common post-glitch feature in our pulsar sample. 
Some of our main findings are summarised in Figure~\ref{fig: PPdot_maxgltd_exp}, in which we place the pulsars of this work on a period-period derivative ($P$-$\dot{P}$) pulsar diagram. The bottom panel illustrates the fraction of glitches of a given pulsar for which model selection favoured the presence of exponential recovery. The top panel shows the maximum relaxation timescale detected for each pulsar in the total of its glitches. We also display the same maximum exponential timescale as reported in previous studies for different pulsar samples. Despite the large uncertainty in measuring exponential timescales, for these pulsars that are shared between our study and others, the results are encouragingly consistent.

It is notable that for five of the oldest pulsars ($\tau_{\mathrm{c}}\gtrsim6\times10^{5}$ yr) we found no evidence of exponential recoveries, but for the next seven old pulsars ($8\times10^{4}\;\mathrm{yr}\lesssim\tau_{\mathrm{c}}\lesssim6\times10^{5}$ yr) all glitches we tested showed exponentials. For the younger pulsar population ($\tau_{\mathrm{c}}\lesssim8\times10^{4}$ yr) the fraction of glitches with detectable exponentials differs from source to source. They moreover present a wide variability of glitch recoveries: for a given pulsar, some glitches might be associated with up to two detectable exponential transients, others with none. We note that it is hard to get away from selection effects when considering the evolution of the fraction of glitches with exponential recoveries. In our work we only consider pulsars with large glitches, and the result may differ if pulsars with smaller glitches are included. However, it can be harder to detect exponential terms in small glitches, so additional care would be needed to include them.

Whilst the relaxation timescales vary from a few days up to years, the amplitudes $\Delta\nu_{\rm d}$ are generally small compared to the total glitch size, as best depicted by the recovery ratio $Q$ in the last column of Table.~\ref{tab: glitch_parameters}. In the majority of glitches the exponentially relaxing amplitudes are at least one order of magnitude smaller than the permanent ones, the two being comparable only in a few glitches such as the 1st glitch of PSR B0355$+$54, the 6th and 17th glitch of PSR J0631$+$1036, the 1st glitch of PSR J1907$+$0631, and the 9th glitch of PSR J2229$+$6114.

\begin{figure}
    \centering
    \begin{subfigure}[b]{0.49\textwidth}
        \centering
        \includegraphics[scale=0.31]{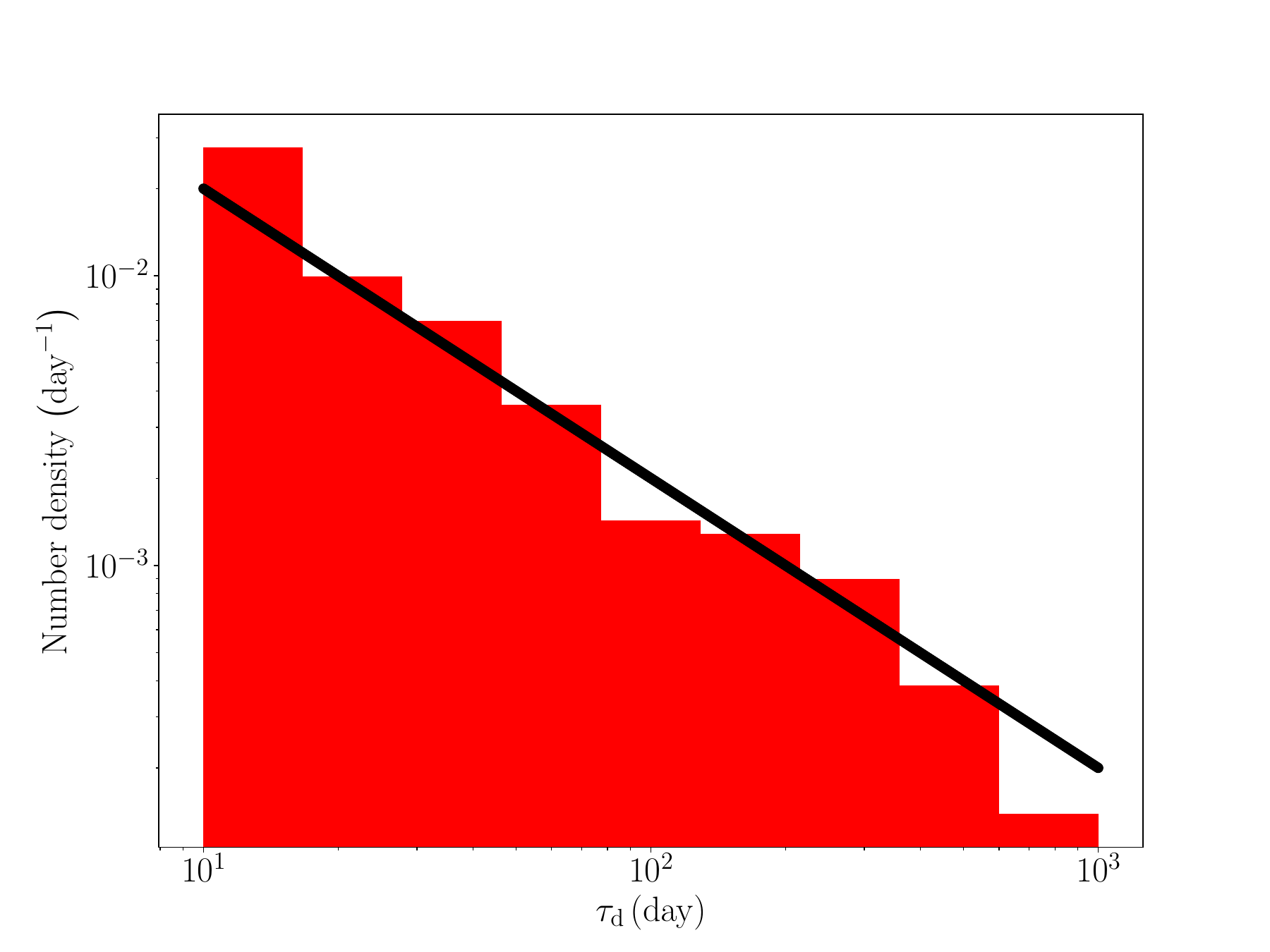}
    \end{subfigure}
    \caption{The probability density function (PDF) histogram of the timescale of all exponential recoveries in the 37 glitches with recoveries in our sample. The black-solid line represent a power law index of $-1$.}
    \label{fig: GLTD_hist}
\end{figure}

In Figure~\ref{fig: GLTD_hist} we construct a histogram of all exponential decaying timescales found in this study. The distribution resembles a power-law, though we remind the reader that our prior range for $\tau_{\rm d}$ sets a 10-day minimum so we do not resolve the distribution for very short timescales. The majority of tested glitches in pulsars with $\tau_{\mathrm{c}}\leq50$ Kyr had decay timescales shorter than 200 days. Longer recovery timescales are more common in older pulsars ($8\times10^{4}\;\mathrm{yr}\lesssim\tau_{\mathrm{c}}\lesssim6\times10^{5}\;\mathrm{yr}$), for which exponential timescales covering all range of Figure~\ref{fig: GLTD_hist} have been measured. Since the older pulsars typically have a single glitch in our data span, we cannot draw firm conclusions of a correlation between recovery timescale and pulsar age.

\begin{figure}
    \centering
    \begin{subfigure}[b]{0.49\textwidth}
        \centering
        \includegraphics[scale=0.31]{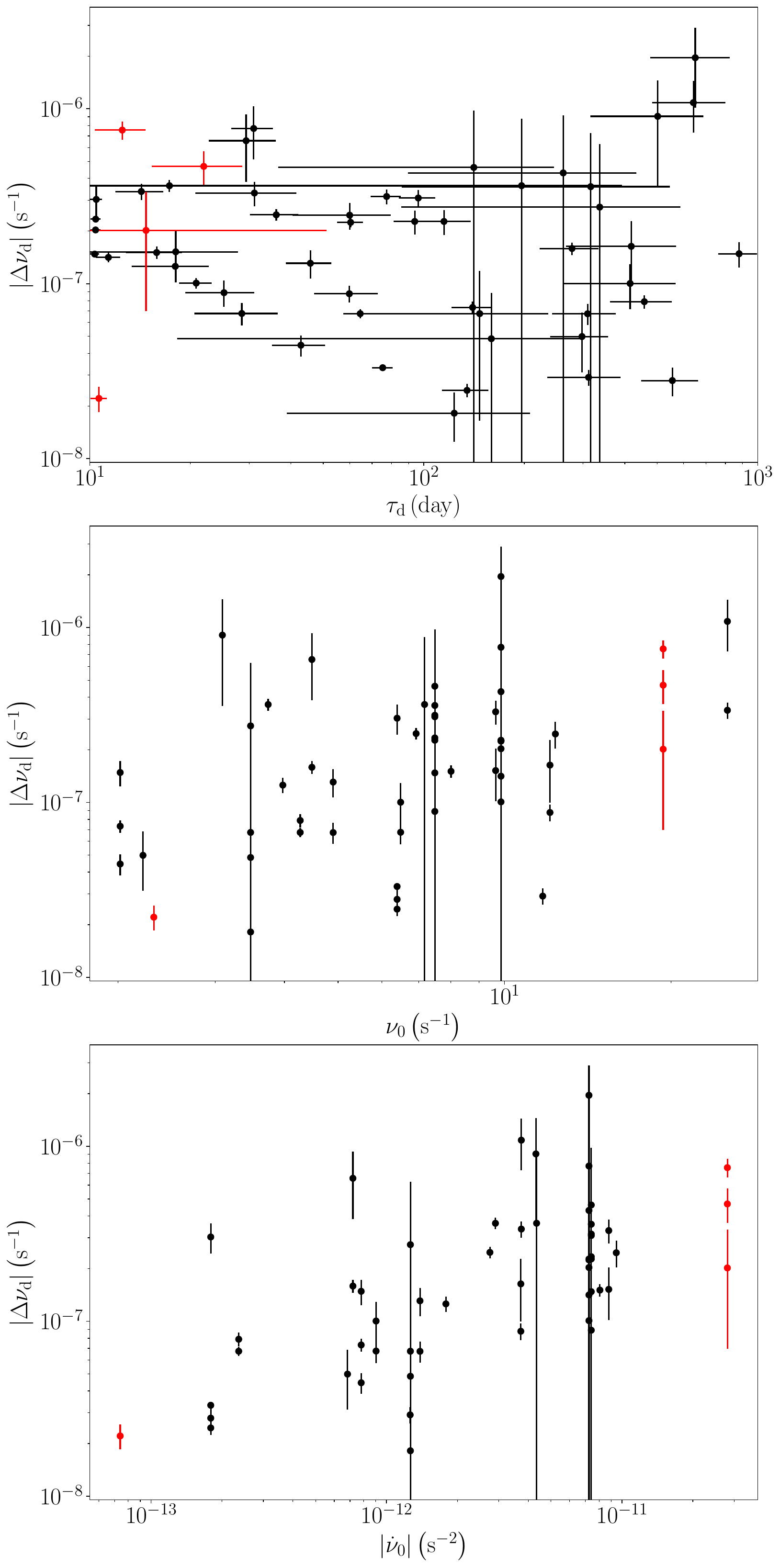}
    \end{subfigure}
    \caption{The correlation between exponential decaying amplitude and other glitch parameters. Black points are the positive exponential terms, and red points represent exponentials with negative amplitudes. Top panel: The timescale of exponential recoveries $\tau_{\mathrm{d}}$ in pulsars from our sample and their corresponding amplitude $|\Delta\nu_{\mathrm{d}}|$. Middle panel: The spin frequency $\nu_{0}$ and the amplitude of exponential recoveries $|\Delta\nu_{\mathrm{d}}|$. Bottom panel: The absolute value of frequency derivative $|\dot{\nu}_{0}|$ and the amplitude of exponential recoveries $|\Delta\nu_{\mathrm{d}}|$.}
    \label{fig: GLF0D}
\end{figure}

To strengthen our understanding of the exponential recoveries, we tested further relations between the exponential recovery parameters and other glitch and pulsar parameters, however no strong trends emerged. The most interesting results are presented in Figure~\ref{fig: GLF0D}, which shows how $\Delta\nu_{\mathrm{d}}$ varies with $\tau_{\mathrm{d}}$ (top panel), with the pulsar spin frequency $\nu_{0}$ (middle panel), and with the spin-down rate $|\dot{\nu}_{0}|$ (bottom panel). There is no clear correlation between the amplitude of an exponential component and its decay timescale across the pulsar sample.  The weak tendency seen in the top panel of Figure~\ref{fig: GLF0D}, for shorter timescales to be associated with larger $\Delta{\nu}_{\mathrm{d}}$ could be possibly attributed to the degeneracy between these parameters in the fitting procedure. There is, however, some correlation between $\Delta{\nu}_{\mathrm{d}}$ and $\nu_{0}$ ($\rho_{s}=0.42$ of $p$-value 0.009), and an even clearer correlation between $\Delta{\nu}_{\mathrm{d}}$ and $|\dot{\nu}_{0}|$ ($\rho_{s}=0.58$  with $p$-value 0.0002).

\section{Conclusions}\label{sec: conclu}
In this paper, we examine 157 glitches in 35 pulsars among the 800 pulsars routinely monitored by the JBO timing program. Our sample comprises of pulsars with at least one large glitch ($\Delta\nu/\nu\geq10^{-6}$). For glitches with $\Delta\nu/\nu\geq10^{-7}$ (and for some smaller glitches when appropriate) we used Bayesian inference and model selection to analyse the post-glitch recoveries. The remaining small glitches were characterised as pure steps in $\nu$ and $\dot{\nu}$ and their parameters measured by a linear least square fit. We present the updated glitch epochs, fractional sizes $\Delta\nu/\nu$, and $\Delta\dot{\nu}/\dot{\nu}$ for all glitches, along with exponentially decaying amplitudes $\Delta\dot{\nu}_{\mathrm{d}_\mathrm{i}}$ and decay timescales $\tau_{\mathrm{d}_\mathrm{i}}$ for glitches with evidence of exponential recoveries. 

We also provide updated measurements of the second frequency derivative $\ddot{\nu}$ that describes the linear evolution of the spin-down rate between glitches. From these, interglitch braking indices $n$ can be calculated and contrasted with the $n_{\mathrm{b}}=3$ prediction for pure electromagnetic dipole braking. In general, internal torques appear to dominate the interglitch evolution resulting in values of $n$ considerably larger than 3. We also observe a strong correlation between $\ddot{\nu}$ and $\dot{\nu}$. Upon fitting $\ddot{\nu}$ as a power law function of $\nu$ and $\dot{\nu}$ we find that only the youngest pulsars (with $\tau_{\mathrm{c}}\approx10^{2}\sim10^{4}\mathrm{yr}$) such as the Crab pulsar and PSR B1509$-$58 may be expected to show $n_{\mathrm{b}}\simeq3$. 

Out of the 85 glitches for which the Bayesian model selection was performed, 37 show Bayes factors in support of at least one exponentially decaying component. In most cases the best-fit model includes a single exponential term, but for 15 glitches two exponential terms were preferred, and for 2 glitches three exponential terms were preferred. Pulsars with higher spin-down rate $|\dot{\nu}|$ typically present larger decaying amplitudes $\Delta\nu_{\mathrm{d}}$ at their glitches. Even though our sample is small, the emerging distribution of exponential timescales appears powerlaw-like, therefore the common monitoring programs that observe pulsars biweekly or monthly are not adept at capturing the full post-glitch recovery signature.

We confirmed the power law relation between $\ddot{\nu}_{\mathrm{i}}$ and $|\Delta\dot{\nu}_{i}|/\Delta t^{+}_{i}$, where $\Delta t^{+}_{i}$ is the `forward' waiting time after glitch $i$ until glitch $i+1$ takes place, proposed by \citet{lower_impact_2021}. However we also find that the correlation holds nearly as well when the `backward' waiting time $\Delta t^{-}_{i}$ is used instead. According to our results, a major contribution to the instantaneous changes $|\Delta\dot{\nu}|$ comes from the persisting component $|\Delta\dot{\nu}_{\mathrm{p}}|$ which correlates with its rate of recovery $\ddot{\nu}$. An even stronger correlation exists between $|\Delta\nu_{\mathrm{p}}|$ and $|\dot{\nu}|$. To cancel out the effects caused by the $|\ddot{\nu}|$-$|\dot{\nu}|$ and $\Delta\nu_{\mathrm{p}}$-$|\dot{\nu}|$ correlations we compare the ratio $\left|\Delta \dot{\nu}_{\mathrm{p}}/\ddot{\nu}\right|$ to waiting times $\Delta t$. The data for the entire pulsar sample loosely follows the $\left|\Delta \dot{\nu}_{{\mathrm{p}},i}/\ddot{\nu}^{\pm}_{i}\right|=\Delta t^{\pm}_{i}$ line, however the scatter is considerable and most individual pulsars have too few glitches to confirm if the correlation holds for each one of them. Therefore, currently, the power of this relation to predict glitch epochs with accuracy is limited; it can however serve as a general guide for observations.

Glitch characterisation and parameters definition with the use of Bayesian model selection continues, as new glitches are being detected by the JBO timing programme. Our results revealed a range of post-glitch recovery timescales, from few days up to several hundred days. Regular pulsar monitoring is necessary for detecting new glitches and to resolve their subsequent long-term recoveries. It is also, however, critical to design observing programs able to detect glitches promptly and adapt the monitoring strategy in real-time in order to resolve the early phase of post-glitch relaxation.

\section*{Data availability}\label{sec: avail}
The data underlying the work in this paper are available upon reasonable request.

\section*{Acknowledgements}\label{sec: acknow}
Pulsar research at Jodrell Bank is supported by a consolidated grant (ST/T000414/1; ST/X001229/1) from the UK Science and Technology Facilities Council (STFC). D.A. acknowledges support from an EPSRC/STFC fellowship (EP/T017325/1). YL acknowledges Xiaoxi Song for discussion of coding.




\bibliographystyle{mnras}
\bibliography{reference} 




\appendix

\section{Parameters used in figures}\label{sec: appendix}

\begin{table}
	\centering
	\caption{The values of reference epoch ($t_{\mathrm{f}}$), reference spin frequency ($\nu_{\mathrm{f}}$), and reference derivatives ($\dot{\nu}_{\mathrm{f}}$, and $\ddot{\nu}_{\mathrm{f}}=\ddot{\nu}_{0}$), used for plotting Figures~\ref{fig: nu_nudot_J0205-B0919} to \ref{fig: nu_nudot_B1930-B2334}.}
	\label{tab: plot_parameters}
    \begin{tabular}{lrrrr}
\hline
\hline
Pulsar name & $t_{\mathrm{f}}$ & $\nu_{\mathrm{f}}$ & $\dot{\nu}_{\mathrm{f}}$ & $\ddot{\nu}_{\mathrm{f}}$ \\
PSR & (MJD) & ($\mathrm{s}^{-1}$) & ($10^{-12}\;\mathrm{s}^{-2}$) & ($10^{-23}\;\mathrm{s}^{-3}$) \\
\hline
 J0205$+$6449 &    54684.9 & 15.214801 & $-$44.778 &  75.07509 \\
   B0355$+$54 &    45119.4 &  6.394623 &  $-$0.179 &   0.02778 \\
 J0611$+$1436 &    56070.3 &  3.699189 &  $-$0.052 &  $-$1.04269 \\
 J0631$+$1036 &    51041.6 &  3.475034 &  $-$1.265 &   0.45351 \\
 J0729$-$1448 &    54229.7 &  3.973193 &  $-$1.788 &   0.34882 \\
   B0919$+$06 &    52117.3 &  2.322211 &  $-$0.074 &   0.01950 \\
   B1727$-$33 &    52009.4 &  7.169822 &  $-$4.351 &   6.86759 \\
   B1737$-$30 &    47228.6 &  1.648570 &  $-$1.266 &   1.57019 \\
 J1737$-$3137 &    51694.4 &  2.220062 &  $-$0.684 &   0.30977 \\
 J1740$+$1000 &    54728.7 &  6.489593 &  $-$0.903 &   0.17690 \\
   B1754$-$24 &    55491.9 &  4.271555 &  $-$0.236 &   0.01475 \\
   B1757$-$24 &    48804.1 &  8.007414 &  $-$8.175 &  30.57507 \\
   B1800$-$21 &    48102.3 &  7.484960 &  $-$7.489 &  22.58134 \\
 J1806$-$2125 &    54266.5 &  2.075460 &  $-$0.504 &   0.31466 \\
 J1809$-$1917 &    53417.0 & 12.084432 &  $-$3.724 &   2.80846 \\
   B1821$-$11 &    52404.6 &  2.294842 &  $-$0.019 &   0.00425 \\
   B1823$-$13 &    49823.6 &  9.856202 &  $-$7.277 &  16.27654 \\
   B1830$-$08 &    46517.8 & 11.725904 &  $-$1.260 &  $-$0.16820 \\
 J1837$-$0604 &    56119.4 & 10.383004 &  $-$4.855 &   4.32971 \\
 J1841$-$0345 &    57108.7 &  4.899669 &  $-$1.387 &   0.65018 \\
 J1841$-$0524 &    53293.2 &  2.243321 &  $-$1.175 &   1.07386 \\
 J1842$+$0257 &    57113.3 &  0.323806 &  $-$0.003 &   0.00025 \\
 J1850$-$0026 &    55534.3 &  6.001024 &  $-$1.407 &   1.21201 \\
   B1853$+$01 &    52798.5 &  3.738403 &  $-$2.904 &   3.52870 \\
 J1856$+$0245 &    56345.8 & 12.358758 &  $-$9.457 &  24.63916 \\
   B1859$+$01 &    52277.1 &  3.469575 &  $-$0.028 &  $-$0.00240 \\
 J1907$+$0631 &    57044.3 &  3.089528 &  $-$4.313 &   6.34664 \\
 J1909$+$0749 &    57353.6 &  4.215960 &  $-$2.696 &   2.29685 \\
 J1909$+$0912 &    54845.2 &  4.485117 &  $-$0.720 &   0.11405 \\
 J1921$+$0812 &    55614.4 &  4.747222 &  $-$0.121 &   0.05491 \\
   B1930$+$22 &    52848.3 &  6.921494 &  $-$2.749 &   1.99691 \\
   B1951$+$32 &    47890.0 & 25.297111 &  $-$3.742 &   0.86522 \\
 J2021$+$3651 &    55916.0 &  9.638468 &  $-$8.793 &  41.63839 \\
 J2229$+$6114 &    52545.5 & 19.369566 & $-$29.175 & 336.44432 \\
   B2334$+$61 &    53452.8 &  2.018725 &  $-$0.781 &   0.33489 \\
\hline
    \end{tabular}
\end{table}

\section{Evidence of model selection}\label{sec: appdx}

\begin{table*}
\begin{center}
\centering
\caption{The logarithm of Bayes factor (ratio of the Bayesian evidence) between the 7 models we tested and the best model we adopted for the 85 glitches in 35 pulsars included in model selection. See Table~\ref{tab: models} for definitions of Model 1-7. The last column is the model we take for each glitch in the analysis of this work. The models with the largest evidence for each glitch are marked in bold. The five small glitches with asterisks are also included in the model selection process.}
\label{tab: model_evidence}
\begin{threeparttable}
\begin{tabular}{lrrrrrrrrrr}
\hline
\hline
Pulsar name & Gl & Model 1 & Model 2 & Model 3 & Model 4 & Model 5 & Model 6 & Model 7 & Best model \\
PSR & No. &  &  &  &  &  &  &  &   \\
\hline
J0205+6449 & 1  & $-$5.88(1) & 0 & \textbf{0.45(1)} & $-$1.69(2) & -- & -- & -- & Model 2 \\
         & 2  & 0 & \textbf{0.34(4)} & $-$0.87(4) & $-$0.74(3) & -- & -- & -- & Model 1 \\
         & 3  & 0 & 1.33(2) & \textbf{2.35(1)} & 0.15(2) & -- & -- & -- & Model 1 \\
         & 5  & \textbf{0} & $-$2.46(6) & $-$1.44(4) & $-$3.99(2) & -- & -- & -- & Model 1 \\
         & 7  & $-$5.97(9) & \textbf{0} & $-$1.47(7) & $-$2.75(5) & -- & -- & -- & Model 2 \\
B0355+54 & 1*  & $-$47.21(4) & $-$46.61(3) & 0 & 0.12(4) & \textbf{0.1(3)} & $-$0.41(3) & -- & Model 3 \\
         & 2  & $-$66.10(8) & $-$70.90(7) & $-$13.02(8) & $-$39.34(7) & $-$5.47(7) & $-$5.57(7) & \textbf{0} & Model 7 \\
J0611+1436 & 1  & 0 & \textbf{0.1(4)} & $-$5.0(2) & $-$5.2(2) & -- & -- & -- & Model 1 \\
J0631+1036 & 4  & 0.49(1) & \textbf{0.8(3)} & 0 & $-$0.42(3) & -- & -- & -- & Model 3 \\
         & 6*  & $-$2.85(2) & $-$2.54(2) & $-$0.78(7) & \textbf{0} & -- & -- & -- & Model 4 \\
         & 12* & 0 & 0.17(2) & \textbf{1.13(7)} & 0.94(2) & -- & -- & -- & Model 1 \\
         & 15 & $-$3.20(5) & $-$3.61(5) & \textbf{0.12(5)} & 0 & -- & -- & -- & Model 4 \\
         & 17* & $-$2.3(1) & $-$2.2(1) & \textbf{0} & $-$0.6(1) & -- & -- & -- & Model 3 \\
J0729$-$1448 & 5  & $-$41.11(1) & $-$42.53(1) & \textbf{0} & $-$2.82(1) & $-$0.4(6) & $-$3.81(2) & -- & Model 3 \\
B0919+06 & 1  & $-$20.44(3) & $-$20.7(1) & \textbf{0} & $-$2.2(1) & -- & -- & -- & Model 3 \\
B1727$-$33 & 1  & \textbf{0} & $-$4.09(9) & $-$3.74(9) & $-$7.30(8) & -- & -- & -- & Model 1 \\
         & 2  & $-$2.77(1) & $-$3.91(2) & \textbf{0} & $-$1.97(2) & $-$3.29(2) & $-$5.06(4) & -- & Model 3 \\
B1737$-$30 & 1  & \textbf{0} & $-$1.19(2) & $-$1.59(3) & $-$2.37(2) & -- & -- & -- & Model 1 \\
         & 5  & \textbf{0} & $-$0.98(4) & $-$1.64(5) & $-$2.18(4) & -- & -- & -- & Model 1 \\
         & 7  & \textbf{0} & $-$2.18(1) & $-$1.00(2) & $-$2.53(4) & -- & -- & -- & Model 1 \\
         & 11 & \textbf{0} & $-$2.80(8) & $-$1.9(1) & $-$4.45(5) & -- & -- & -- & Model 1 \\
         & 14 & \textbf{0} & $-$2.5(1) & $-$2.11(9) & $-$2.45(2) & -- & -- & -- & Model 1 \\
         & 15 & \textbf{0} & $-$1.83(1) & $-$4.25(2) & $-$2.85(3) & -- & -- & -- & Model 1 \\
         & 21 & 0 & $-$2.10(3) & \textbf{2.01(4)} & 0.76(4) & -- & -- & -- & Model 1 \\
         & 26 & \textbf{0} & $-$2.28(9) & $-$3.15(7) & $-$5.1(1) & -- & -- & -- & Model 1 \\
         & 32 & \textbf{0} & $-$2.25(9) & $-$3.63(9) & $-$5.0(4) & -- & -- & -- & Model 1 \\
         & 35 & \textbf{0} & $-$2.26(2) & $-$0.50(4) & $-$0.54(2) & -- & -- & -- & Model 1 \\
         & 36 & \textbf{0} & $-$1.65(1) & $-$3.31(1) & $-$3.48(4) & -- & -- & -- & Model 1 \\
J1737$-$3137 & 2  & $-$6.88(4) & $-$6.03(8) & \textbf{0} & $-$3.24(4) & -- & -- & -- & Model 3 \\
         & 3  & 0 & $-$1.48(3) & \textbf{0.5(1}) & $-$0.53(3) & -- & -- & -- & Model 1 \\
         & 5  & \textbf{0} & $-$1.4(6) & $-$4.86(1) & $-$6.55(5) & -- & -- & -- & Model 1 \\
J1740+1000 & 2  & $-$41.08(3) & $-$42.53(3) & $-$3.73(4) & $-$7.07(4) & \textbf{0} & $-$4.50(4) & $-$4.74(9) & Model 5 \\
B1754$-$24 & 1  & $-$156(1) & $-$158.7(4) & $-$18.0(4) & $-$14.1(4) & \textbf{0} & $-$1.1(4) & $-$2.0(4) & Model 5 \\
B1757$-$24 & 1  & \textbf{0} & $-$1.44(7) & $-$2.12(4) & $-$4.16(4) & -- & -- & -- & Model 1 \\
         & 2  & \textbf{0} & $-$4.0(7) & $-$4.3(7) & $-$6.5(7) & -- & -- & -- & Model 1 \\
         & 3  & 0 & \textbf{3(1)} & $-$3.08(6) & $-$3.73(4) & -- & -- & -- & Model 1 \\
         & 5  & $-$39.70(3) & $-$41.25(4) & 0 & $-$1.33(6) & \textbf{0.50(7)} & $-$3.01(2) & -- & Model 3 \\
         & 6  & \textbf{0} & $-$2.2(1) & $-$3.6(1) & $-$6.3(1) & -- & -- & -- & Model 1 \\
B1800$-$21 & 1  & $-$22.92(2) & $-$22.4(3) & \textbf{0} & $-$5.98(3) & $-$4.34(2) & $-$8.33(5) & -- & Model 3 \\
         & 3  & $-$52.69(2) & $-$50.86(2) & $-$5.39(3) & $-$18.40(3) & \textbf{0} & $-$3.67(3) & $-$2.9(5) & Model 5 \\
         & 4  & $-$336.95(1) & $-$334.23(1) & $-$107.62(4) & $-$136.16(4) & \textbf{0} & $-$4.10(1) & $-$2.2(1) & Model 5 \\
         & 5  & $-$379.30(2) & $-$374.88(7) & $-$97.90(2) & $-$59.54(3) & \textbf{0} & $-$2.61(1) & $-$2.77(1) & Model 5 \\
         & 6  & $-$14.91(1) & $-$3.28(1) & \textbf{0} & $-$0.03(3) & $-$1.8(2) & $-$2.44(6) & -- & Model 3 \\
J1806$-$2125 & 1  & \textbf{0} & $-$0.21(9) & $-$5.29(9) & $-$5.27(9) & -- & -- & -- & Model 1 \\
J1809$-$1917 & 1  & $-$74.53(4) & $-$68.17(4) & $-$3.78(5) & $-$5.2(5) & \textbf{0} & $-$2.57(4) & $-$5.06(6) & Model 5 \\
B1821$-$11 & 1  & \textbf{0} & $-$2.15(4) & $-$5.78(5) & $-$7.3(1) & -- & -- & -- & Model 1 \\
B1823$-$13 & 1  & $-$78.75(4) & $-$73.49(2) & $-$19.91(3) & $-$39.79(3) & \textbf{0} & $-$0.62(4) & $-$4.32(5) & Model 5 \\
         & 4  & $-$290.98(2) & $-$290.39(4) & $-$78.44(4) & $-$76.97(3) & \textbf{0} & $-$0.20(2) & $-$3.82(4) & Model 5 \\
         & 5  & $-$159.46(2) & $-$152.51(2) & $-$7.36(5) & $-$6.39(8) & \textbf{0} & $-$3.08(3) & $-$2.9(5) & Model 5 \\
         & 6  & $-$136.23(1) & $-$132.12(4) & $-$30.27(2) & $-$21.30(8) & \textbf{0} & $-$1.98(4) & $-$2.68(3) & Model 5 \\
B1830$-$08 & 2  & $-$12.3(2) & $-$12.6(2) & \textbf{0} & $-$0.7(2) & $-$0.9(2) & $-$1.2(2) & -- & Model 3 \\
J1837$-$0604 & 1  & $-$4.1(1) & \textbf{0} & $-$3.11(2) & $-$2.8(1) & -- & -- & -- & Model 2 \\
J1841$-$0345 & 1  & $-$98.6(2) & $-$95.7(4) & $-$11.8(2) & $-$15.4(2) & \textbf{0} & $-$3.3(2) & $-$5.0(2) & Model 5 \\
J1841$-$0524 & 3  & \textbf{0} & $-$2.0(1) & $-$4.42(9) & $-$6.79(9) & -- & -- & -- & Model 1 \\
         & 4  & \textbf{0} & $-$3.22(6) & $-$3.53(2) & $-$5.83(2) & -- & -- & -- & Model 1 \\
         & 6  & \textbf{0} & $-$2.64(8) & $-$3.74(9) & $-$5.27(8) & -- & -- & -- & Model 1 \\
         & 7  & \textbf{0} & $-$1.59(4) & $-$2.43(4) & $-$3.30(4) & -- & -- & -- & Model 1 \\
         & 8  & \textbf{0} & 0.08(7) & $-$0.67(8) & $-$0.6(1) & -- & -- & -- & Model 1 \\
J1842+0257 & 1  & \textbf{0} & $-$5.37(4) & $-$20.77(4) & $-$15.75(4) & -- & -- & -- & Model 1 \\
J1850$-$0026 & 4  & \textbf{0} & $-$1.93(8) & $-$4.99(4) & $-$6.34(3) & -- & -- & -- & Model 1 \\
B1853+01 & 1  & $-$108.1(2) & $-$108.1(2) & 0 & $-$2.1(2) & \textbf{1.5(2)} & $-$0.3(2) & -- & Model 3 \\
         & 2  & 0 & $-$1.06(1) & \textbf{0.86(7)} & $-$0.58(2) & $-$2.38(5) & $-$3.22(3) & -- & Model 1 \\
\end{tabular}
\end{threeparttable}
\end{center}
\end{table*}
\begin{table*}
\begin{center}
\centering
\contcaption{}
\begin{threeparttable}
\begin{tabular}{lrrrrrrrrrrrr}
\hline
\hline
Pulsar name & Gl & Model 1 & Model 2 & Model 3 & Model 4 & Model 5 & Model 6 & Model 7 & Best model \\
PSR & No. &  &  &  &  &  &  &  &   \\
\hline
J1856+0245 & 1  & $-$21.2(3) & $-$23.1(3) & \textbf{0} & $-$4.2(3) & $-$2.0(3) & $-$5.5(3) & -- & Model 3 \\
B1859+01 & 3  & \textbf{0} & $-$0.09(5) & $-$6.10(6) & $-$6.28(5) & -- & -- & -- & Model 1 \\
J1907+0631 & 1  & $-$2.12(4) & $-$2.15(4) & \textbf{0} & $-$1.54(5) & $-$4.51(6) & $-$6.30(4) & -- & Model 3 \\
         & 2  & \textbf{0} & $-$0.9(2) & $-$3.0(2) & $-$3.4(2) & -- & -- & -- & Model 1 \\
J1909+0749 & 1  & \textbf{0} & $-$2.88(3) & $-$2.90(9) & $-$4.17(4) & -- & -- & -- & Model 1 \\
J1909+0912 & 2  & $-$43.36(8) & $-$43.5(1) & $-$7.23(7) & $-$5.64(6) & 0 & \textbf{0.63(9)} & $-$6.07(7) & Model 5 \\
J1921+0812 & 1  & 0 & \textbf{1.02(1)} & $-$5.06(3) & $-$5.14(5) & -- & -- & -- & Model 1 \\
B1930+22 & 1  & $-$59.75(5) & $-$59.0(5) & \textbf{0} & $-$2.6(5) & $-$2.6(2) & $-$5.9(1) & -- & Model 3 \\
B1951+32 & 6  & $-$63.00(3) & $-$63.44(2) & $-$4.90(4) & $-$6.3(1) & \textbf{0} & $-$0.01(2) & $-$3.79(3) & Model 5 \\
J2021+3651 & 1  & 0 & \textbf{0.0(2)} & $-$0.65(4) & $-$0.77(4) & -- & -- & -- & Model 1 \\
         & 2  & \textbf{0} & $-$1.69(9) & $-$1.01(6) & $-$3.21(8) & -- & -- & -- & Model 1 \\
         & 3  & $-$8.00(1) & $-$8.64(1) & 0 & \textbf{1.21(2)} & $-$1.51(2) & $-$1.50(1) & -- & Model 3 \\
         & 4  & $-$17.58(4) & $-$18.78(2) & \textbf{0} & $-$2.96(2) & $-$2.07(2) & $-$3.6(1) & -- & Model 3 \\
         & 5  & \textbf{0} & $-$1.1(2) & $-$1.6(2) & $-$2.7(2) & -- & -- & -- & Model 1 \\
J2229+6114 & 1  & \textbf{0} & $-$1.63(1) & $-$0.6(1) & $-$2.99(6) & -- & -- & -- & Model 1 \\
         & 2  & $-$17.4(6) & $-$16.7(6) & \textbf{0} & $-$3.2(6) & -- & -- & -- & Model 3 \\
         & 4  & $-$4.06(1) & $-$5.47(1) & 0 & \textbf{1.44(2)} & -- & -- & -- & Model 3 \\
         & 5  & \textbf{0} & $-$1.74(1) & $-$1.46(8) & $-$3.11(5) & -- & -- & -- & Model 1 \\
         & 6*  & 0 & $-$2.38(2) & \textbf{0.11(4)} & $-$0.39(4) & -- & -- & -- & Model 1 \\
         & 7  & \textbf{0} & $-$2.1(3) & $-$1.5(2) & $-$3.7(2) & -- & -- & -- & Model 1 \\
         & 8  & \textbf{0} & $-$1.59(1) & $-$1.53(2) & $-$3.07(4) & -- & -- & -- & Model 1 \\
         & 9  & $-$28.40(6) & $-$29.0(1) & 0 & $-$1.33(7) & \textbf{1.09(7)} & $-$0.72(7) & -- & Model 3 \\
B2334+61 & 1  & $-$250.35(3) & $-$244.77(3) & $-$42.99(3) & $-$36.88(5) & $-$2.77(3) & $-$3.36(3) & \textbf{0} & Model 7 \\
\hline
\end{tabular}
\end{threeparttable}
\end{center}
\end{table*}




\bsp	
\label{lastpage}
\end{document}